\documentclass[iop,numberedappendix,twocolappendix]{emulateapj}

\usepackage{float} 
\usepackage{natbib}
\usepackage{graphicx}
\usepackage{color}

\usepackage{rotating}
\usepackage{lineno}
\graphicspath{{plots_dcpg/}}

\usepackage[breaklinks,colorlinks,citecolor=blue,backref=section]{hyperref} 

\def\,{\thinspace}
\def\lsim{\mathrel{\raise .4ex\hbox{\rlap{$<$}\lower 1.2ex\hbox{$\sim$}}}}
\def\gsim{\mathrel{\raise .4ex\hbox{\rlap{$>$}\lower 1.2ex\hbox{$\sim$}}}}

\def\simprop{\mathrel{\raise .4ex\hbox{\rlap{$\propto$}\lower 1.2ex\hbox{$\sim$}}}}
\def\deg{\ifmmode^\circ\else$^\circ$\fi}
\def\pdeg{\ifmmode $\setbox0=\hbox{$^{\circ}$}\rlap{\hskip.11\wd0 .}$^{\circ}
          \else \setbox0=\hbox{$^{\circ}$}\rlap{\hskip.11\wd0 .}$^{\circ}$\fi}
\def\arcs{\ifmmode {^{\scriptstyle\prime\prime}}
          \else $^{\scriptstyle\prime\prime}$\fi}
\def\arcm{\ifmmode {^{\scriptstyle\prime}}
          \else $^{\scriptstyle\prime}$\fi}
\newdimen\sa  \newdimen\sb
\def\parcs{\sa=.07em \sb=.03em
     \ifmmode \hbox{\rlap{.}}^{\scriptstyle\prime\kern -\sb\prime}\hbox{\kern -\sa}
     \else \rlap{.}$^{\scriptstyle\prime\kern -\sb\prime}$\kern -\sa\fi}
\def\parcm{\sa=.08em \sb=.03em
     \ifmmode \hbox{\rlap{.}\kern\sa}^{\scriptstyle\prime}\hbox{\kern-\sb}
     \else \rlap{.}\kern\sa$^{\scriptstyle\prime}$\kern-\sb\fi}
\def\GHz{\ifmmode $\,GHz$\else \,GHz\fi}
\def\MJysr{\ifmmode \,$MJy\,sr\mo$\else \,MJy\,sr\mo\fi}
\def\microns{\ifmmode \,\mu$m$\else \,$\mu$m\fi}
\def\micron{\microns}
\def\kms{\ifmmode $\,km\,s$^{-1}\else \,km\,s$^{-1}$\fi}
\providecommand{\sorthelp}[1]{}

\newcommand{\fwhm}{{\rm FWHM}}
\newcommand{\nep}{NEP}

\newcommand{\Tra}{T_{\rm a}}
\newcommand{\Trb}{T_{\rm b}}
\newcommand{\Trs}{T_{\rm s}}
\newcommand{\Trsys}{T_{\rm sys}}
\newcommand{\DeltaV}{\Delta v_{\rm shift}}

\newcommand{\vrl}{v_{\rm LSR}}

\newcommand{\lon}{\ell}

\newcommand{\expon}{\gamma}

\newcommand{\iarcm}{arcmin$^{-1}$}
\newcommand{\psf}{\phi}
\newcommand{\psfk}{\tilde{\phi}}
\newcommand{\nch}{n_{\rm c}}
\newcommand{\nmap}{n_{\rm m}}

\newcommand{\Kkms}{K\,\kms}
\newcommand{\Planck}{\textit{Planck}}
\newcommand{\Spitzer}{\textit{Spitzer}}
\newcommand{\Herschel}{\textit{Herschel}}
\newcommand{\Chandra}{\textit{Chandra}}
\newcommand{\Einstein}{\textit{Einstein}}

\newcommand{\IRAS}{{IRAS}}
\newcommand{\COBE}{{COBE}}
\newcommand{\ISO}{{ISO}}
\newcommand{\AKARI}{{AKARI}}
\newcommand{\ROSAT}{{ROSAT}}

\newcommand{\nh}{$N_{\rm H I}$}
\newcommand{\wh}{$W_{\rm H I}$}

\newcommand{\hi}{\ion{H}{1}}

\newcommand{\cmm}{${\rm cm}^{-2}$}

\newcommand{\ncpl}{NCPL} 
\newcommand{\ncpeb}{NCPLEB} 
\newcommand{\nfd}{NFD} 
\newcommand{\lhm}{{LHM}} 

\newcommand{\ghigls}{GHIGLS}

\newcommand{\beamfrac}{0.07}
\newcommand{\kmax}{$k_{\rm max}$}
\newcommand{\kmaxval}{0.12\,arcmin$^{-1}$}

\newcommand{\nepLVCslope}{$-2.86\pm0.04$}
\newcommand{\nepIVCslope}{$-2.69\pm0.04$}
\newcommand{\nepHVCslope}{$-2.59\pm0.07$}
\newcommand{\nepLVCnarrow}{$-1.9\pm0.1$}
\newcommand{\nepLVCbroad}{$-2.7\pm0.1$}

\def\GASSmGBT{GASS $-$ \ghigls}
\def\EBHISmGBT{EBHIS $-$ \ghigls}

\newcommand{\GPs}{GPs}

\newcommand{\ghiglsarchive}{www.cita.utoronto.ca/GHIGLS}

\newcommand{\LH}{UM2M} 
\newcommand{\GR}{{09A079}} 
\newcommand{\grossan}{09A079} 
\newcommand{\NGCLH}{NGC3310} 
\newcommand{\NEPft}{UMIN} 

\newcommand{\nfifty}{{75}}
\newcommand{\vfifty}{{60}}
\newcommand{\feffifty}{{0.14}}
\newcommand{\fbfifty}{{0.3}}

\newcommand{\reso}{1.0}

\newcommand{\garea}{1000} 
\newcommand{\gcount}{37} 
\newcommand{\spcount}{11} 

\newcommand{\spacs}{0.97} 

\newcommand{\hpeter}[1]{}
\newcommand{\hkevin}[1]{}
\newcommand{\hmamd}[1]{}
\newcommand{\hdaniela}[1]{}
\newcommand{\hjay}[1]{}

\slugcomment{Accepted for publication in The Astrophysical Journal, 2015 July 16}

\shorttitle{\ghigls: Deep \hi\ Surveys with the GBT}
\shortauthors{Martin et al.}

\begin{document}

\title{\ghigls: \hi\ Mapping at Intermediate Galactic Latitude\\ using the Green Bank Telescope}

\author{
P.~G.~Martin\altaffilmark{1},
K.~P.~M.~Blagrave\altaffilmark{1},
Felix J.~Lockman\altaffilmark{2},
D.~Pinheiro Gon\c{c}alves\altaffilmark{1,3},\\
A.~I.~Boothroyd\altaffilmark{1},
G.~Joncas\altaffilmark{4},
M.-A.~Miville-Desch\^enes\altaffilmark{1,5},
G.~Stephan\altaffilmark{1,5,6}
}

\altaffiltext{1}{Canadian Institute for Theoretical Astrophysics,
  University of Toronto, 60 St. George Street, Toronto, ON M5S~3H8,
  Canada; \email{pgmartin@cita.utoronto.ca}}

\altaffiltext{2}{National Radio Astronomy Observatory, Green Bank, WV
  USA 24944}

\altaffiltext{3}{Department of Astronomy \& Astrophysics, University
  of Toronto, 50 St. George Street, Toronto, ON  M5S~3H4, Canada}

\altaffiltext{4}{Universit\'{e} Laval, Qu\'{e}bec, PQ, Canada}

\altaffiltext{5}{Institut d'Astrophysique Spatiale, CNRS (UMR8617) Universit\'{e} Paris-Sud 11, B\^{a}timent 121, Orsay, France}

\altaffiltext{6}{I. Physikalisches Institut, Universit\"{a}t zu
  K\"{o}ln, Z\"{u}lpicher Str. 77, 50937, K\"{o}ln, Germany}

\begin{abstract}
This paper introduces and describes the data cubes from \ghigls, deep
Green Bank Telescope (GBT) surveys of the 21-cm line emission of \hi\
in \gcount\ targeted fields at intermediate Galactic latitude.
The \ghigls\ fields together cover over \garea\ deg$^2$ at
$9\farcm55$ spatial resolution.  The \hi\ spectra have an effective
velocity resolution about \reso~\kms\ and cover at least
$-450<\vrl<+250$ \kms, extending to $\vrl<+450$ \kms\ for most fields.
As illustrated with various visualizations of the \hi\ data cubes,
\ghigls\ highlights that even at intermediate Galactic latitude the
interstellar medium is very complex.
Spatial structure of the \hi\ is quantified through power spectra of
maps of the integrated line emission or column density, \nh. 
For our featured representative field, centered on the North Ecliptic
Pole, the scaling exponents in power-law representations of the power
spectra of \nh\ maps for low, intermediate, and high velocity gas
components (LVC, IVC, and HVC) are \nepLVCslope, \nepIVCslope, and
\nepHVCslope, respectively.
After Gaussian decomposition of the line profiles, \nh\ maps were also
made corresponding to the narrow-line and broad-line components in the
LVC range; for the narrow-line map the exponent is \nepLVCnarrow,
reflecting more small scale structure in the cold neutral medium
(CNM).
There is evidence that filamentary structure in the \hi\ CNM is oriented
parallel to the Galactic magnetic field.
The power spectrum analysis also offers insight into the various
contributions to uncertainty in the data, yielding values close to
those obtained using diagnostics developed in our earlier
independent analysis.
The effect of 21-cm line opacity on the \ghigls\ \nh\ maps is
estimated.
Comparisons of the GBT data in a few of the \ghigls\ fields with data
from the EBHIS and GASS surveys explore potential issues in
data reduction and calibration and reveal good agreement.
The high quality of the \ghigls\ data enables a variety of studies in
directions of low Galactic column density, as already demonstrated by
the \Planck\ collaboration.
Fully-reduced \ghigls\ \hi\ data cubes and other data products are
available at \url{\ghiglsarchive}.
\end{abstract}

\keywords{ISM: clouds -- ISM: structure -- radio lines: ISM}

\maketitle

\section{Introduction}
\label{intro}

The 21-cm emission line of \hi\ is the most commonly used tracer of
the three-dimensional structure of the diffuse interstellar medium
(ISM).  Original studies were focused on structures and kinematics
within the Galactic plane of the Milky Way
\citep{Burton1976,kulk1982,Stil2006}, and have been expanded to
studies of the vertical structure of the disk and, more generally, gas
at intermediate Galactic latitudes extending into the Galactic halo
\citep{Heiles1976,dickey90,Kalberla2009,Putman2012}

The discoveries and insights gained in these studies have benefited
tremendously from the all-sky LAB survey \citep{kalb05}, which is beam
sampled at 36\arcmin\ angular resolution.  Stimulated by these
results, the range of motivations for further \hi\ surveys is quite
sweeping.

We have used the 100-m Robert C. Byrd Green Bank Telescope (GBT,
\citet{Prestage2009}) at the National Radio Astronomy Observatory
(NRAO\footnote{
The National Radio Astronomy Observatory is a facility of the National
Science Foundation operated under cooperative agreement by Associated
Universities, Inc.}) 
for complementary studies at intermediate Galactic latitude, where
often the line of sight column density is low so that the emission is
not strong.
This paper is to introduce, describe, and disseminate the data from
our deep GBT surveys of the 21-cm line emission of \hi\ in \gcount\
fields at intermediate Galactic latitude, obtained mainly with the
Auto-Correlation Spectrometer (ACS) over the period 2005 to 2010 (see
\citealp{blag10} for a preliminary report).
We refer to the project by the acronym \ghigls\ (GBT \hi\
Intermediate Galactic Latitude Survey).
The total area mapped is over \garea\ deg$^2$ and although this
comprises only about 2.5\,\% of the sky, the judicious choice of
environments sampled means that a broad range of scientific questions
can be addressed, as outlined below.
Compared to LAB, the \ghigls\ data have a higher angular resolution of
about 9\arcmin\ and are Nyquist sampled.  We have developed observing
and reduction techniques for the GBT that result in high quality
spectral line data cubes \citep{boot11} with the requisite sensitivity
for studies at low column density.

An \hi\ spectrum, whether for a single line of sight or averaged over
a region, generally has emission spread over a range of frequency,
which through the Doppler effect is interpreted as radial velocity, in
this paper $\vrl$ relative to the Local Standard of Rest (LSR)
and hereafter denoted simply $v$.
\renewcommand{\vrl}{v}

\subsection{Insights from \hi\ data}\label{insights}

This kinematic information in \hi\ spectra provides essential
diagnostics of various physical properties of the gas.
\hi\ gas in the local neighborhood is identifiable by its low $|\vrl
|$, while gas in the halo appears at both intermediate and high
velocities, for example in \nep\ in the ranges $-80\le \vrl \le
-20$~\kms\ and $-140 \le \vrl \le-80$~\kms, respectively (for the
\ghigls\ fields, it turns out that the non-local gas is only at
negative velocities.)
\citet{wak2013book} denote the latter two both as ``high-velocity
clouds."  However, for the three ``components" that can be
distinguished via velocity we prefer the standard terminology LVC,
IVC, and HVC, respectively, because there is an underlying physical
distinction between these components of interstellar gas, irrespective
of the apparent kinematics.  There is a component of Galactic gas with
an interesting history, circulating in a ``Galactic fountain"
\citep{shap1976,breg1980}.  Its distinctive motion projects into a
radial velocity that often sets it apart from local LVC emission, in
the IVC range.  On the other hand, there is gas that appears to be of
extragalactic origin (perhaps it has been in other galaxies) and is
now accreting on the Galaxy and destined for interaction with Galactic
gas in the halo and disk.  Its radial velocity is often in the HVC
range.  Complex C is one such example, with a low metallicity and high
deuterium-to-hydrogen ratio that point to a non-Galactic origin
\citep{tripp03,sembach04}.

Surveys like \ghigls\ can provide new insight into these three
components, LVC, IVC, and HVC.  For example, correlated dust emission
corroborates this distinction.  Analysis with the \ghigls\ data
\citep{planck2011-7.12} shows that the IVC gas has an emission
signature from embedded dust with a dust-to-gas ratio comparable to
the LVC, whereas for HVC there is no detectable dust emission
signature, consistent with a dust-to-gas ratio at least as low as
implied by the low metallicity.  We note that depending on the
geometry, gas with physical properties similar to IVC or HVC might
appear within the LVC velocity range and so not be distinguishable by
its velocity \citep{wak2013book}; some other clue is then needed such
as, in the case of HVC-like gas, low \hi-correlated dust emission
\citep{planck2013-XVII}.

There is a long history of using the power spectrum of an image to
quantify the statistical properties of intensity fluctuations and
structural information, including \hi\ \citep{crov1983} and thermal
emission by dust \citep{gaut1992}.  This structure is linked to a
turbulent cascade of kinetic energy.
For \hi\ \citet{laza2000} showed that the power spectrum of a velocity
channel is a complex mixture of velocity and density fluctuations. For
the typical steep power spectra in the ISM, the three-dimensional
spectral index of density can be obtained from power spectrum analysis
at high spatial frequencies by averaging enough velocity channels that
the brightness temperature fluctuations are dominated by density
fluctuations; otherwise the power spectrum is too shallow by up to one
in the power-law exponent.
As pointed out and illustrated by \citet{dick2001} using \hi\
observations in the inner Galactic plane, different regions of the ISM
could have different statistical properties projected on the sky
because of different fractions of gas in the cold and warm phases of
the neutral medium (CNM and WNM, respectively), different optical
depth effects, and geometry.
In Figure~10 from their study of turbulent molecular clouds
\citet{henne12} provide a comprehensive summary of the power law
exponents for different tracers of the ISM at different scales.  Even
for the same tracer, there is a considerable range.
Compared to fields near the Galactic plane or in molecular clouds, the
\ghigls\ fields at intermediate Galactic latitude are relatively
simple lines of sight, and yet they offer the opportunity to explore
both environmental differences and the structural properties of LVC,
IVC, and HVC components separately.

The \hi\ line profile, which might consist of a number of peaks with
different centroid velocities, can be segmented by Gaussian
decomposition \citep{haud00,vers04} (this can be applied to \hi\
absorption spectra as well, \citealp{roy13}).
Gaussian decomposition methods offer the opportunity to differentiate
between components with different line widths \citep{haud07}.  In
combination with absorption-line studies \citep{dickey03,heiles03},
components with broad or narrow line widths have been found to arise
from the WNM and CNM, respectively \citep{dickey90,wolfire03,henne12}.
For the CNM the line width is larger than simply the thermal line
width because of turbulent broadening.  Such an analysis of \hi\ data
from surveys like \ghigls\ can illuminate theoretical numerical
modelling of the phases in the ISM \citep[e.g.,][]{saur14}, and vice
versa, leading to a better understanding of the dynamical
formation of CNM gas in a thermally bistable medium and the filling
factors of the phases.

\subsection{Dust and Gas}
\label{dustgas}

Observations of dust emission integrate over all dust along the line
of sight, regardless of velocity.  The strong dust-gas correlation
seen empirically in projection at high Galactic latitude
\citep{boul88} is consistent with a strong correlation spatially in
three dimensions.
The dust-gas correlation has been used extensively to infer the
physical properties of dust in various environments \citep{deul92,
boul96, jone95}.
\citet{reac94,reac98} have argued that excess infrared emission
relative to \hi\ (an excess ``dust emissivity") toward brighter cloud
structures signals a phase transition to H$_{\rm 2}$, untraced by CO
at those column and volume densities.  Subsequent all-sky observations
\citep{planck2011-7.0} indicate that such ``CO-dark" gas is an
important component widespread in the Galaxy.

The kinematic content of the \hi\ spectra is important for decoding
information in dust emission maps.
Morphological spatial detail in maps of \hi\ varies as a function of
velocity and so any dust closely correlated with a velocity component
of the gas leaves a related morphological imprint in the dust emission
map.  This approach has been used to show that there is dust of
significant emissivity associated with IVC gas \citep{mart94}.

The selection of fields targeted by \ghigls\ enables exploration of
the different kinematics and spatial distributions of \hi\ gas and the
dust evolution at diverse stages of Galactic evolution.
\ghigls\ data have been used in combination with \Planck\ data on
thermal dust emission to find the emissivity, opacity, and temperature
of dust associated with both LVC and IVC gas \citep{planck2011-7.12}.
\citet{pinh2013,pinh13} extend the analysis to emissivities
characterizing non-equilibrium dust emission in the mid-infrared
\IRAS\ bands
(see also Pinheiro Gon{\c c}alves, D. et al.\ 2015, in preparation).

\citet{mamd05fls} have reported thermal dust emission associated with
the HVC gas in the \Spitzer\ First Look Survey (FLS) field.  (The GBT
Spectral Processor (SP) \hi\ data used are described in
\citealp{lock05}; they have been reprocessed here.)  However, this is
challenging because the dust emissivity of this lower metallicity
extragalactic gas has apparently been depressed by a lower dust to gas
ratio and also potentially by a lower dust temperature.  Furthermore,
chance correlation of the foreground LVC and IVC dust emission with
the cosmic infrared background fluctuations or ``anisotropies"
(thus CIBA), which so far have not been separated from the dust maps,
is a significant source of systematic uncertainty and no
HVC-correlated dust emission has been found in the \ghigls\ fields
\citep{planck2011-7.12}.

From another perspective, dust emission in the Galaxy is a significant
foreground contamination of the CIBA signal.  However, by exploiting
the tight relation between Galactic dust and gas emission at low
column densities, \hi-correlated dust emission can be removed from the
infrared and submillimeter maps, as done for \COBE\ \citep{aren98}.
In combination with \Planck\ and \Spitzer\ data, \ghigls\ data have
been used for ``cleaning" in this way to facilitate analysis of the
residual CIBA \citep{planck2011-6.6,peni12,planck2013-pip56}.

Dust emission also contaminates measurements of the cosmic
microwave background, for example compromising detection of a
polarized B-mode signal from inflation \citep{planck2014-XXX,pb2015}.
The Galactic magnetic field tends to be oriented parallel to the
elongation of filamentary dust structures
\citep{planck2014-XXXII,planck2015-XXXV}.  Filamentary gas structures
in the cold neutral medium might also be useful in tracing the
orientation of the magnetic field, another way in which \hi\ data
might contribute to a more complete understanding of the ISM both
phenomenologically and physically. 
\citet{clark14} have presented observations of slender, linear
\hi\ features in the diffuse ISM at high Galactic latitude and found
them to be oriented along the interstellar magnetic field as probed by
starlight polarization.

\vskip 0.2 true cm
The plan of the paper is as follows.
The selection of the \ghigls\ fields surveyed is presented in
Section~\ref{fields}.
Appendix~\ref{arch} discusses data obtained using the GBT SP 
while Appendix~\ref{arch1} discusses reprocessing of archival data
overlapping one of these fields obtained using the GBT ACS.
Section~\ref{obs} describes spectral-line mapping using the GBT and
the data reduction pipeline developed to produce a data cube.
Various ways of visualizing the data in a cube are reviewed briefly in
Section~\ref{illu} to illustrate the \ghigls\ data.
Separation of emission from distinct components of gas at different
velocities, LVC, IVC, and HVC, is discussed in Section~\ref{lih}.
Section~\ref{column} discusses properties of maps of the line
integral \wh\ (proportional to column density \nh) for LVC, IVC, and
HVC components.
The effect of 21-cm line opacity on \nh\ is addressed in Appendix~\ref{self}.
Appendix~\ref{quality} evaluates the uncertainties in the data,
particularly as applied to maps of the column density \nh.
To explore possible issues in data calibration and reduction, GBT
data in a few of the targeted \ghigls\ fields are compared to data
from a new generation of wide-area \hi\ surveys, in the north EBHIS
(Appendix~\ref{ebhis}) and in the south GASS (Appendix~\ref{gass}).
Section~\ref{power} examines angular power spectra of the \nh\ maps
of the three components.
Gaussian decomposition of line profiles is explored in
Section~\ref{profile1}.
The structure of the cold neutral medium and
its relationship to the Galactic magnetic field are investigated in
Section~\ref{magnetic}.
Section~\ref{conclusions} summarizes our conclusions.

\def\sigmaeff{1}
\def\SP{2}
\def\boa{3}
\def\SPb{4}
\def\bob{5}
\def\ddf{6}
\def\n1{7}
\def\pn{8}
\def\spi{9}
\def\spc{10}
\def\grm{11}

\begin{deluxetable*}{lllllll}
\tablecolumns{7} 
\tablewidth{0pc}
\tablecaption{The Constituent Fields in \ghigls}
\tablehead{ 
\colhead{Field} & \colhead{Name} & \colhead{Size} & \colhead{Subfield} & \colhead{Repeats}  & \colhead{$\sigma_{\rm ef}$ \tablenotemark{\sigmaeff}} & \colhead{Scan and} \\ 
\colhead{} & \colhead{} & \colhead{} & \colhead{Layout} &\colhead{} & \colhead{[mK]} & \colhead{Coordinates} }
\startdata
G056.98$-$81.50 & MC & $6\arcdeg\times5\arcdeg$ & & 2   &  81    & Galactic \\
G058.10+68.55 & BOOTES & $12\fdg4\times4\arcdeg$ & $5 \times 1$ & 5(2) \tablenotemark{\SP,\boa}& 70(110) \tablenotemark{\SPb} & ICRS \\
                                                                                                    &  & & & 3/ 1(2) \tablenotemark{\bob} & 50/ 80(60) \tablenotemark{\SPb} \\ 
G067.74+67.73 & Necklace & $2\arcdeg\times2\arcdeg$ & & 2   &  73    & Galactic \\
G071.00+41.47 & OX3 & $1\fdg9\times2\fdg4$ & & 1 \tablenotemark{\SP} &  90 \tablenotemark{\SPb}  & Galactic \\
G085.33+44.28 & N1 \tablenotemark{\ddf} & $5\arcdeg\times5\arcdeg$ &  & 2 \tablenotemark{\n1}& 71 & Galactic \\
G087.95+59.05 & G86 & $5\arcdeg\times5\arcdeg$ &  & 3 & 59 & ICRS \\
G088.32+34.89 & FLS  & $3\arcdeg\times3\arcdeg$ &  & 8 \tablenotemark{\SP}  & 82 & ICRS \\
G091.38+47.95 & MRK290 & $4\arcdeg\times4\arcdeg$ & & 1   &  95    & Galactic \\
G092.24+38.43 & DRACO & $5\arcdeg\times5\arcdeg$ &  & 3 & 60 & ICRS \\
G096.27+59.91 & GROTH & $2\arcdeg\times2\arcdeg$ & & 3 \tablenotemark{\SP} &  77 \tablenotemark{\SPb}  & ICRS \\
G096.40+30.03 & \nep & $12\arcdeg\times12\arcdeg$ & $3 \times 3$  & 3 & 64 & Galactic \\
G115.62+30.40 & \NEPft & $4\arcdeg\times3\arcdeg$ & & 1   & 100    & Galactic \\
G125.00+27.42 & POL & $6\arcdeg\times10\arcdeg$ & & 1   & 111    & Galactic \\
G125.03+37.36 & POLNOR & $6\arcdeg\times10\arcdeg$ &  & 2(1) \tablenotemark{\pn} & 75(110) & Galactic  \\ 
G125.37+41.67 & MRK205 & $4\arcdeg\times4\arcdeg$ & & 1   & 104    & Galactic \\
G125.89+54.84 & DFN & $2\arcdeg\times2\arcdeg$ & & 5 \tablenotemark{\SP} &  45 \tablenotemark{\SPb}  & ICRS \\
G132.37+47.50 & SP & $5\arcdeg\times5\arcdeg$ &  & 2 & 71 & Galactic \\
G134.95+54.13 & UM1 & $4\arcdeg\times4\arcdeg$ & & 1 \tablenotemark{\SP} &  83 \tablenotemark{\SPb}  & ICRS \\
G134.98+39.97 & SPIDER \tablenotemark{\ddf} & $10\arcdeg\times10\arcdeg$ & $5 \times 5$ & 2/ 1 \tablenotemark{\spi}& 75/ 105 & Galactic \\
G135.36+30.29 & SPC & $12\arcdeg\times 9\fdg5$ & $3 \times 2$  & 1/ 2 \tablenotemark{\spc}& 103/ 72 & Galactic \\
G143.94+28.02 & 1H0717 & $4\arcdeg\times4\arcdeg$ & & 1   &  97    & Galactic \\
G144.25+38.56 & UMA & $9\arcdeg\times9\arcdeg$ & $3 \times 3$ & 1   & 107    & Galactic \\
G145.68+23.35 & HS0624 & $4\arcdeg\times4\arcdeg$ & & 1   & 101    & Galactic \\
G147.46+44.09 & UM3 & $4\arcdeg\times4\arcdeg$ & & 1 \tablenotemark{\SP} &  82 \tablenotemark{\SPb}  & ICRS \\
G148.65+52.21 & \grossan\ \tablenotemark{\grm} & $5\arcdeg\times4\arcdeg$ & irregular & 1 to 4   &  62    & Galactic \\
G152.31+53.31 & UM2M & $4\arcdeg\times4\arcdeg$ & & 1 \tablenotemark{\SP} &  86 \tablenotemark{\SPb}  & ICRS \\
G152.44+25.63 & MS0700 & $4\arcdeg\times4\arcdeg$ & & 1   & 101    & Galactic \\
G155.76+37.00 & UMAEAST & $10\fdg5\times6\arcdeg$ & $3 \times 1$  & 1 & 107 & Galactic \\
G156.38+32.57 & LOOP4 & $4\arcdeg\times4\arcdeg$ & & 1   & 122    & Galactic \\
G156.45+54.06 & NGC3310 & $4\arcdeg\times4\arcdeg$ & & 1   &  95    & Galactic \\
G158.32+28.75 & MRK9 & $4\arcdeg\times4\arcdeg$ & & 1   &  96    & Galactic \\
G164.84+65.50 & AG & $5\arcdeg\times5\arcdeg$ & & 2   &  71    & Galactic \\
G170.02$-$59.90 & SUBA & $2\arcdeg\times2\arcdeg$ & & 2 \tablenotemark{\SP} &  83 \tablenotemark{\SPb}  & ICRS \\
G172.01+26.84 & MBM23 & $4\arcdeg\times4\arcdeg$ & & 1   &  99    & Galactic \\
G175.36+43.38 & 091346A & $4\arcdeg\times2\arcdeg$ & & 1 \tablenotemark{\SP} &  83 \tablenotemark{\SPb}  & ICRS \\
G179.50+65.03 & MRK421 & $4\arcdeg\times4\arcdeg$ & & 1   &  95    & Galactic \\
G223.57$-$54.44 & CDFS & $2\arcdeg\times2\arcdeg$ & & 5 \tablenotemark{\SP} &  93 \tablenotemark{\SPb}  & ICRS
\enddata 
\label{f_table}

\tablenotetext{\sigmaeff}{Average noise in emission-free channels of width 0.8\,\kms\ (except 1.0\,\kms\ where noted\tablenotemark{\SPb}).}
\tablenotetext{\SP}{Using the GBT SP (Appendix~\ref{arch}).}
\tablenotetext{\boa} {Central $4\arcdeg \times 4\arcdeg$ subfield with two repeats.  
Inner $3\fdg5 \times 3\fdg5$ three further repeats, totalling five.}
\tablenotetext{\SPb}{Channel spacing 1.0\,\kms.}
\tablenotetext{\bob} {Four overlapping $2\fdg5 \times 4\arcdeg$
subfields flanking the central GBT SP subfield symmetrically. Three
repeats, except western-most only once with its central 0.24 fraction
in $\delta$ repeated twice. All regridded to 1.0\,\kms\ channel spacing of GBT SP data.}
\tablenotetext{\ddf}{A DRAO Deep Field (Blagrave, K. et al.\  2015, in preparation).} 
\tablenotetext{\n1} {From data taken at times that minimize the stray radiation correction (Appendix~\ref{snoise}).}
\tablenotetext{\pn} {Upper 0.53 fraction of field repeated twice.}
\tablenotetext{\spi} {Two repeats on the inner $3\times3$ of the $5\times5$ subfields.}
\tablenotetext{\spc} {Three $4\arcdeg\times6\arcdeg$ subfields and
above these three $4\arcdeg\times2\fdg5$ subfields; only the western
pair of subfields observed twice. Additional $1\arcdeg$ strips on north
and west to provide an overlap with SPIDER and with POL and POLNOR,
respectively.}
\tablenotetext{\grm} {Reprocessed archival GBT ACS data toward the Lockman Hole (Appendix~\ref{arch1}).}

\end{deluxetable*} 

\section{The \ghigls\ Fields}
\label{fields}

Table~\ref{f_table} presents the fields that we surveyed with the GBT
SP (Appendix~\ref{arch}) and the GBT ACS (since 2005), in order of
increasing Galactic longitude.  We also reprocessed archival
data obtained with the ACS for \GR\ (Appendix~\ref{arch1}).
For each field the table lists
the center coordinates, adopted name, size,
subfield layout where relevant (Section~\ref{obs}),
the number of repeated observations,
the noise $\sigma_{\rm ef}$ as measured in emission-free channels
(Appendix~\ref{enoise}) per 0.8~\kms\ or 1.0~\kms\ channel
(Section~\ref{spectrometer}),
and the scan orientation (Section~\ref{mapping}), along lines of
constant Galactic latitude (Galactic) or Declination (ICRS).

The raw \hi\ spectral line data using the GBT SP are available under
the following proposal numbers:
GBT/02A-007, 02A-023, 02A-031, and 03B-030.
Those using the GBT ACS are under 
GBT/05C-009, 05C-021, 06B-030, 06C-032,
07A-104, 08A-083, 08B-038, 09A-079, 09B-042, 10A-012, and 10A-078.
The field names adopted here are recognizable designations for the
data in the NRAO archive,\footnote{ \url{archive.nrao.edu}
}
except for DFN (both CDFN and HDFN), SPIDER (DDI), \NEPft\ (NEP42),
and \grossan\ for the archival data described in Appendix~\ref{arch1}.

Figure~\ref{ffig} shows the fields in the northern Galactic
intermediate latitude sky and inserts for three southern fields,
including the one ACS field in the southern sky, MC, which covers part
of the Magellanic Stream.  The North Celestial Pole (NCP) is marked by
an ``$\times$" and those fields scanned in Equatorial coordinates can
be identified by their different orientation.  A ``$+$" marks the
North Ecliptic Pole (\nep).  Although many fields are close neighbors,
even intentionally overlapping, given their size a considerable
variety of physical conditions has been probed, sometimes
serendipitously but particularly because the fields were chosen with
certain science goals in mind.

\begin{figure*} 
\centering
\includegraphics[width=0.8\linewidth]{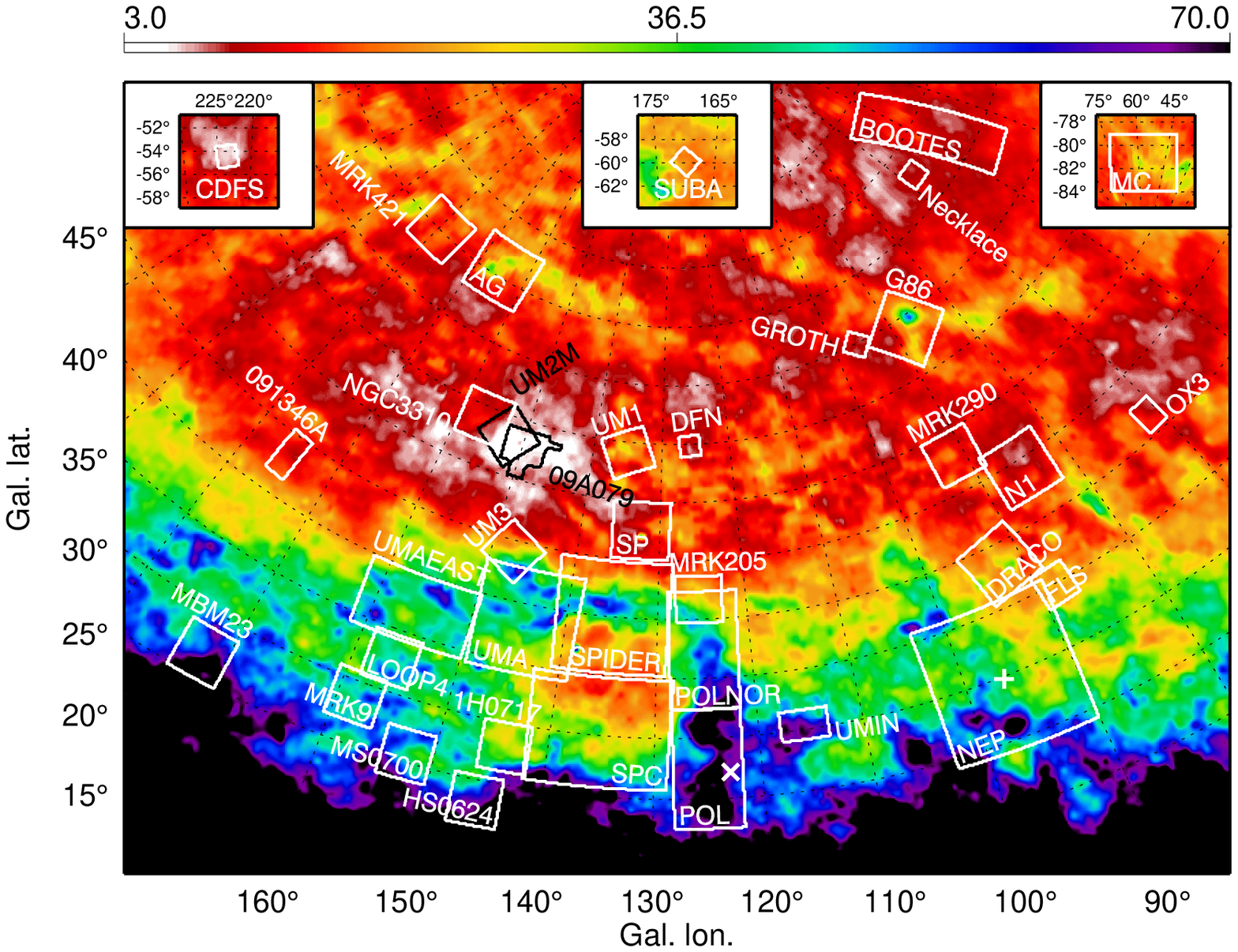}
\caption{
Locations of \ghigls\ fields.  Fields are rectangular in the local GLS
projection appropriate to the OTF mapping but in this stereographic
(STG) projection the outlines are slightly distorted.  Background
image is the velocity-integrated \hi\ emission from the all-sky LAB
survey \citep{kalb05}, expressed as column density \nh\ in units
$10^{19}$\,\cmm.
The North Celestial Pole (NCP) is marked by an ``$\times$" at
$122\fdg93$, $27\fdg13$.  The GBT fields cover the entire NCP Loop and
its interior.  The North Ecliptic Pole (\nep) is marked by a ``$+$" at
$96\fdg38$, $29\fdg81$.
MC and two other southern fields are shown to scale in the
insets on GLS grids.
}
\label{ffig}
\end{figure*}

\subsection{Science Goals}
\label{criteria}

Below we describe some motivations for targeting particular fields for
our \hi\ surveys, building on the general discussion in
Section~\ref{intro}.  Abstracts of our proposals making use of the GBT
ACS are also available at \url{library.nrao.edu/proposals}.

However, it can be noted, certainly in retrospect, that data obtained
in a focused proposal are often useful for addressing several of the
distinct science goals highlighted in others.  This is illustrated by
the data for many \ghigls\ fields that proved useful for the
above-mentioned \Planck\ studies of the CIBA and Galactic dust.

\subsubsection{Extragalactic Windows}\label{windows}

\hi\ surveys are used first to identify regions of low Galactic column
density and then to characterize or remove foregrounds that otherwise
compromise extragalactic science.  With the GBT SP, the foreground
\hi\ has been explored in a number of extragalactic or ``cosmic"
windows for deep multiwavelength observations of the distant universe.
The fields \LH\ together with \GR\ and NGC3310 (both with the GBT ACS)
overlap the Lockman Hole, the region with the lowest integrated \hi\
emission (Figure~\ref{ffig}).
The Lockman Hole has been studied previously in \hi\ at 10\arcmin\
resolution using the NRAO 300-ft telescope \citep{lock86,jahoda1990}
and was an important field for the \Spitzer\ SWIRE survey
\citep{lons03} and many other studies, e.g., with XMM-Newton
\citep{hasi2001}.  UM1 has similarly low LVC emission, but has much
brighter IVC emission and some HVC as well.

In addition to those discussed further below, BOOTES (extended with
the ACS) and N1 (with the ACS), \ghigls\ GBT SP fields span many other
notable windows:
OX3 -- the Hercules field in the Oxford-Dartmouth Thirty Degree Survey
\citep{macd2004};
GROTH -- the All-Wavelength Extended Groth Strip International Survey
(AEGIS) region \citep{davis2007};
DFN -- \Chandra\ Deep Field North/Hubble Deep Field North/GOODS-N
\citep{horn2001};
SUBA -- the Subaru/XMM-\textit{Newton} Deep Survey Field
\citep{ouch2001} (also SWIRE);
and CDFS -- \Chandra\ Deep Field South/GOODS-S \citep{giac2001} (also
SWIRE).
We note that these areas include four of the five (tiny) CANDELS
fields \citep{grogin2011}.

\subsubsection{High Velocity Infall}\label{infall}

The European Large-Area \ISO\ Survey (ELAIS) N1 field is an
extragalactic window also targeted by SWIRE whose low column density
can be appreciated in Figure~\ref{ffig}.  The complex \hi\ structure
has been studied at 1\arcmin\ resolution using the Synthesis Telescope
(ST) at the Dominion Radio Astrophysical Observatory (DRAO\footnote{
\url{www.nrc-cnrc.gc.ca/eng/solutions/facilities/drao.html}}),
mosaicing data from 76 closely-packed pointings to produce a sensitive
``DRAO Deep Field" for which our GBT ACS data, the \ghigls\ N1 field,
provide the short spacing information (Blagrave, K. et al.\ 2015, in
preparation).
Unlike the Lockman Hole, this region has striking HVC emission,
enabling a search for HVC-correlated dust emission.  The HVC gas
morphology is not immediately obviously traced by dust, and in fact it
turns out quantitatively, at least at the GBT angular resolution, that
the HVC has a low dust emissivity \citep{planck2011-7.12}.

Along with N1 and FLS, we extended our study of dust emissivity in HVC
complexes by selecting regions in which the HVC to IVC plus LVC column
density in the LAB survey showed a high contrast: MC, SP, and AG.
(There are also significant knots of high contrast HVC emission
contained within SPC, UMA, and UMAEAST.)  Their low total column
density makes these fields suitable for CIBA studies as well
\citep{planck2011-6.6,peni12,planck2013-pip56}, but we found that the
CIBA significantly contaminates studies of (HVC) \hi-correlated dust
emission \citep{planck2011-7.12}.

\subsubsection{The Galactic Fountain}\label{fountain}

DRACO \citep{herb93} and G86 \citep{mart94} were selected because of
their prominent and distinctive IVC gas which has a clear dust
signature.  Targeted \Herschel\ dust emission and DRAO ST \hi\
observations were carried out for these two fields as well. 
Such fields are ideal for searching for differences in dust properties
between local and IVC gas, evidence of dust evolution in different IVC
environments.  The dynamics is fascinating and in DRACO the transition
from atomic to molecular gas can be studied.

The central region of our BOOTES field targeted the extragalactic
window known as the NOAO Deep Wide Field Survey \citep{jann99} and the
focus of a \Spitzer\ MIPS wide/shallow survey \citep{dole04}.  It is
actually crossed by a band of Galactic dust emission within which we
subsequently discovered anomalously high dust emissivity in one faint
component of IVC gas \citep{lock05b}; to explore this further, with
the GBT ACS, we added flanking subfields along the direction of the
band.
In \IRAS\ 60 and 100~\micron\ dust maps of the B{\"o}{\"o}tes region
there appears a remarkable feature that we dubbed the Necklace and for
which we carried out a small \hi\ survey to study the dust emissivity.

\subsubsection{North Ecliptic Pole Foreground}\label{nepforeground}

We mapped \hi\ in a large region centered on the \nep.  As far as gas
and dust content is concerned, there is nothing special a priori about
this intermediate latitude location.  However, for many all-sky
surveys by satellites that scan along great circles passing near the
ecliptic poles, the \nep\ region is special in having more coverage
and hence data products with lower uncertainty.  This is the case for
\Einstein, \IRAS, \COBE, \ROSAT, \AKARI, and recently \Planck, as
illustrated in the coverage map (see, e.g., figure~5 in
\citealp{planck2013-p01}) and related variance maps.
Thus the \nep\ region is the focus for many studies, e.g.,
recently with \Chandra\ \citep{krum2015}.  For several science
applications using these data it is important to understand the
foreground from the Galaxy.

\hi\ in this region was surveyed previously using the 140-ft telescope
at NRAO \citep{lock86}.  Our \nep\ GBT ACS data cover a larger region
with higher angular resolution and much reduced noise and systematic
uncertainty.  The observations, repeated three times, have a good
signal to noise ratio, there is significant LVC, IVC, and even HVC
emission, and the field is large, all reasons why we chose \nep\ data
for illustrations in this paper.

\subsubsection{The North Celestial Pole Loop}\label{ncpspider}

The first large field targeted with the GBT ACS was SPIDER, at the top
of the arch of the North Celestial Pole Loop (\ncpl), a giant gas
structure north of the Galactic Plane with a cylindrical morphology
\citep[see][]{meyer91}.  In SPIDER the complex LVC \hi\ structure has
been studied with the DRAO ST, mosaicing data from 91 pointings to
produce a second DRAO Deep Field.  We have also mapped the stunning
dust emission there at even higher resolution with \Herschel.

Ultimately we explored \hi\ throughout the spider's web, along the
entire \ncpl\ extending from POL to UMAEAST, and also \hi\ in its
``interior" below the arch with SPC and several fields toward higher
longitude.  See Figure~\ref{ffig}.  POL covers the Polaris flare,
which is actually dominated by molecular gas, not \hi.  \Herschel\
dust emission maps and DRAO ST \hi\ observations were acquired for
parts of POL and UMA (Blagrave, K. et al.\ 2015, in preparation).

\subsubsection{Molecular Hydrogen}\label{ismphases}

In some of the \ghigls\ fields analyzed in \citet{planck2011-7.12}
there was evidence from excess submillimeter dust emission for gas not
traced by \hi, suggesting the presence of H$_2$.  To explore
evolutionary aspects of this phenomenon in relation to different
environmental and kinematic factors at intermediate Galactic latitude,
we focused on fields containing an AGN for which FUSE observations
have been used to characterize H$_2$ \citep{gill06,wakk06} and one
field containing a small, non star forming molecular cloud, MBM~23
\citep{magn1985}.
Many of these fields have been observed with the DRAO ST as well, for
which our GBT observations provide the short spacing data.

\section{Spectral-line Mapping with the GBT}
\label{obs}

Surveys of \hi\ spectra in sizeable fields were carried out using
on-the-fly (OTF) mapping.
As described in detail by \citet{boot11}, data were acquired with the
GBT ACS by in-band frequency switching, with spectra recorded every
4~s in two independent polarizations.
The velocity coverage was $-1700 \leq \vrl \leq +900$~\kms\ in
independent 0.161~\kms\ channels, from which was extracted the central
$-450 \leq \vrl \leq 400$~\kms.  With the broad spectral coverage,
emission from LVC, IVC, and HVC gas is all accessible.

In our \hi\ surveys, scans of the sky up to $6\deg$ long were made.
For practical purposes larger fields were broken up into smaller
subfields with scan dimension between 2\deg\ and 4\deg\ that were
mapped separately.  The scans are still sufficiently long that the
choice of OTF mapping greatly reduces the system overhead.  The
mapping speed is typically one deg$^2$ per hour.  As described below,
we mapped over \garea\ deg$^2$ using the GBT ACS. Observations
of many fields were repeated both to increase the sensitivity and to
examine the reproducibility of the data, leading to better maps.

Aspects of the mapping, processing of the spectra, and data cubes are
discussed in this section.  The accuracy of the data has been
discussed by \citet{boot11} and this is complemented by the analysis
in Appendix~\ref{quality}.
 
\subsection{Processing of the Spectral Data}
\label{spectrometer}

The desired quantity to be measured, designated simply by $\Trb$ in
this paper, is the \hi\ brightness temperature averaged over the main
beam.
As described by \citet{boot11}, the data reduction process involves
removal of radio frequency interference (RFI),
producing the frequency-switched spectra including
a cubic spline interpolation of the data onto a common grid in $\vrl$,
spectral smoothing and resampling to 0.807~\kms\ channels\footnote{
We filtered the spectra with an eleven-channel Hanning kernel (nine
non-zero weights) and then sampled every fifth channel to produce
independent data. The effective resolution is \reso~\kms.}
(or 1.03~\kms),
averaging the two polarizations,
calibration of the intensity scale to antenna temperature, $\Tra$,
calculation and subtraction of the stray radiation spectrum,
correction for the main beam efficiency and
the atmosphere,
interpolation of the sampled spectra from all repeated observations
into a data cube,
and finally removal of a small (low-order and low-amplitude)
instrumental baseline for each interpolated spectrum in the data cube.
In some cases, the presence of a galaxy makes it impossible to fit a
reliable baseline and thus no baseline is removed.  A mask developed
for the data product release reflects these cases.

The repeated observations were also used to produce independent data
cubes for diagnostic analysis (see \citealp{boot11} and
Appendix~\ref{quality}).  Similarly, we have compared data products
obtained by processing the two independent polarizations separately
\citep{boot11} to check for reliability.

\subsection{Mapping, Gridding, and Resulting Angular Resolution}
\label{mapping}

The native primary beam of the GBT is only mildly elliptical with full
width half maximum (\fwhm) 9\farcm1 and 9\farcm0 in the
cross-elevation (azimuthal) and elevation direction, respectively.
The integration time (4~s) and telescope scan rate were chosen to
sample the spectrum every $3\farcm5$ in the in-scan direction, more
finely than the Nyquist interval, $3\farcm86$.
Beam broadening in OTF mapping can be made negligible by sampling at a
smaller fraction of the Nyquist rate \citep{mang07}, ideally at half
that interval.
Our actual sampling amounts to a convolution of the beam with a
$3\farcm5$ boxcar, extending the beam to an effective \fwhm\ of
$9\farcm4$ in the in-scan direction (not aligned along either azimuth
or elevation).

Fields were mapped boustrophedonically, scanning the telescope in one
direction (Galactic longitude or Right Ascension), with steps of
$3\farcm5$ in the orthogonal (cross-scan) coordinate direction before
the subsequent reverse scan.  The quality of the spectra was monitored
closely during most sessions.  Occasionally artifacts from the
spectrometer would compromise a spectrum and it was flagged; usually
it was possible to schedule an observation to replace the entire scan.

The individual spectra were gridded into data cubes.
For these large fields, the equal-area Global Sinusoidal (GLS)
coordinate projection \citep{cala02} was chosen to complement our
equal-area sampling strategy.  GLS is now replaced by the
Sanson-Flamsteed (SFL) projection, which can be considered equivalent
as used here.  Gridding data from a large region to another
projection, like TAN, would produce pixels (spectra in the cube) with
non-uniform coverage.
The data are mapped to the grid following a convolution.  The simplest
convolution would be a pillbox, the naive-mapmaking strategy of
assigning each sample to its nearest neighboring grid cell.  With GLS
projection, a grid can be defined with $3\farcm5$ pixels corresponding
to where the data were sampled, in which case each pixel contains the
actual spectrum recorded.  This is very useful for diagnostic purposes
as we assess the accumulation of error in individual spectra.

However, pillbox is not the ideal gridding for our data.  Instead we
used an optimal tapered Bessel function for interpolation
\citep{mang07}, available in Classic AIPS with the task SDGRD.  This
approach also allowed us to account for small telescope pointing
deviations in the in-scan direction relative to the expected uniform
raster.
Convolution with a Bessel function, being the inverse Fourier
transform of a top hat, is the equivalent of equally sampling all
spatial scales in the Fourier domain with a two-dimensional boxcar
function of size extending to the resolution limit $D/\lambda =
1/7.25$\,\iarcm.  It preserves the true power on small spatial scales
while avoiding the introduction of noise on scales smaller than the
beam.  The corresponding ``width" of the Bessel function is
$3\farcm75$.
Because the observed fields are finite (i.e., no information exists
beyond the edge of the observed field), the Bessel function is tapered
by a Gaussian.  The support size (the radius beyond which the tapered
Bessel function is truncated) was chosen to correspond to a zero
intercept of the Bessel function.  Both the first and second of these
intercepts, corresponding to $7\farcm5$ and $11\farcm25$, were
considered.  A larger support size allows for a broader Gaussian taper
resulting in less modification of the beam profile; while this results
in loss of information at the map edge, for our large maps this loss
is inconsequential.  Thus a support size of $11\farcm25$ was selected
along with a Gaussian taper ``width'' (as defined in the AIPS task
SDGRD) of $9\farcm1\dot{6}$, which corresponds to a Gaussian with
$\fwhm = 15\farcm26$.  The width of the Gaussian taper is chosen to be
as broad as possible, while also maintaining a smooth transition from
the (otherwise infinite in nature) Bessel function to a value of zero
beyond its support size (i.e., its truncation radius).

As is the case with any non-pillbox grid system, the beam ends up
being broadened slightly in both in-scan and cross-scan directions.
The resulting effective beam for our modified Bessel function gridded data cubes
can be approximated by an elliptical Gaussian of size $9\farcm55
\times 9\farcm24$ (\fwhm).

Data cubes were constructed in Classic AIPS, combining spectra from
the two polarizations and from repeated observations as appropriate.
A weight map was created using SDGRD during convolution of the spectra
into the GLS grid and this map was used to remove a few pixels along
the edge of the map, leaving a cube with uniform coverage.
Spectra from the subfields mapped were combined directly into a common
GLS grid (with the same $3\farcm5$ pixel size), providing a mosaic of
the entire region. Because the areas covered by the subfields are
rotated slightly compared to this common grid, there is a saw-tooth
pattern to the weighting function along the mosaic edges.  This too
was removed for the final data products.  These cropped mosaics have
near-uniform coverage (weight), increasing where the subfields
overlap.

These cubes were examined for any evidence of the effects of an
anomalous spectrum that might have been missed in the previous
flagging, in which case the bad spectrum was identified and flagged
and the cube remade, where possible including a replacement scan; any
remaining locations affected by missing/flagged spectra are
identifiable in the weight and noise maps.  Finally a baseline was fit
to the gridded data in each pixel and subtracted.  A mask was
developed to record the few pixels for which a baseline fit to the
spectrum was deemed not feasible, usually due to the presence of
emission from a galaxy; the uncorrected spectra are still very good
because of the intrinsically flat baselines of the GBT ACS.

On the archive of \ghigls\ data,\footnote{\url{\ghiglsarchive}} for
each field there is a FITS file with extensions as follows: 0, the
cube of spectra $\Trb$ in K; 1, a mask recording the few pixels with
no baseline fit removed (many fields do not require this extension);
2, a map of the noise $\sigma_{\rm ef}$ as measured in emission-free
channels of each spectrum, in K, and 3, the (relative) weight map. In
the noise map can be seen the reduction in noise from
overlapping/repeat observations, the subfield layout where relevant
(e.g., Figure~\ref{fig:efchannelebhis}), and the rare increases in
noise because of flagged spectra.  These are a direct consequence of
the weighting and so appear in the weight map as well.

\subsection{Mosaiced Data}
\label{makemosaics}

The data for the tabulated \ghigls\ fields have been analysed
separately, but to explore larger-scale connections between features
in individual fields we have combined the data in some
adjacent/overlapping fields into mosaics, in the same way that
subfields were combined for a given target.  In this reprocessing, we
first ensured that the individual spectra had a common velocity grid.
Then we executed the last two steps of our pipeline,
interpolation of the spectra onto a common GLS grid,
and removal of a small instrumental baseline fit at each pixel.  FITS
files of the mosaics are available on the \ghigls\ archive.

We have made a small mosaic \lhm\ covering the low column density area
in the Lockman Hole region by combining the data from \NGCLH, \LH, and
\grossan.  Because of the overlap the sensitivity is improved over the
central region, as can be seen in the noise map.  Close inspection
shows that of the three fields, UM2M has slight residuals from the
stray radiation correction.

We have combined data for \nep, FLS, and DRACO into a mosaic named
\nfd.  This reprocessing also includes data for a small 3 deg$^2$ map
KnotN under proposal GBT/06B-030 exploring the IVC and an embedded 16
deg$^2$ field H1821 under GBT/09C-042, not documented in
Table~\ref{f_table} or Figure~\ref{ffig} but evidenced in the noise
map.
In this mosaic it can be seen how the distinctive LVC filament in FLS
extends into DRACO and is roughly parallel to the filaments already
mentioned in \nep.

As Figure~\ref{ffig} shows, \ghigls\ has extended coverage of the
\ncpl\ and so we have made a large mosaic thus named.  This
reprocessing also includes data from a few embedded surveys, namely
KnotA, B, and C under GBT/05C-021, our exploratory focus on
high-contrast HVC emission, plus fields called PG0804 under
GBT/09B-042 and LISZTA, LOOP1+2, and LOOP4B under GBT/10A-012.
This shows the spectacularly complex structures in the LVC emission
collectively defining the arch, but also fascinating IVC and HVC
emission as well.

\section{Visualization of a Data Cube}
\label{illu}

\begin{figure}
\centering
\includegraphics[width=1.08\linewidth]{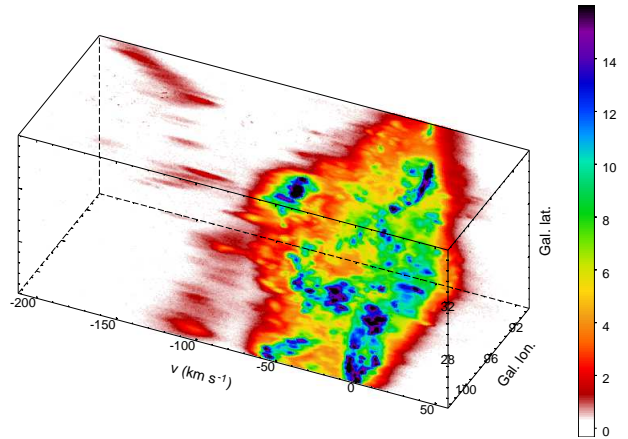}
\caption{
Rendering of the $\Trb(x,y,v)$ data in the spectral cube for the \nep\
field between $-210$ and $60$~\kms.  The colorbar is linear in $\Trb$
(K).  Changing the viewing angle can reinforce the sense of
connectedness and separation of the structures,
as illustrated in a movie of a tumbling cube beginning and
ending at the frame shown here
(\url{www.cita.utoronto.ca/GHIGLS/MOVIES/TUMBLE/GHIGLS_NEP_tumble.mp4}\footnote{All movies are also available in ``ogg" format.}).
} 
\label{fig:render}
\end{figure}

The complex \hi\ spectral information $\Trb(x,y,v)$ assembled in a
data cube can be visualized in a number of ways.
The rendering\footnote{
Created with SAOImage DS9 (\url{http://ds9.si.edu}) using a
simple ray-trace algorithm with the Maximum Intensity Projection
method.}
in Figure~\ref{fig:render} for the \nep\ field reveals the
characteristic clumpy structure of $\Trb$ in the three dimensions of
the cube.  We note, however, that $v$ does not necessarily map
directly into a third spatial dimension.
The sense of connectedness and separation of the structures can
be reinforced by interactively changing the viewing angle (azimuth and
elevation).  We have created a movie to illustrate this, smoothly
varying the viewing angle so that the cube appears to tumble before
eventually returning to the frame shown in Figure~\ref{fig:render}.

\begin{figure}
\centering
\includegraphics[width=1.0\linewidth]{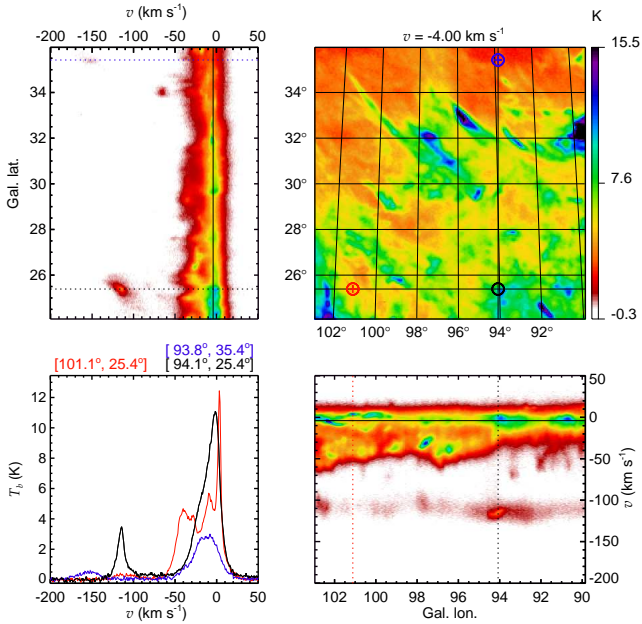}
\caption{
Maps of $\Trb$ on three complementary orthogonal slices of the \hi\
spectral cube for the \nep\ field.
Upper right: Channel map $(\lon, b$) at $v= -4.0$~\kms, viewed from
the high $v$ face of the cube as rendered in Figure~\ref{fig:render}.
This is a single frame from a movie running through channel maps
of the entire cube, starting from the highest $v$
(\url{www.cita.utoronto.ca/GHIGLS/MOVIES/LB/GHIGLS_NEP.mp4}).
Striking filamentary structures cross the field diagonally, roughly
parallel to the direction of the Galactic magnetic field
(Section~\ref{magnetic}).  Solid vertical or horizontal lines
represent the cross-section cut from which the adjacent complementary
position-velocity diagrams are built.
The two position-velocity diagrams are as follows.
Lower right: $(\lon, v$) at fixed $b = 25\fdg4$; the view is from the
low-$b$ face of the cube in Figure~\ref{fig:render} with the $v$ axis
foreshortened.
This is a single frame from a movie running through $(\lon, v$)
slices of the entire cube, starting from the lowest $b$
(\url{www.cita.utoronto.ca/GHIGLS/MOVIES/LV/GHIGLS_NEP_lv.mp4}).
Vertical dashed lines mark the longitudes along which profiles of the
intensity in this diagram are shown in the panel at the lower left;
these profiles are the spectra at two of the pixels marked by
bulls-eyes in the channel map.
Upper left: $(v, b)$ for $\lon \sim 94\deg$, more precisely the plane
through ($\lon, b) = (94\fdg1, 25\fdg4)$; the view is from the
high-$\lon$ face of the cube in Figure~\ref{fig:render} with the $v$
axis foreshortened.  
This is a single frame from a movie running through $(v, b)$
slices of the entire cube, starting from the highest $\lon$
(\url{www.cita.utoronto.ca/GHIGLS/MOVIES/VB/GHIGLS_NEP_vb.mp4}).
Horizontal dashed lines mark the latitudes of the spectra shown.
Lower left: Distinctive individual spectra for the three pixels marked
by the bulls-eyes in the channel map, along the dashed lines in the
position-velocity diagrams.  The spectrum at the special pixel
$(94\fdg1, 25\fdg4)$ can be seen in both position-velocity diagrams.
}
\label{fig:pvd}
\end{figure}

Figure~\ref{fig:pvd}, also for \nep, shows three complementary slices
of the cube: ``a channel map" and two ``position-velocity diagrams."
Position-velocity diagrams reveal not only the complex structure in
the gas, but also regularity such as distinct ranges in $v$ in which
the emission is concentrated (e.g., distinct LVC, IVC, and HVC are
clearly seen in the $(\lon, v)$ diagram shown at the lower right).
Connectivity in these diagrams, or more generally in the rendered
cube, can suggest a physical relationship between the different
velocity components.
A sequence of channel maps (or orthogonal planes) can be combined as
the frames of a movie file.  
For the \nep\ we have made three such movies, within which the
images in Figure~\ref{fig:pvd} are single frames (see caption).
%
For the other \ghigls\ fields, equivalent movies including
tumbling cubes can be found on the \ghigls\ archive.

The individual spectra at the lower left in Figure~\ref{fig:pvd} can
be compared to the mean spectrum for the cube in
Figure~\ref{fig:avg_std}; across the field the spectrum changes
dramatically in the LVC, IVC, and HVC ranges.

A striking phenomenon seen over the LVC range of the \nep\ cube is a
series of filamentary structures all running roughly diagonally across
the field.  Some of these can be seen in the single channel map in
Figure~\ref{fig:pvd}.  As discussed in Section~\ref{magnetic}, these
appear to be aligned with the local Galactic magnetic field.

\section{Separating Low, Intermediate, and High Velocity Gas}
\label{lih}

\begin{figure}
\centering
\includegraphics[width=1.0\linewidth]{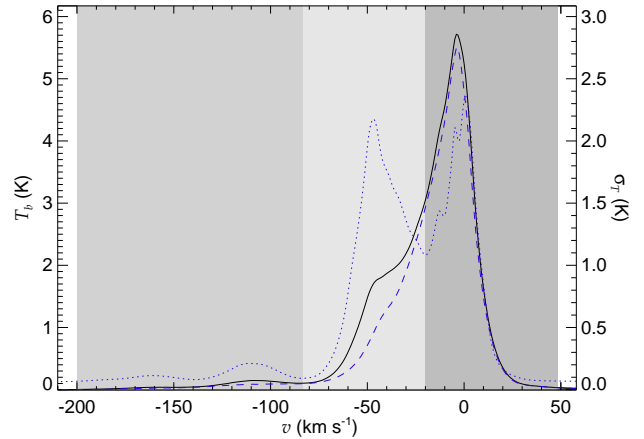}
\caption{
Spectral profiles of the mean (solid) and median (dashed) values of
$\Trb$ for the \nep\ field (left axis) and of the standard deviation
about the mean of $\Trb(\vrl)$ (dotted, $\sigma_T$, right axis).  The
complexity of \hi\ structure along the line of sight complicates the
separation of the different radial velocity components.  Although more
blended in the mean and median profiles, the LVC, IVC, and HVC
components become more distinct in the standard deviation spectrum.
}
\label{fig:avg_std}
\end{figure}


Separating the \hi\ emission into contributions from LVC, IVC, and HVC
gas is challenging when the velocity ranges of the components overlap.
For \nep\ the HVC emission is relatively weak but in
Figures~\ref{fig:render} and \ref{fig:pvd}, and even in the average
spectrum in Figure~\ref{fig:avg_std}, it is fairly well separated.
The IVC emission is not as strong as LVC and while the existence of
IVC is apparent in the average spectrum in Figure~\ref{fig:avg_std} it
is not obvious exactly where to make the separation from LVC.

A useful diagnostic is based on the standard deviation about the mean
of a channel map, because it depends not just on the presence of
signal but also on the fluctuations.  Thus the standard deviation
spectrum takes advantage of the rich structure within the cube
and so when the emission corresponding to LVC, IVC, and HVC components
is not immediately distinguishable in the mean spectrum it offers the
potential to separate the cube into distinct velocity ranges
\citep{planck2011-7.12}.

We have evaluated this approach using two cubes from the
hydrodynamical simulations of the structure of the thermally bistable
and mildy turbulent atomic gas in the local ISM \citep{saur14}.  These
include turbulent forcing with a mixture of compressive and solenoidal
modes whose partition is set by the spectral weight $\zeta$ (ranging
from 0 for compressive to 1 for solenoidal), a turbulent forcing
amplitude $v_s$, and initial density $n_0$, and for investigating the
approach to component separation here, the exact choice of
parameters/simulations does not matter.
Each of these cubes consists only of LVC.  However, we combined them
by summing $\Trb(x,y,v)$ after translating one cube by $\DeltaV$, thus
producing an IVC component as well.  We added noise characteristic of
observations of the GBT fields.
When $\DeltaV$ exceeds the typical \fwhm\ of the line profile, the
separation of components is unambiguous, whether using the mean or
standard deviation spectrum.  But the separation becomes more
challenging for smaller $\DeltaV$ and the superior utility of the
standard deviation spectrum is clear.
We quantified this by comparing the column densities
(Section~\ref{column}) computed for the two ranges to the actual
column densities in the original cubes.  
For conditions like encountered in the actual \ghigls\ fields,
the internal boundaries between LVC and IVC and between IVC and HVC
can be determined to better than a few \kms.

The application to \nep\ is shown in Figure~\ref{fig:avg_std}, from
which the selected bounding velocities are 
[47.9, $-20.5$, $-83.3$,$-199.9$]\,\kms.
Table~\ref{compvel_table} shows the adopted velocities for component
boundaries in the \ghigls\ fields.
These values correspond precisely to particular channels
boundaries in the data cube, but depending on the field the actual
internal boundaries can be uncertain by a few channels.

\begin{deluxetable}{lrrrr}
\tablecolumns{5}
\tablewidth{0pc}
\tablecaption{Component Velocity Boundaries (km s$^{-1}$)}
\tablehead{
\colhead{Name} & \multicolumn{4}{c}{LVC{\hskip 2.5em}IVC{\hskip 2.5em}HVC}
}
\startdata
MC & $  50.4$ & $  -6.0$ & $ -39.8$ & $-240.1$ \\
BOOTES & $  29.5$ & $ -17.9$ & $ -69.4$ & $-111.6$ \\
Necklace & $  14.9$ & $  -9.2$ & $ -50.3$ & \nodata \\
OX3 & $  50.2$ & $ -28.2$ & $ -95.1$ & $-179.6$ \\
N1 & $  57.6$ & $ -10.8$ & $ -59.9$ & $-151.6$ \\
G86 & $  23.8$ & $ -26.9$ & $ -64.8$ & $-137.2$ \\
FLS & $  27.8$ & $ -19.7$ & $-114.6$ & $-210.3$ \\
MRK290 & $  35.1$ & $ -47.9$ & $ -80.9$ & $-175.0$ \\
DRACO & $  23.8$ & $  -8.4$ & $ -72.8$ & $-209.5$ \\
GROTH & $  27.5$ & $ -37.4$ & $ -66.3$ & $-125.0$ \\
NEP & $  47.9$ & $ -20.5$ & $ -83.3$ & $-199.9$ \\
\NEPft & $  50.4$ & $ -22.9$ & $ -58.3$ & $-129.9$ \\
POL & $  31.8$ & $ -48.7$ & $ -97.0$ & $-140.4$ \\
POLNOR & $  31.8$ & $ -26.9$ & $ -97.0$ & $-220.0$ \\
MRK205 & $  39.9$ & $ -35.0$ & $-100.2$ & $-220.0$ \\
DFN & $  50.2$ & $ -28.2$ & $-105.4$ & $-165.2$ \\
SP & $  31.8$ & $ -30.2$ & $-105.0$ & $-190.2$ \\
UM1 & $  60.5$ & $ -24.1$ & $ -93.1$ & $-175.4$ \\
SPIDER & $  39.9$ & $ -14.9$ & $ -88.1$ & $-159.7$ \\
SPC & $  32.6$ & $ -47.9$ & $-100.2$ & $-216.8$ \\
1H0717 & $  50.4$ & $ -36.6$ & $-109.8$ & $-210.3$ \\
UMA & $  50.4$ & $ -26.1$ & $ -84.9$ & $-193.5$ \\
HS0624 & $  39.9$ & $ -33.4$ & $ -68.0$ & $-150.0$ \\
UM3 & $  29.5$ & $ -25.1$ & $ -84.8$ & $-170.3$ \\
\grossan\ & $  60.0$ & $ -30.2$ & $ -80.1$ & $-159.7$ \\
UM2M & $  39.8$ & $ -30.2$ & $ -79.7$ & $-125.0$ \\
MS0700 & $  36.7$ & $ -31.8$ & $-100.2$ & \nodata \\
UMAEAST & $  40.7$ & $ -18.1$ & $ -88.1$ & $-216.8$ \\
LOOP4 & $  39.9$ & $ -22.9$ & $-100.2$ & $-204.7$ \\
NGC3310 & $  30.2$ & $ -30.2$ & $ -76.0$ & $-150.0$ \\
MRK9 & $  35.1$ & $ -33.4$ & $ -70.4$ & \nodata \\
AG & $  47.9$ & $ -31.8$ & $ -72.8$ & $-153.3$ \\
SUBA & $  50.2$ & $ -12.7$ & $ -79.7$ & $-259.8$ \\
MBM23 & $  50.4$ & $ -13.3$ & $ -72.8$ & \nodata \\
091346A & $  39.8$ & $ -23.0$ & $ -90.0$ & $-140.4$ \\
MRK421 & $  35.1$ & $ -44.6$ & $-120.3$ & \nodata \\
CDFS & $  41.9$ & $ -19.9$ & $ -60.1$ & $-119.8$
\enddata
\label{compvel_table}
\end{deluxetable}


If the separation is ambiguous, it becomes a potential systematic
error in whatever dependent analysis is being carried out and so a
sensitivity analysis needs to be done.
Consider the case of distinct dust emissivities associated with LVC
and IVC components.  Our approach has been to correlate
simultaneously the LVC and IVC column density with a dust emission map
from the all-sky \Planck\ survey \citep{planck2013-p06b}, with the two
emissivities as free parameters.  When the component separation seemed
somewhat arbitrary, the velocity boundary between the LVC and IVC
components was changed incrementally over the range of possible
velocities, producing a pair of emissivities and an rms of the
residual dust map for each case.  The velocity boundary corresponding
to the minimum value of the rms was consistent with that determined
independently from the standard deviation spectrum.  The derived
values of the emissivities were robust and the uncertainty of the
velocity boundary did not significantly increase the uncertainty of
the emissivities.  We note that this is not a test of uniqueness, but
of consistency.  But additionally, if the two emissivities were
the same, then the motivation for finding a precise separation of LVC
and IVC would be moot.

For many emissivity analyses \citep[e.g.,][]{planck2011-7.12}, some of
the spectral data end up masked -- mostly due to the presence of
unaccounted hydrogen in the form of H$_{2}$ or sometimes H$^{+}$.  In
such cases, it might be beneficial to iterate on the standard
deviation spectrum to determine the ideal separation between
components including only the relevant data, those in the retained
(not-masked) region.  This has not been implemented here.

Another consideration for closely adjacent ranges is that some signal
from the LVC gas can contaminate the IVC component and vice versa,
because of the overlap of the extended wings of their respective line
profiles.  An alternative method that takes into account these
``intruding'' extended wings implicitly is based on Gaussian
decomposition of the individual spectra (Section~\ref{profile1}).

\section{Maps of the Line Integral}
\label{column}

It is informative to calculate the integral $W_{\rm H I}$ of the \hi\
emission spectrum over velocity (or over distinct velocity ranges --
e.g., LVC, IVC, HVC), because \wh\ is related to the column density
\nh.
At high latitudes, where the brightness of the infrared dust emission
-- the ``cirrus" -- is low because of low dust column density, the
high-latitude 21-cm emission is also faint, with the peak temperature
of the spectral lines usually small compared to the likely spin
temperature $\Trs$, so that the emission is optically thin.  In this
limit, the column density \nh\ derived directly from \wh\ is that
corresponding to an infinite spin temperature:
\begin{equation}
N_{\rm H I}(\infty)/C = \int \Trb\ \mathrm{d}v \equiv W_{\rm H I} \,,
\label{thin}
\end{equation}
where the conversion factor $C = 1.823 \times 10^{18}$~cm$^{-2}$(K km
s$^{-1}$)$^{-1}$ and the integral is over the velocity range specific
to a given component.
This can be carried out for each spectrum in the data cube, producing
a column density map for each of the chosen components.  For optically
thin emission, this is of course accomplished equivalently by summing
channel maps.

\begin{figure}
\centering
\includegraphics[width=0.8\linewidth]{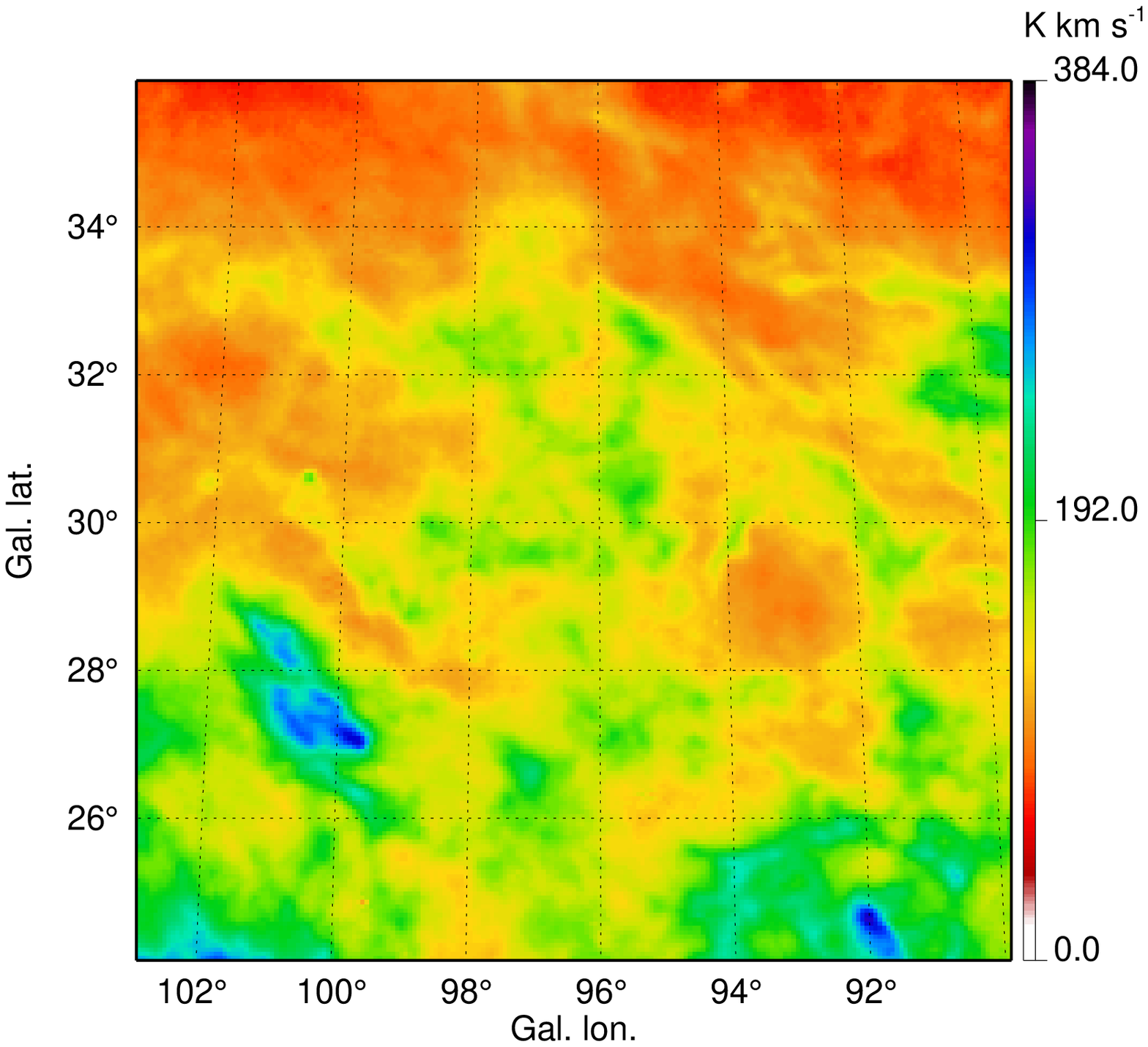}
\includegraphics[width=0.8\linewidth]{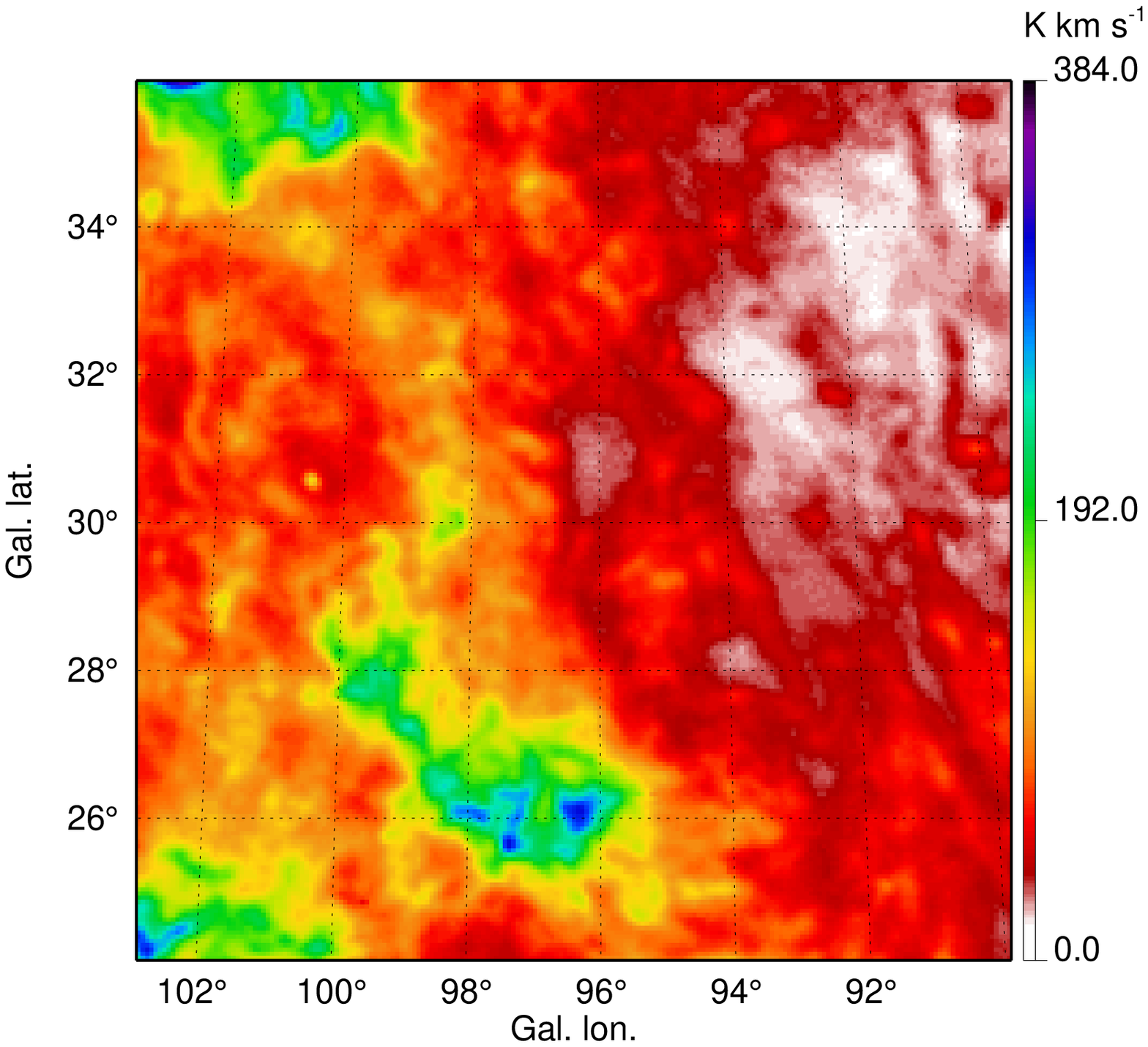}
\includegraphics[width=0.8\linewidth]{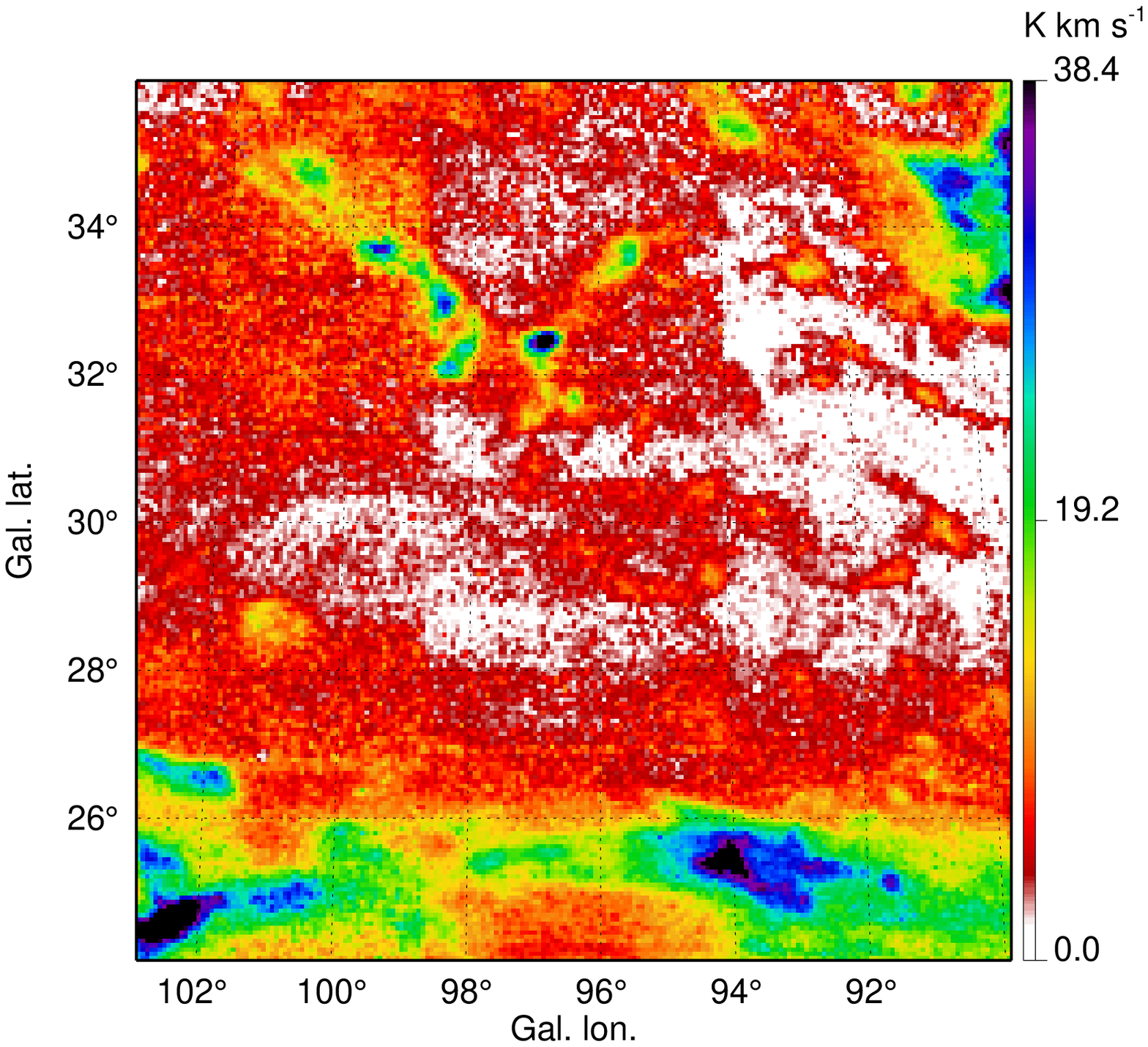}
\caption{
\wh\ maps for \nep\ (upper, middle, lower: LVC, IVC, HVC). Note that the
range in the colorbar in the HVC map is a factor of ten smaller so
that the noise level is apparent in the map.
}
\label{fig:nepWmaps}
\end{figure}

We have made \wh\ maps for each \ghigls\ field using the velocity cuts
in Table~\ref{compvel_table}.  Representative maps of \wh\ are shown
in Figure~\ref{fig:nepWmaps} for the three velocity components in
\nep.

\begin{deluxetable*}{lrrrrr@{\hskip 1.7em}rrrrr@{\hskip 1.7em}rrrrr}
\tablecolumns{16}
\tablewidth{0pc}
\tablecaption{Characteristics of \nh\ Component Maps (in $10^{19}$~cm$^{-2}$)}
\tablehead{
\colhead{Name} & \multicolumn{5}{c}{LVC} & \multicolumn{5}{c}{IVC} & \multicolumn{5}{c}{HVC}\\
\colhead{} & \colhead{$\langle N_{\rm H I} \rangle$} & \multicolumn{2}{c}{$N_{\rm H I}$ {Range}\tablenotemark{1}} & \multicolumn{2}{c}{$\sigma_{\rm N_{H I}}$\,\tablenotemark{2}}&
\colhead{\phm{aa}$\langle N_{\rm H I} \rangle$ } & \multicolumn{2}{c}{$N_{\rm H I}$ Range} & \multicolumn{2}{c}{$\sigma_{\rm N_{H I}}$} &
\colhead{\phm{aa}$\langle N_{\rm H I} \rangle$ } & \multicolumn{2}{c}{$N_{\rm H I}$ Range} & \multicolumn{2}{c}{\phm{11}$\sigma_{\rm N_{H I}}$}
}
\startdata
MC &   6.6 & $  3.8 $ & $ 11.4$ &   0.2 &    0.1 &   7.1 & $  4.4 $ & $ 13.2$ &   0.1 &    0.1 &   6.2 & $  0.0 $ & $ 21.2$ &   0.4 &    0.3 \\
BOOTES &   7.0 & $  5.2 $ & $ 11.4$ &   0.2 &    0.1 &   3.9 & $  1.3 $ & $ 12.1$ &   0.2 &    0.1 &   0.2 & $  0.0 $ & $  1.9$ &   0.1 &    0.1 \\
Necklace &   4.5 & $  3.7 $ & $  5.6$ &   0.1 &    0.1 &   4.4 & $  2.6 $ & $  5.9$ &   0.2 &    0.1 & \nodata & \nodata & \nodata & \nodata & \nodata \\
OX3 &  10.9 & $  7.4 $ & $ 15.9$ &   0.4 &    0.4 &   1.6 & $  0.0 $ & $  4.4$ &   0.4 &    0.5 &   1.6 & $  0.0 $ & $  7.6$ &   0.4 &    0.7 \\
N1 &   6.3 & $  4.2 $ & $ 14.3$ &   0.2 &    0.2 &   3.1 & $  1.5 $ & $ 15.1$ &   0.2 &    0.1 &   3.2 & $  0.0 $ & $ 14.2$ &   0.2 &    0.2 \\
G86 &  11.0 & $  4.5 $ & $ 25.1$ &   0.1 &    0.1 &   8.0 & $  1.3 $ & $ 37.3$ &   0.2 &    0.1 &   0.4 & $  0.0 $ & $  2.9$ &   0.1 &    0.1 \\
FLS &  17.6 & $ 11.9 $ & $ 26.9$ &   0.4 &    0.2 &   4.2 & $  2.0 $ & $  9.2$ &   0.4 &    0.3 &   2.0 & $  0.0 $ & $  7.5$ &   0.4 &    0.3 \\
MRK290 &  10.0 & $  7.0 $ & $ 18.8$ &   0.4 &    0.3 &   0.8 & $  0.0 $ & $  9.6$ &   0.2 &    0.1 &   4.1 & $  0.0 $ & $ 14.2$ &   0.5 &    0.3 \\
DRACO &   6.5 & $  4.6 $ & $ 16.3$ &   0.1 &    0.1 &  11.9 & $  4.2 $ & $ 35.4$ &   0.2 &    0.2 &   3.7 & $  0.2 $ & $ 16.0$ &   0.4 &    0.3 \\
GROTH &   6.5 & $  4.8 $ & $ 13.4$ &   0.3 &    0.2 &   3.1 & $  0.8 $ & $  4.7$ &   0.2 &    0.1 &   0.6 & $  0.0 $ & $  3.3$ &   0.3 &    0.2 \\
NEP &  25.9 & $ 12.5 $ & $ 56.6$ &   0.3 &    0.3 &  14.4 & $  2.7 $ & $ 56.7$ &   0.2 &    0.2 &   1.4 & $  0.0 $ & $  9.3$ &   0.4 &    0.3 \\
\NEPft  &  52.0 & $ 33.4 $ & $ 67.9$ &   0.5 &    0.4 &   6.1 & $  3.5 $ & $ 13.0$ &   0.2 &    0.2 &   2.3 & $  0.2 $ & $ 11.7$ &   0.4 &    0.2 \\
POL &  56.5 & $ 15.8 $ & $ 97.3$ &   0.9 &    0.5 &  10.3 & $  4.1 $ & $ 21.0$ &   0.5 &    0.3 &   0.4 & $  0.0 $ & $  5.0$ &   0.6 &    0.3 \\
POLNOR &  35.1 & $  8.8 $ & $ 84.3$ &   0.4 &    0.4 &   7.8 & $  1.8 $ & $ 16.9$ &   0.4 &    0.3 &   0.2 & $  0.0 $ & $  8.0$ &   0.6 &    0.2 \\
MRK205 &  23.4 & $  9.3 $ & $ 57.2$ &   0.4 &    0.3 &   5.0 & $  1.6 $ & $ 13.4$ &   0.4 &    0.2 &   0.2 & $  0.0 $ & $  9.9$ &   0.6 &    0.2 \\
DFN &   4.4 & $  2.0 $ & $  6.5$ &   0.4 &    0.2 &   6.9 & $  4.0 $ & $ 10.0$ &   0.4 &    0.3 &   2.6 & $  0.0 $ & $  9.7$ &   0.3 &    0.3 \\
SP &   6.2 & $  3.8 $ & $ 12.6$ &   0.2 &    0.2 &   3.7 & $  1.1 $ & $  9.6$ &   0.2 &    0.2 &   1.9 & $  0.0 $ & $  9.3$ &   0.3 &    0.2 \\
UM1 &   3.2 & $  1.0 $ & $  6.1$ &   0.5 &    0.4 &  13.2 & $  3.8 $ & $ 38.4$ &   0.4 &    0.4 &   1.8 & $  0.0 $ & $ 12.9$ &   0.4 &    0.3 \\
SPIDER &  19.9 & $  6.2 $ & $ 69.9$ &   0.3 &    0.3 &   7.3 & $  2.1 $ & $ 16.9$ &   0.3 &    0.3 &   0.2 & $  0.0 $ & $  4.2$ &   0.2 &    0.2 \\
SPC &  28.0 & $  9.9 $ & $ 83.7$ &   0.5 &    0.3 &   2.6 & $  0.2 $ & $ 10.9$ &   0.2 &    0.2 &   0.8 & $  0.0 $ & $ 25.9$ &   0.4 &    0.2 \\
1H0717 &  32.4 & $ 21.0 $ & $ 70.8$ &   0.5 &    0.4 &   3.5 & $  0.9 $ & $ 18.6$ &   0.4 &    0.2 &   0.7 & $  0.0 $ & $ 12.1$ &   0.5 &    0.2 \\
UMA &  27.5 & $ 10.2 $ & $ 74.4$ &   0.4 &    0.3 &   9.4 & $  3.7 $ & $ 24.2$ &   0.4 &    0.2 &   0.9 & $  0.0 $ & $ 31.6$ &   0.6 &    0.2 \\
HS0624 &  62.3 & $ 31.5 $ & $109.0$ &   0.5 &    0.8 &   2.6 & $  0.7 $ & $  6.6$ &   0.2 &    0.2 &   2.5 & $  0.0 $ & $  9.3$ &   0.4 &    0.3 \\
UM3 &  24.0 & $  4.5 $ & $ 59.7$ &   0.4 &    0.3 &   6.6 & $  1.7 $ & $ 19.7$ &   0.3 &    0.3 &   0.4 & $  0.0 $ & $  6.0$ &   0.4 &    0.2 \\
\grossan &   4.1 & $  2.4 $ & $  7.8$ &   0.5 &    0.2 &   1.6 & $  0.6 $ & $  3.0$ &   0.3 &    0.1 &   0.3 & $  0.0 $ & $  2.0$ &   0.4 &    0.2 \\
\LH &   3.3 & $  1.4 $ & $  7.2$ &   0.4 &    0.3 &   2.2 & $  0.4 $ & $  6.2$ &   0.3 &    0.2 &   0.1 & $  0.0 $ & $  1.9$ &   0.3 &    0.1 \\
MS0700 &  49.8 & $ 30.0 $ & $ 76.5$ &   0.5 &    0.4 &   2.3 & $  0.7 $ & $  6.5$ &   0.4 &    0.2 & \nodata & \nodata & \nodata & \nodata & \nodata \\
UMAEAST &  31.4 & $ 17.7 $ & $ 55.3$ &   0.4 &    0.3 &   9.7 & $  2.7 $ & $ 23.4$ &   0.4 &    0.3 &   2.9 & $  0.0 $ & $ 31.2$ &   0.7 &    0.4 \\
LOOP4 &  34.5 & $ 22.7 $ & $ 47.7$ &   0.4 &    0.4 &   8.7 & $  2.0 $ & $ 17.7$ &   0.4 &    0.3 &  0.02 & $  0.0 $ & $  3.5$ &   0.6 &    0.2 \\
NGC3310 &   5.8 & $  2.1 $ & $ 16.8$ &   0.3 &    0.2 &   4.1 & $  1.5 $ & $  7.7$ &   0.3 &    0.2 &   0.5 & $  0.0 $ & $  2.4$ &   0.4 &    0.2 \\
MRK9 &  43.9 & $ 25.1 $ & $ 67.6$ &   0.4 &    0.3 &   2.5 & $  0.6 $ & $  9.0$ &   0.2 &    0.1 & \nodata & \nodata & \nodata & \nodata & \nodata \\
AG &   5.2 & $  2.6 $ & $ 11.8$ &   0.3 &    0.2 &   9.5 & $  4.3 $ & $ 16.7$ &   0.2 &    0.1 &   3.9 & $  0.0 $ & $ 28.4$ &   0.2 &    0.2 \\
SUBA &  13.9 & $ 11.4 $ & $ 17.2$ &   0.3 &    0.3 &   6.8 & $  3.9 $ & $ 11.1$ &   0.4 &    0.3 &   1.2 & $  0.0 $ & $  8.9$ &   0.9 &    0.6 \\
MBM23 &  57.0 & $ 31.1 $ & $ 89.8$ &   0.5 &    0.5 &   8.2 & $  3.7 $ & $ 18.6$ &   0.3 &    0.2 & \nodata & \nodata & \nodata & \nodata & \nodata \\
091346A &  12.9 & $  9.3 $ & $ 17.7$ &   0.3 &    0.3 &   1.9 & $  0.7 $ & $  4.7$ &   0.4 &    0.3 &   0.2 & $  0.0 $ & $  1.5$ &   0.3 &    0.1 \\
MRK421 &   8.2 & $  4.7 $ & $ 14.4$ &   0.4 &    0.3 &   6.7 & $  2.4 $ & $ 23.0$ &   0.4 &    0.2 & \nodata & \nodata & \nodata & \nodata & \nodata \\
CDFS &   5.6 & $  4.0 $ & $  9.7$ &   0.3 &    0.4 &   0.8 & $  0.2 $ & $  2.8$ &   0.2 &    0.2 &   0.2 & $  0.0 $ & $  3.5$ &   0.3 &    0.3
\enddata
\label{n_table}
\tablenotetext{1}{The two columns give the lower and upper ends of the range of $N_{\rm H I}$ as defined by the 0.1\,\%
and 99.9\,\% percentiles of values in the map.
Negative \nh, which shows up at a level equivalent to the \nh\ noise in faint HVC fields, is excluded in this calculation.}
\tablenotetext{2}{First column: The uncertainty $\sigma_{\rm N_{H I}}$ is calculated from repeat observations where available (Table~\ref{f_table}), or from estimates of noise in the line emission, baseline, stray radiation, and scaling uncertainties following \citet{boot11}.  Second column: Complementary calculation from power spectrum analysis (Appendix~\ref{psanalysis}).}
\end{deluxetable*}

\subsection{\nh}
\label{nhres}

The quantity \wh\ is a direct observable.  Computation of maps of
column density \nh\ with allowance for the effects of optical depth is
discussed in Appendix~\ref{self}.  For the low column densities
characteristic of most \ghigls\ fields, the corrections are small and
so the \wh\ maps scaled by $C$ are very close to those shown as \nh\
in \citet{planck2011-7.12} for many of the \ghigls\ fields, aside from cropping, gridding differences
(Section~\ref{mapping}), and the effect the latter has on the baseline
removal.  In the results below and on the \ghigls\ archive \nh\ has been
calculated assuming $\Trs=80$~K (Equation~(\ref{correct2})).
On the archive we have made available FITS files of \nh\ maps
for each field.  Each file is a cube with five planes (0 to 4)
corresponding to the velocity components HVC, IVC, and LVC and in
addition IVC+LVC and HVC+IVC+LVC, respectively.  We note that all of
these maps are in units $10^{19}$~cm$^{-2}$. 

For each component map of the \ghigls\ fields, Table~\ref{n_table}
gives 
the mean \nh,
the lower and upper ends of the range of $N_{\rm H I}$ as defined
by the 0.1\,\% and 99.9\,\% percentiles,
and the uncertainty $\sigma_{\rm N_{H I}}$.

Our estimates of $\sigma_{\rm N_{H I}}$ in the first of two columns
are derived from differences in repeat observations.  Where only
single observations are available, $\sigma_{\rm N_{H I}}$ is the
summation in quadrature of noise, baseline, stray radiation, and
scaling uncertainties \citep{boot11}.  In Appendix~\ref{quality} we
present a complementary assessment of the uncertainties in \nh,
evaluating a second estimate of $\sigma_{\rm N_{H I}}$ with a power
spectrum analysis (Appendix~\ref{psanalysis}).  These values, in the
second of the two columns, are in good agreement with those from the
first approach.
Uncertainties that could arise from correction for opacity of the line
are discussed in Appendix~\ref{self}.

In the tabulated units of $10^{19}$~\cmm\ the mean column density in
LVC ranges from 3.2 in UM1 to 62.3 in HS0624;
in IVC, from 0.8 in MRK290 to 14.4 in \nep;
and in HVC, from undetectable to 0.02 in LOOP4, to 0.1 in \LH, and to
6.2 in MC.
Summing these, the mean total column density \nh\ ranges from 5.6 in
\LH\ to 67.4 in HS0624.

Even for the low column densities typical of the \ghigls\ fields the
data are of high sensitivity, with noise much less than the range of
\nh\ within the component maps.
Those with the highest range of \nh\ are the most useful for
correlating with dust maps, because the morphological match will be
better defined statistically.

To identify possible issues in data calibration and reduction, GBT ACS
data in a few of the targeted \ghigls\ fields are compared to data
from a new generation of wide-area \hi\ surveys, in the north EBHIS
(Appendix~\ref{ebhis}) and in the south GASS (Appendix~\ref{gass}).
As shown by \citet{boot11}, \ghigls\ data agree well with the LAB \hi\
survey data, in scale to within a few percent.  We find that the
agreement with EBHIS data is equally good.  A more limited comparison
with GASS reveals a calibration difference of about 6\,\%.
We note that comparison of GBT 21-cm values of \nh\ with those derived
from measurement of Ly$\alpha$ absorption toward high-latitude
quasars are in good agreement, in the ratio $1.00\pm0.11$
\citep{Wakker2011}.

\section{Angular Power Spectrum of \nh}
\label{power}

The two-dimensional angular power spectrum of a map (image of the sky)
$f(x,y)$ is the square of the modulus of the Fourier transform
$\tilde{f}(k_x,k_y)$ where $k$ is the spatial frequency (wavenumber)
in the Fourier plane \citep{mamd2007}:
\begin{equation}
P(k_x,k_y)  = \vert \tilde{f}(k_x,k_y) \vert^2 \,. 
\label{amplitude}
\end{equation}
We apodize the image $f(x,y)$, from which the median is removed, with
a cosine function along its boundary prior to the Fourier transform
operation in order to reduce edge effects, which otherwise produce a
centered cross in the $P(k_x,k_y)$ image \citep{MivilleDeschenes2002}.
In practice, apodizing over five pixels at each edge is normally
sufficient.

The collapsed one-dimensional power spectrum $P(k)$ is the azimuthal
average of $P(k_x,k_y)$ on a series of annuli of constant
$k=(k_x^2+k_y^2)^{1/2}$.  Uncertainties are assigned based on the
standard deviation of the mean within each of these annuli.
If the above-mentioned cross is problematic, then adopting the median
rather than the average is an effective alternative; in fact to be
conservative we always adopted the median.  
To mitigate against these effects further, the data point at the
lowest $k$ was excluded in the analysis below.

\begin{figure}
\centering
\includegraphics[width=1.0\linewidth]{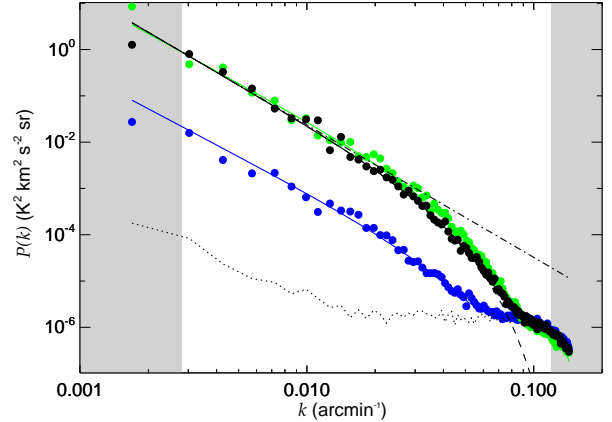}
\caption{
Median power spectra of component maps of \nh$/C$ in the \nep\ (black:
LVC; green: IVC; and blue: HVC).  Overlaid are fits (solid lines) to
the model in Equation~(\ref{plaw}).
As shown only for the case of LVC, this model consists of a power law
(dash dot line), modified by the GBT beam (resulting in dashed line),
plus the scaled noise template (dotted line).  Shaded regions define
ranges of $k$ excluded from the fit.
}
\label{anatomy}
\end{figure}

The dots in Figure~\ref{anatomy} illustrate the basic anatomy of the
power spectrum of our data.
At small $k$ (large spatial scales in the map), $P(k)$ roughly follows
a power law:
$P_{o} \; (k/k_o)^{\expon}$,
where $P_{o}$ is the amplitude of the power spectrum at some
representative scale $k_{o}$ and $\expon$ is the scaling exponent.
The exponent is alternatively called the spectral index or the slope
(in a log$-$log representation as in Figure~\ref{anatomy}).  The size
of the exponent and its variations from one type of map to another
provide insight into the turbulent structure of the ISM
\citep{henne12}.

At large k the power law is modified by the effect of the point spread
function (beam) of the telescope and the noise.
For a symmetrical Gaussian beam $\psf$ described in one spatial
dimension by its \fwhm\ we have
\begin{equation}
\psf(x)= \exp{ \left(- \frac{x^2}{2 \sigma^2} \right)}  \;   {\rm{and} }   \; \psfk(k) =  \exp{\left( -\frac{k^2}{2 \sigma^2_k}\right)} \,,
\label{psf}
\end{equation}
where the dispersion $\sigma = \fwhm/(2 \sqrt{2 \ln{2}})$ is related
to $\sigma_k$ by the Fourier relation $\sigma_k = 1/(2 \pi \sigma)$.
Above $k \approx 0.04$, where spatial scales are comparable to the
size of the GBT beam, the intrinsic $P(k)$ is reduced multiplicatively
by $\psfk^2(k)$.  This can be seen in the power spectra in
Figure~\ref{anatomy}.  
At even higher $k$, the noise dominates; the distinctive shape of the
noise power spectrum is discussed in Appendix~\ref{psanalysis}.

We have investigated the effect of the asymmetrical beam.  For the
range of spatial scales sampled before the signal disappears into the
noise, the optimal beam size is 9\farcm24 when treated as a
Gaussian.  As an alternative, starting with an image of the effective
beam calculated for our modified Bessel function gridded data cubes
(Section~\ref{mapping}), we computed from its power spectrum the
azimuthally averaged $\psfk^2(k)$.  From our model of the GBT beam we
estimate an additional uncertainty
\begin{equation}
\delta P(k) = b\, P(k)\, (1 - \psfk^2(k)) \,,
\label{beamerror}
\end{equation}
where $b$ is an adjustable fractional error that we set to \beamfrac.
This is added in quadrature to the uncertainty derived for each
annulus of constant $k$.  While it is only an approximation, $\delta
P$ has the desired effect of assigning lower weight to data strongly
affected by the beam and still fitting the noise adequately.  The
parameters derived from the fits below are not sensitive to the
precise choice of $b$.
We also mitigated against further uncertainty by excluding data
above a value \kmax; again the results below are not sensitive to the
precise value so long as the noise can be adequately assessed, and we
adopted \kmaxval.

\subsection{Power Spectrum Model}
\label{pswh}

The computed power spectrum is fit to data in the restricted
range of $k$ with a parameterized model consisting of a power law,
modified by the beam, plus noise:
\begin{equation}
P_{\rm model}(k) = \psfk^2(k) \; P_{o} \; (k/k_o)^{\expon} + \eta N(k) \,.
\label{plaw}
\end{equation}
The origin of the noise model, $N(k)$, scaled by fitting factor
$\eta$, is discussed in Appendix~\ref{psanalysis}.
The best-fit parameters are found using the above uncertainties as
weights in the IDL routine {\tt mpfit.pro}
\citep{markwardt2009}.  The derived model is plotted in
Figure~\ref{anatomy}.  We also inspected a plot of $P(k)/P_{\rm
model}(k)$ to verify that there was no bias hidden by the logarithmic
display.

For the three component maps of \nh\ of the \nep\ field the model fit
to the data (see Figure~\ref{anatomy}) yields exponents
\nepLVCslope, \nepIVCslope, and \nepHVCslope\ for LVC, IVC, and HVC,
respectively.  The uncertainties cited are the formal $1\sigma$ errors
from the fits and do not include any systematic uncertainties.  
From extensive testing of alternative choices in the fitting
analysis, we estimate that systematic uncertainties of the exponent
are of order 0.1 for the LVC map and probably somewhat smaller for the
IVC and HVC maps.  We also found that relative differences between the
exponents for different components are robust against the systematic
effects.  The IVC component has a marginally shallower spectrum
compared to the LVC component, as can be seen directly from the data
in Figure~\ref{anatomy}.

With respect to the summary by \citet{henne12}, the exponent for \nep\
LVC is roughly consistent with the value of $-2.75$ for \hi\ seen in
absorption \citep{desh2000}, but by contrast much shallower than the
exponent of $-3.6$ for \hi\ in emission found by \citet{mamd2003} in
an intermediate latitude field in Ursa Major.
The comparison of power spectra component by component among
different \ghigls\ fields is an interesting topic taken up in a
forthcoming paper.

\begin{figure*}
\centering
\begin{tabular}{ll}
\includegraphics[height=6.5cm, keepaspectratio]{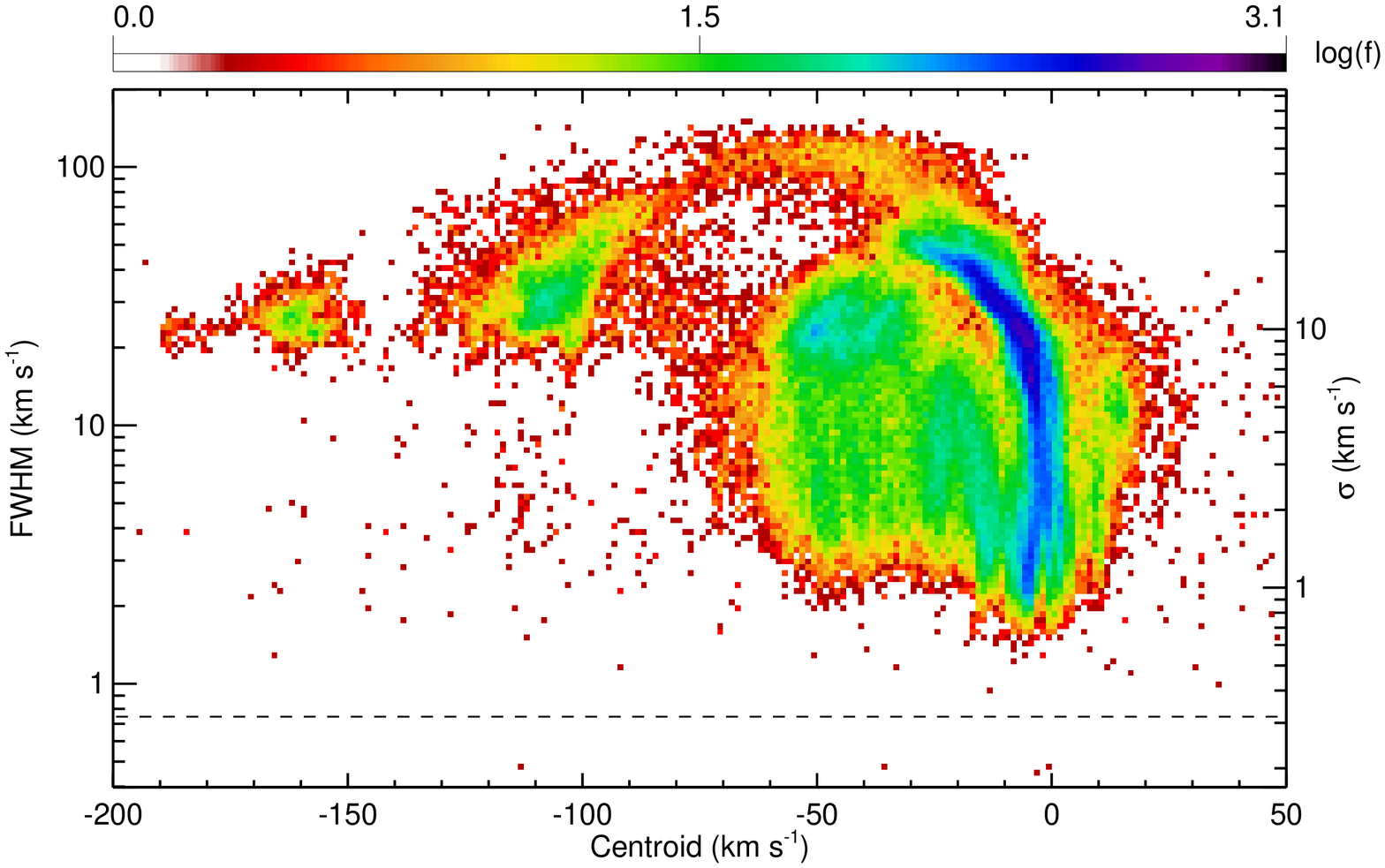}  &
\includegraphics[height=6.5cm, keepaspectratio]{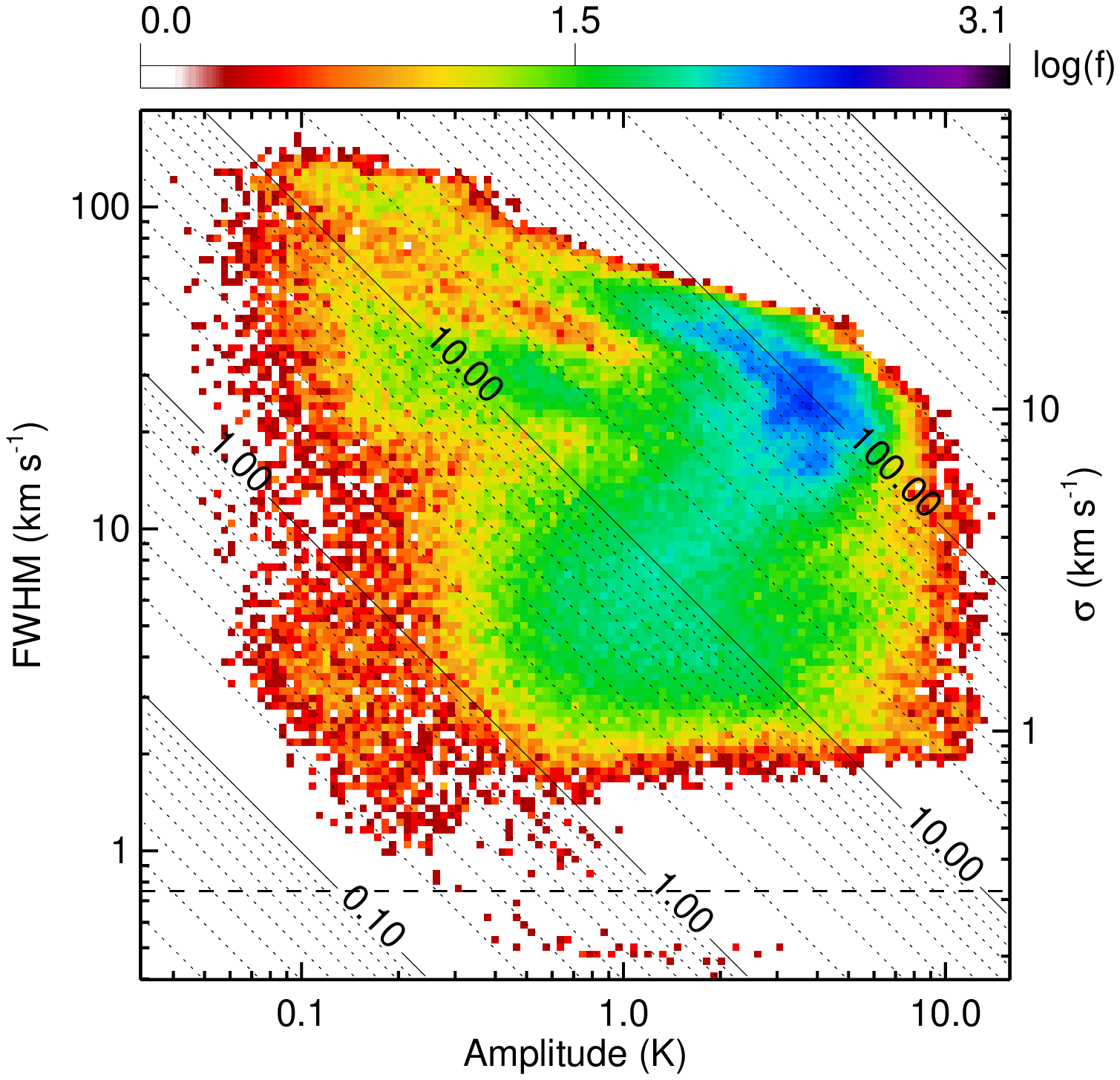}
\end{tabular}
\caption{
Histograms with respect to parameters of Gaussian components fit
to spectra in the \nep\ field.  The number of components in each
two-parameter bin, $f$, is shown on a logarithmic scale quantified by
the colorbar.
Left: \fwhm$-$centroid histogram.  Components of the LVC gas with
centroids near 0\,\kms\ have a large range of \fwhm\ and the ridge in
the histogram forms a distinctive vertical ``pillar."  At large \fwhm\
the ridge in the histogram ``arches" toward lower centroid velocities
because of components with unphysically large \fwhm\ fit to low
amplitude emission from both IVC and LVC gas (see text).
Right: \fwhm$-$amplitude histogram.  Lines of constant \wh\ (K~\kms)
are shown as solid and dotted diagonal lines.
Components with \fwhm\ $ < 1$~\kms\ (dashed line) are flagged as
unphysical; these typically have low \wh.  Any component with \wh\ $ <
1.0$ K~\kms\ is below the noise limit.
}
\label{fig:sigmav}
\end{figure*}

The values from \hi\ column density can be compared to those from dust
emission maps, for example the value of $-2.9$ for dust calculated for
four regions at $100~\mu$m \citep{gaut1992}.  Extending analysis to
the entire high latitude sky \citet{mamd2007} found a median $-2.93$
with considerable dispersion (0.3) from region to region and a trend
of flattening for fainter regions.  A value of $-2.7\pm0.1$ was
obtained for dust mapped in the Polaris flare with \Herschel/SPIRE
250~$\mu$m and IRIS 100~$\mu$m \citep{mamd2010}.
The \Planck\ 857\GHz\ mask-differenced power spectrum
\citep{planck2013-p11}, characterizing Galactic dust in the
intermediate latitude sky, has an exponent reaching $-2.6$
asymptotically at high $\ell$ multipoles, in the range relevant to our
analysis.

We note that the exponent for the \Planck\ 353\GHz\ EE and BB
polarized dust power spectra in the intermediate latitude sky
\citep{planck2014-XXX} is quite similar, $-2.42$, even though these
spectra depend additionally on the linkage of the statistical
properties of density and magnetic field geometry.

A detailed discussion of the relationship between the power spectra of
\nh\ maps and dust emission maps of the same \ghigls\ fields will also
be presented in the forthcoming paper.

\section{Gaussian Decomposition of Line Profiles}
\label{profile1}

With a goal of understanding the physical conditions and velocity
structure in the diffuse regions where these spectra originate, we
have analysed the \hi\ line profiles by decomposition into a series of
Gaussian functions using a method similar to that of \cite{haud00},
details of which can be found in another forthcoming paper (Blagrave,
K. et al.\ 2015, in preparation).  Each spectrum is fit within the
noise with multiple Gaussian components, with Gaussian parameters
amplitude, centroid velocity, and \fwhm\ (or dispersion $\sigma$).
Each spectrum is fit individually independent of its neighbors, but
further iterative modifications could include information from
neighboring solutions, resulting in a more spatially coherent set of
components.

As described further below (see also Blagrave, K. et al.\ 2015, in
preparation), using various simulations based on \citet{saur14} we
have found that the decomposition distinguishes effectively different
parcels of gas at different temperatures.  From simulated IVC plus LVC
cubes (as in Section~\ref{lih}), we have also examined various sets of
two-dimensional histograms of the Gaussian parameters in order to
differentiate IVC and LVC gas.

\subsection{Distributions of Recovered Gaussian Parameters}
\label{gparm}

We illustrate the results on \ghigls\ data using the \nep\ field for
which the average number of Gaussians fit to a given spectrum is 4.8.
Figure~\ref{fig:sigmav} shows a pair of two-dimensional histograms of
the parameters for all Gaussian components in the \fwhm$-$centroid and
\fwhm$-$amplitude planes.  The \fwhm-centroid histogram (left)
typically reveals what can be called ``pillars,'' vertically-aligned
features with an approximately constant centroid over a range of
\fwhm\ values. A particularly distinct example of such a pillar
appears at $v\sim0$~\kms\ in the range of the \nep\ LVC gas.  In other
\ghigls\ fields, pillars can be seen at other LVC and IVC centroid
velocities, with varying degrees of contrast.

There is a sharp decrease in the number of lines as the
recovered \fwhm\ decreases below 2~\kms\ (e.g.,
Figure~\ref{fig:sigmav}, right), similar to what has been found for
both the LAB \citep{haud07} and GASS \citep{kalb15} surveys: \hi\
emission lines narrower than \fwhm\ 2~\kms\ are rarely found.  A
\fwhm\ of 2~\kms\ could arise as the thermal line width of 100~K gas
or might reflect the contribution of turbulence if the gas is even
colder.  The few components with \fwhm\ $< 1$~\kms\ are unphysical,
the result of the automated fitting routine attempting to improve the
model by fitting Gaussians to rare remaining noise spikes.

There are a few other features in three-dimensional parameter space
that we have identified as unphysical Gaussian components that arise
because of the limitations of automated unconstrained Gaussian fitting
routines.  For example, a number of features with very low \wh\
improve the model by fitting Gaussians to residual baselines.  These
can be excluded in Figure~\ref{fig:sigmav} right; any component with
\wh\ $ < 1.0$ K~\kms\ is below the detection limit as determined by
the error analysis in \citet{boot11} and uncertainties in
Table~\ref{n_table}.

At the other extreme, there are very broad components with low
amplitude that show up consistently with centroids between LVC and IVC
in the \fwhm$-$centroid histograms (Figure~\ref{fig:sigmav} left),
creating ``arches" between the pillars.  Similarly, arches of
extremely broad components also appear in the \fwhm$-$centroid
histogram with centroids between LVC and HVC.
We have demonstrated that these too are unphysical, as follows.  
As in Section~\ref{lih}, we used the hydrodynamical simulations of
\citet{saur14} to create an IVC plus LVC cube.  Although this cube
contains no intrinsic very broad components, we found on Gaussian
decomposition (Blagrave, K. et al.\ 2015, in preparation) that such
very broad components nevertheless do arise, again appearing at
intermediate velocities between the LVC and IVC pillars in the
\fwhm$-$centroid histogram.  In the simulation and actual data these
are artifacts of the Gaussian fitting routine, a result of fitting a
single Gaussian to a merger of two (or more) weaker, narrower Gaussian
components.

\begin{figure}
\centering
\includegraphics[width=0.8\linewidth]{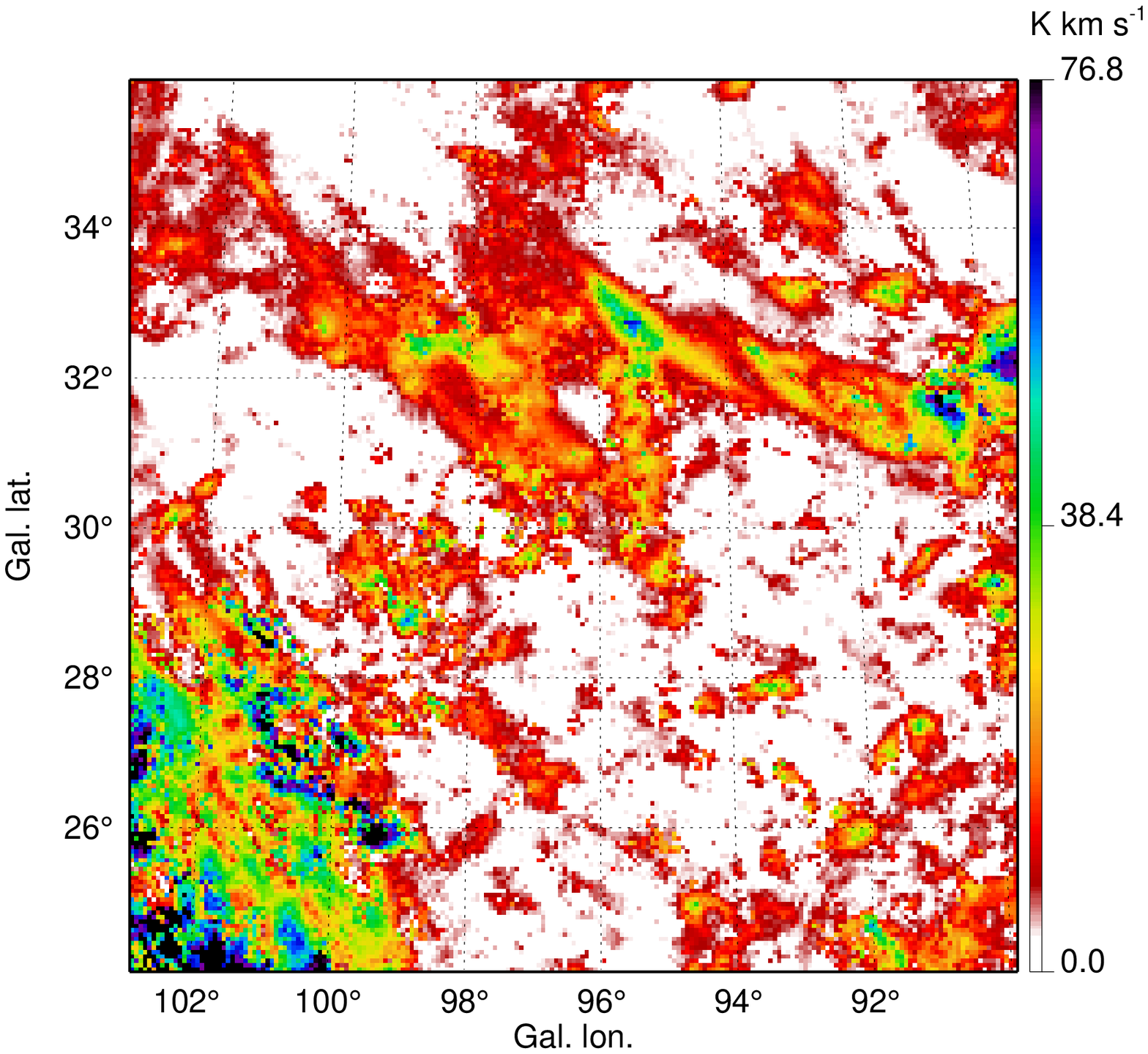}
\includegraphics[width=0.8\linewidth]{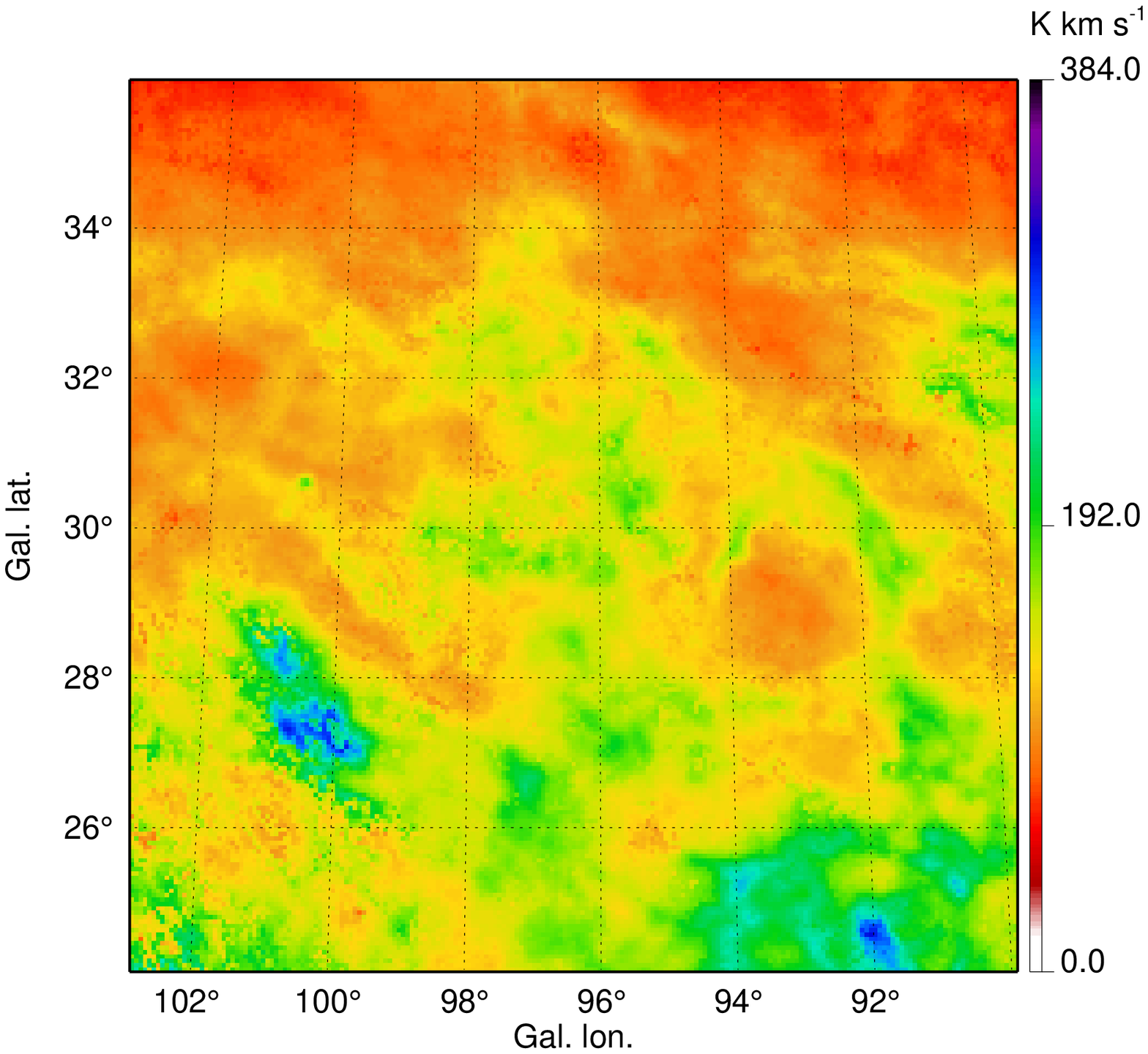}
\caption{
Column density maps for \nep\ for gas in the LVC range in units \nh$/C$
from Equation~(\ref{correct2}) after Gaussian decomposition of the
line profiles.
Upper: Column density from narrow components ($\fwhm < 7.1$~\kms) and
assuming $\Trs = 80$~K.  Note that the range in the colorbar is a
factor of five smaller than for the map below.  This map can be
interpreted as from the CNM.  Filamentary structure is seen with a
roughly diagonal orientation as in individual channel maps (cf.\
Figure~\ref{fig:pvd}, upper right).
Lower: Column density map obtained by subtracting the above CNM
map from the map of the total LVC component in
Figure~\ref{fig:nepWmaps}, upper.  This is interpreted as being from
the WNM and in this field is very similar to the total.
}
\label{fig:nepWmapsGauss}
\end{figure}

\subsection{Maps of \nh\ for Components with Different Line Widths}
\label{linew}

The aforementioned LVC, IVC, and HVC velocity-selected components
would be closely related to the Gaussian components forming vertical
pillars within a range of centroid velocity in \fwhm$-$centroid
histograms.  For example we would select Gaussian components as LVC if
their centroids fall in the velocity range for LVC from
Table~\ref{compvel_table}, which for \nep\ is $-20.5 < \vrl <
+47.9$~\kms.

These histograms suggest a possible new complementary direction for
the subdivision of a cube into components on the basis of distinctive
\fwhm.
As mentioned above narrow and broad components are commonly associated
with two phases of the diffuse neutral atomic ISM: the CNM and the
WNM, respectively.
Analysis of histograms from all of the \ghigls\ Gaussian components
suggests that the natural division $\fwhm_d$ between narrow and broad
Gaussian components for the LVC emission occurs at about 7.1~\kms.
With this division we can make \nh\ maps corresponding to CNM and WNM
by summing up the contributions from the appropriate components.  To
allow for uncertainties introduced by the Gaussian decomposition model
and particularly by a sharp divide at $\fwhm_d = 7.1$~\kms, we have
employed a Monte Carlo approach to produce and average $\sim100$
versions of the map using $\log \fwhm_d$ drawn from a normal
distribution with a $1\sigma$ dispersion of 0.05~dex about the
indicated mean.

The CNM--WNM separation using the Gaussian component approach can be
tested using simulations in which the true results are known \emph{a
priori} because the gas temperature is known.  For example, we have
studied (Blagrave, K. et al.\ 2015, in preparation) a simulation from
\citet{saur14} for which $\zeta=0.2$, $v_s=12.5$~\kms, and
$n_0=1.0$~cm$^{-3}$, resulting in $f_{\rm CNM}\sim0.29$.
As we did above for the \ghigls\ observations, in that work we
selected a division in \fwhm\ based on the distribution of Gaussian
components for the simulation, in that case finding
$\fwhm_d=4.5$~\kms.  This results in $f_{\rm CNM}\sim0.32$, consistent
with the actual $f_{\rm CNM}$ for the simulation.  This test also
showed reasonable agreement between narrow and broad-component column
density maps created from the Gaussian components and corresponding
maps created from gas in known temperature ranges.

The CNM \nh\ map thus made from the selection of Gaussian components
for \nep\ LVC is shown in Figure~\ref{fig:nepWmapsGauss}, upper, as
\nh$/C$.
The CNM map has a low column density compared to the total LVC but
also highlights the filamentary structure running roughly diagonally
across the field.
The fraction of CNM in the LVC of \nep\ by column density is $f_{\rm
CNM}\sim0.08$, reflecting the dearth of narrow Gaussian components
identified along many lines of sight in the NEP.  We note that the CNM
fraction by mass could be different, depending on the relative
distances of the CNM and WNM gas.
This value of $f_{\rm CNM}$ can be compared to those obtained directly
from absorption line spectra.  \citet{heiles03} found a global ratio
of CNM to total \nh\ of 0.39 for the Arecibo sky.  This is an upper
limit for the mass fraction of CNM because of the systematic
difference in distance between the CNM and WNM.  Likewise,
\citet{dick09} found a fraction $f_{\rm CNM}\sim0.15-0.20$ for the
outer Galaxy.
The result for CNM in the LVC range in \nep\ is therefore lower than
these global values.

We note that the observed low $f_{\rm CNM}$ in \nep\ can be described
well with appropriate tuning of the parameters in the \citet{saur14}
simulations.  For example, some simulations with $\zeta = 0.3$,
$v_s\sim15$~\kms, and $n_0=1.0$~cm$^{-3}$ result in $f_{\rm
CNM}=0.11$.

Also shown in Figure~\ref{fig:nepWmapsGauss}, lower, is the WNM
\nh\ map.  Because of the complications of the unphysical arches at
high \fwhm, this has been produced simply by subtracting the CNM map
from that of the total LVC emission (Figure~\ref{fig:nepWmaps},
upper).  The WNM map is quite similar to the total LVC map, given the
low $f_{\rm CNM}$.

\begin{figure}
\centering
\includegraphics[width=1.0\linewidth]{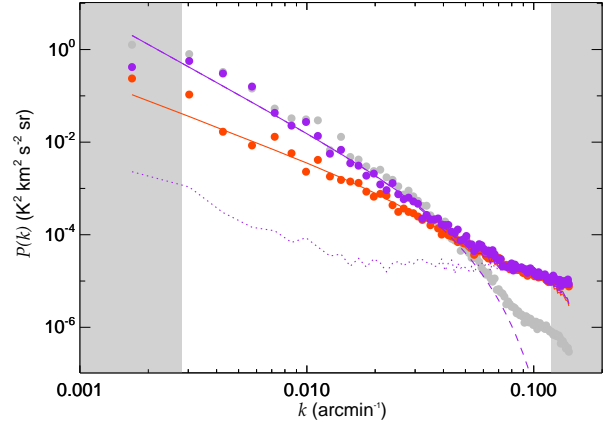}
\caption{
Power spectrum of \nh$/C$ of the \nep\ as in Figure~\ref{anatomy}, but
for LVC emission separated into CNM (red) and WNM (purple) components
(Figure~\ref{fig:nepWmapsGauss}).
Overlaid as gray circles is the power spectrum for the total LVC
\nh$/C$ from Figure~\ref{anatomy}.
The Gaussian decomposition introduces an additional uncharacterized
noise component in the \nh\ maps near the pixel scale, reflected at
high spatial frequencies in the power spectrum above the modelled beam
turnover (compare the purple and gray points); see text.
}
\label{fig:nepGaussPS}
\end{figure}

\subsection{Power Spectra of Maps of \nh\ for Components with Different Line Widths}
\label{plinew}

The power spectra of the \nh\ maps of the LVC CNM and WNM components
for \nep\ are shown in Figure~\ref{fig:nepGaussPS}.  The WNM component
accounts for most of the \nh\ of the LVC emission and so its power
spectrum is similar to that of the total LVC emission in
Figure~\ref{anatomy}.  The CNM spectrum has less power overall and a
shallower dependence on $k$.

Note that the shape of these power spectra at high $k$
(Figure~\ref{fig:nepGaussPS}) is markedly different from that of the
integrated \wh\ maps (Figure~\ref{anatomy}).  Despite attempts that
can be made to keep solutions smoothly varying from pixel to pixel,
the Gaussian fitting introduces an additional
uncharacterized noise contribution near the pixel scale, which
is reflected in the component \nh\ maps and propagates to a larger
noise in the power spectrum at high $k$ above the modelled beam
turnover of the signal.  In the model fitting we allow for this by the
simple device of increasing the beam uncertainty parameter in
Equation~\ref{beamerror} to $b = 0.3$, which in effect gives lower
weight to the many data values at the highest $k$ where noise now
dominates.

The model fits to the LVC WNM and CNM component maps (see
Figure~\ref{fig:nepWmapsGauss}) yield exponents of \nepLVCbroad\ and
\nepLVCnarrow, respectively.
The exponent for the WNM component map is close to the exponent for
the total LVC emission, \nepLVCslope\ (Section~\ref{pswh}), within the
systematic errors, as expected because this component contains most of
the mass.  However, the power spectrum for the CNM component map is
clearly shallower than for the WNM, a differential result that
is robust against systematic effects of how the model is fit.
The significantly flatter spectrum found for the CNM map quantifies
what can be foreseen readily in the maps themselves: the CNM map has
more high-contrast small angular scale filamentary structure than the
WNM map.

Such a shallow dependence in a narrow-line component is not unexpected.
The spectrum of the density on two-dimensional slices through
the simulated cube used for the test in Section~\ref{linew} has an
exponent of $-1.3$, a much shallower dependence than from the
Kolmogorov exponent of $-8/3$.  Thus the spectrum of the
three-dimensional density would be $-2.3$ and furthermore this would
be the exponent of the column density map of the cube
\citep{m-dlf2003}.

\citet{saur14} also expect from their simulations of thermally
bistable gas that the CNM power spectrum will be shallower than for
the warm gas.
Indeed, power spectra of \nh\ maps produced from these simulations by
selecting on the temperature of the gas (Blagrave, K. et al.\ 2015, in
preparation) have much shallower power spectra for the CNM
($\sim-1.9$) as compared to the WNM ($\sim-2.9$).\footnote{
Another way to generate such a shallow
power-law dependence is with cold and supersonic gas, as seen in
simulations of isothermal high Mach number flows
\citep[e.g.,][]{kim2005}; however, these conditions seem less relevant
to the general interstellar medium than those in the simulations of
\citet{saur14}.}

In addition to small scale structure corresponding to enhanced
concentration of \hi\ in the CNM, there are also possible effects from
cold neutral gas becoming molecular and thus leaving structure in the
remaining \hi, albeit probably with lower contrast.  The molecular
transition is not modeled in these particular simulations.

The CNM angular power spectra described here for both the \nep\ LVC
range and the simulated data are significantly shallower than anything
noted previously for \hi\ emission, \hi\ absorption, CO line emission,
or dust emission (see Figure~10 in \citealp{henne12}).

\section{Relationship of \hi\ CNM structure to the orientation of the magnetic field}\label{magnetic}

\begin{figure}
\centering
\hspace*{-0.8cm} 
\includegraphics[width=0.69\linewidth]{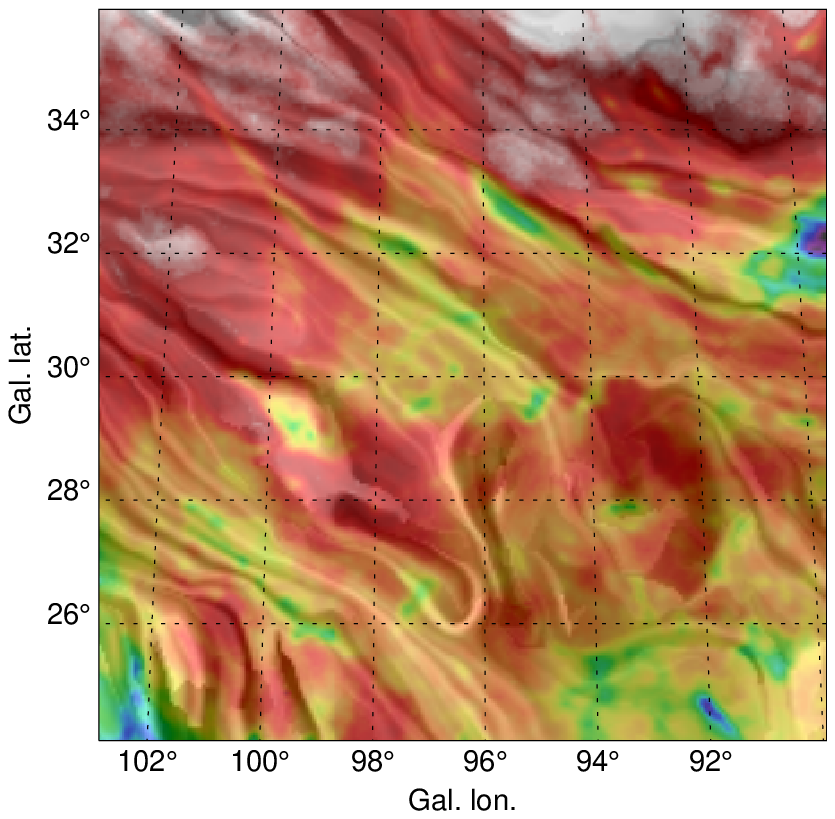}
\caption{
View of the magnetic field measured by \Planck\ overlaid on a map of
\nh\ from selected LVC gas in \nep.  The``drapery" pattern, produced
using the line integral convolution (LIC, \citealp{cabral93}),
indicates the orientation of magnetic field projected on the plane of
the sky, orthogonal to the observed dust polarization
\citep{planck2014-XIX}.  The colors represent \nh\ integrated over
seven channels at the peak of the LVC emission ($-6.4$ to
$-1.6$~\kms), highlighting the structure of CNM gas.  This illustrates
how the orientation of the CNM gas filaments tends to follow the
magnetic field across the upper part of the image, among several
interesting relationships between the gas and the field (see text).
}
\label{fig:LIC}
\end{figure}

In Section~\ref{illu} we commented on the striking filamentary
structures in the LVC channel maps of \nep\ that cross the region
roughly diagonally (lower right to upper left).  See the channel map
in Figure~\ref{fig:pvd}.  These features are fairly narrow in line
width (present over only a few adjacent channel maps, as is evident in
the movie of the cube) and so imprint on the \nh\ map made from the
narrow Gaussian components (Figure~\ref{fig:nepWmapsGauss}, upper).
As is illustrated in Figure~\ref{fig:LIC}, this filamentary \hi\
structure in the condensed CNM in \nep\ is aligned roughly parallel to
the direction of the Galactic magnetic field projected on the plane of
the sky as inferred from the \Planck\ 353\,GHz thermal dust
polarization map \citep{planck2014-XIX}.

This relative orientation is in accord with the recent finding that
the magnetic field tends to be oriented parallel to the elongation of
filamentary dust structures in both the high latitude sky
\citep{planck2014-XXXII} and in nearby Gould Belt molecular clouds
\citep{planck2015-XXXV}.  Such a systematic tendency in relative
orientation is important for understanding the observed power in
B-mode relative to E-mode dust polarization
\citep{planck2015-XXXVIII}.

It is also interesting to note that the Cygnus spur end of Loop~III
seen in synchrotron emission also crosses \nep\ on the same diagonal
and that the polarization of the synchrotron emission indicates that
the magnetic field is parallel to the loop \citep{planck2014-a31}.
Thus in \nep\ the predominant orientation of the field revealed by
thermal dust and synchrotron polarization is similar.

The observed \Planck\ polarization arises from dust not just in the
CNM gas, but in the WNM gas within which the CNM is embedded.  Thus
the field orientation highlighted in the CNM structure is more
pervasive.  (In addition, dust in the IVC contributes to the total
emission, with about half the emissivity and perhaps some
polarization.)  The relationship between gas and magnetic field is
quite interesting in its complexity.  For example, the field appears
to wrap around a depression in the gas emission at $(l, b) = (93\deg,
28\deg)$ (see also the total LVC in Figure~\ref{fig:nepWmaps}, upper),
suggestive of a bubble.  Below that the field appears to wrap around
the edge of the enhanced gas emission.

\section{Conclusions}
\label{conclusions}

In this paper we present data from the GBT \hi\ survey, \ghigls.  This
deep/sensitive survey covers targeted regions of the intermediate
Galactic latitude sky including several with distinctive IVC and/or
HVC features.

The data have been calibrated and corrected for stray radiation
following \citet{boot11}.  Data for most regions has an rms noise
$\lesssim100$~mK in a 0.8~\kms\ channel.  The effect of 21-cm line
opacity on the calculated column density \nh\ and various other small
sources of uncertainty in the \ghigls\ measurements of \nh\ are
discussed.

\ghigls\ data agree well with the LAB \hi\ survey data, in scale to
within a few percent.  We find that the agreement with EBHIS data is
equally good.  A more limited comparison with GASS reveals a
calibration difference of about 6\,\%.

We divide the \hi\ emission into components with different velocities
and produce \nh\ maps of LVC, IVC, and HVC components.  Structure in
these maps is quantified by the angular power spectrum modelled with a
power law.  For the \nep\ field, the power-law exponent for LVC
(\nepLVCslope) is close to that found for dust maps in the
intermediate Galactic latitude sky; the exponents for IVC and HVC
reveal a marginally flatter power-law behavior (exponents
\nepIVCslope\ and \nepHVCslope, respectively).

We fit the spectral line profiles with multiple Gaussian components to
differentiate gas characterized by different line widths, enabling a
subdivision into emission by two ISM phases, the WNM and CNM.  The CNM
map of \nh\ is unique in its angular power spectrum, having a
power-law exponent of \nepLVCnarrow\ for LVC in the \nep\ field,
shallower than seen in any component map but consistent with the
power-law exponent of CNM seen in simulations.  The flatter power-law
behavior reflects more small scale structure associated with this
phase.

There is evidence that filamentary structure in the \hi\ CNM gas is
aligned with the Galactic magnetic field revealed by \Planck\
polarization.

\ghigls\ data have already been used productively in a number of
Galactic and extragalactic applications and should be interesting for
many more.  On the \ghigls\ archive (\url{\ghiglsarchive}),
fully-reduced data cubes, along with movies and \nh\ component maps,
are available for inspection and downloading.

\acknowledgments

We acknowledge support from the Natural Sciences and Engineering
Research Council (NSERC) of Canada.  We thank the NRAO staff for their
outstanding support, in particular G.I.~Langston for providing code
that was important to the data reduction.  We are grateful to Eleonore
Saury for making available data cubes from her hydrodynamical
simulations and to Daniel Lenz, Peter Kalberla, and the EBHIS team for
providing unpublished data cubes from the EBHIS survey.
We thank the referee for constructive comments that have led to
improvements in this presentation of our results.


\appendix 

\section{Data Obtained Using the GBT Spectral Processor } 
\label{arch}

As indicated by the notes in Table~\ref{f_table}, \spcount\ fields
including the central portion of the BOOTES field were observed with
the now-retired GBT Spectral Processor (GBT SP).  As is the case for
the GBT ACS data, these data were collected in in-band
frequency-switched mode, except for the FLS field observed with
out-of-band frequency switching.  The FLS data were regridded from the
observed 0.5~\kms\ spaced channels to the adopted ACS spacing of
0.807~\kms, using a cubic spline interpolation.
For the other fields the channel spacing was maintained as observed at
1.03~\kms\ (for the entire BOOTES data, this was adopted even for the
ACS portion).  The number of repeats and the resulting emission-free
channel noise, $\sigma_{\rm ef}$, are found in Table~\ref{f_table}.

We processed the raw GBT SP data following the GBT ACS
pipeline described in Section~\ref{spectrometer} and \citet{boot11},
including the benefit of the improved stray radiation correction over
\citet{lock05}.  For consistency with the GBT ACS cubes, the spectra
were gridded on a $3\farcm5$ grid using the modified Bessel function.

There is a slightly different calibration unique to the GBT SP because
in that era there was a different noise diode in the GBT 21-cm
receiver.  Therefore, the calibration discussed in \citet{boot11},
specifically the scaling of $\Tra$ by the factor $1.024\pm0.009$, is
not relevant.  The scaling of $\Tra$ appropriate to the GBT
SP data was investigated in two ways.
The first approach used archival GBT SP data on the S6 and S8
calibration standards.  For these, the $XX$ and $YY$ spectra are
slightly inconsistent with each other -- differing by 5 to 10\,\% --
but their average spectrum, $(XX+YY)/2$, is reproducible and indicates
consistently that the GBT SP measure of the main beam temperature
$\Trb$ is 1.05 to 1.08 larger than for the GBT ACS spectra of these
calibrators.
The second approach compared regions of overlap between spectral cubes
made with GBT SP observations with cubes made with GBT ACS
observations: the original SP central field in BOOTES with the ACS
flanking fields, UM2M with NGC3310, UM2M with 09A079, UM3 with UMA
(and LISZTA), and FLS with \nep\ (see Figure~\ref{ffig}).  These
comparisons indicate overestimates of the same order (1.05 to 1.08).
Therefore, in the final pipeline processing the GBT SP $\Tra$ data
were scaled using a factor $\spacs$ $(\pm\, 0.01)$ instead of $1.024$
to bring the GBT SP data onto the same $\Trb$ scale as the GBT ACS
data.

\section{Archival Data Obtained for \GR\ Using the GBT Auto-Correlation Spectrometer} 
\label{arch1}

The low column density region within the Lockman Hole is covered in
part by the \hi\ survey \LH\ using the SP (Appendix~\ref{arch}) and an
overlapping field \grossan\ observed by \citet{Grossan2012} using the
ACS (proposal GBT/09A-079).  The latter data were obtained to estimate
the Galactic dust foreground for studies of the CIBA using a
160~\micron\ map from \Spitzer\ and to probe the relation between dust
and gas in this very low dust regime.
Because of the low column density it is essential to correct the \hi\
spectra for stray radiation before embarking on the science analysis.
Therefore, we have reprocessed these ACS data using the pipeline that
we developed (Section~\ref{obs}).  This cube is available on the
\ghigls\ archive and we have used it in the \lhm\ mosaic described in
Section~\ref{makemosaics}.  These observations were carried out by
mapping several overlapping subfields, some of which involved scans
along constant Galactic longitude rather than the more usual constant
Galactic latitude.  The subfield structure and irregular coverage is
not detailed in Table~\ref{f_table}, but is imprinted in the noise and
weight maps.

\section{Optical Depth Effects}
\label{self} 

A diagnostic warning that an observed \hi\ spectrum might be affected
by opacity is if $\Trb$ approaches the (plausible) spin temperature
$\Trs$.  In the extreme the line profiles might become flat topped,
but that is rarely encountered here.  We wish to explore more
subtle effects.

In the absence of absorption-line measurements to combine with the
emission measurements to distinguish the effects of optical depth
$\tau$ \citep{stra04,dick09}, as is the case for \ghigls\ data,
it is nevertheless possible to make an estimate of the potential
optical depth effects for different assumed $\Trs$.
As a reasonable spin temperature we adopt $\Trs = 80$~K which is the
collisional temperature found from intermediate-latitude $H_2$
observations for column densities near $10^{20}$~\cmm\
\citep{gill06,wakk06}.  This would be appropriate for CNM.
The CNM equilibrium temperature is not constant but depends
inversely on the local density.  Estimates of the CNM temperature in
the Galactic plane do show lower values \citep{dickey03}; however,
\citet{heiles03} find that for $|b| > 10\deg$ and CNM column densities
near $10^{20}$\,\cmm, the mass weighted $\Trs$ is 70\,K (median) and
108\,K (average).

\subsection{Complications in a Multi-phase ISM}\label{multiphase}

Differences among and within individual observed \hi\ line profiles
for different lines of sight and their overall complexity are
reminders that a single $\Trs$ is naive.  The emission is summed over
gas in different stable phases of atomic gas with different
temperatures, the CNM and WNM, and there could be substantial amounts
of warm thermally unstable gas as well
\citep{heiles03,haud07,henne12,saur14}.  While $\Trs$ for the CNM
might be near 80~K, $\Trs$ for the WNM is much higher ($> 300$~K;
\citealp{dick09}).  One could carry out component (phase) separation
by line profile decomposition (see Section~\ref{profile1}), to segment
CNM and WNM and then assign different $\Trs$ in correcting these
classes of profile. 
But even this would not capture the complexities arising from
the relative geometries of the gas in different phases.

However, in \ghigls\ the highest values of $\Trb$ in the observed
profiles are caused by the CNM.  
We have shown this by making a cube of the CNM gas emission
using the Gaussian decomposition method discussed in
Section~\ref{profile1}.  For each voxel we then formed the ratio
$\Trb({\rm CNM})/\Trb$ and found that the ratio approaches unity for
higher $\Trb$.  Because optical depth corrections become important for
$\Trb$ comparable to $\Trs$, which is accentuated in CNM gas where
$\Trs$ is low, such high-$\Trb$ parts of the profile
are most in need of correction, using a $\Trs$ appropriate to CNM.

On the other hand at intermediate latitudes the WNM produces broad
profiles that have low $\Trb$ and so the magnitude of a correction
even using a $\Trs$ that is inappropriately low will be
inconsequential.
Therefore, a simple correction using a single $\Trs$ appropriate to
CNM might not be unreasonable.  That is the approach that is evaluated
quantitatively here.

\subsection{Impact on \nh}
\label{under} 

The actual column density is always larger than the direct line
integral $N_{\rm H I}(\infty)$ in Equation~(\ref{thin}) and within our
assumptions is given by
\begin{eqnarray} \label{correct}
N_{\rm H}(\Trs)/C & = & \int \Trs \tau\ \mathrm{d}v  \label{correct1} \\
& = & \Trs \int -\log(1 - \Trb/\Trs)\ \mathrm{d}v  \label{correct2} \\
& = &  W_{\rm H I} +  1/(2 \Trs) \int \Trb^2\ \mathrm{d}v + \ldots \label{correct3}
\end{eqnarray}
The first-order correction, $1/(2 \Trs) \int \Trb^2\ \mathrm{d}v$ from
Equation~(\ref{correct3}), is quadratic in $\Trb$.  Therefore, the
$\Trb^2$ cube, and maps of integrals made from it, reveal where there
is sensitivity to optical depth.  The form of the first-order
correction has the great utility of showing the explicit scaling with
$1/(2 \Trs)$, quantifying the importance of an appropriate choice of
$\Trs$.

\begin{figure}
\centering
\includegraphics[width=0.9 \linewidth]{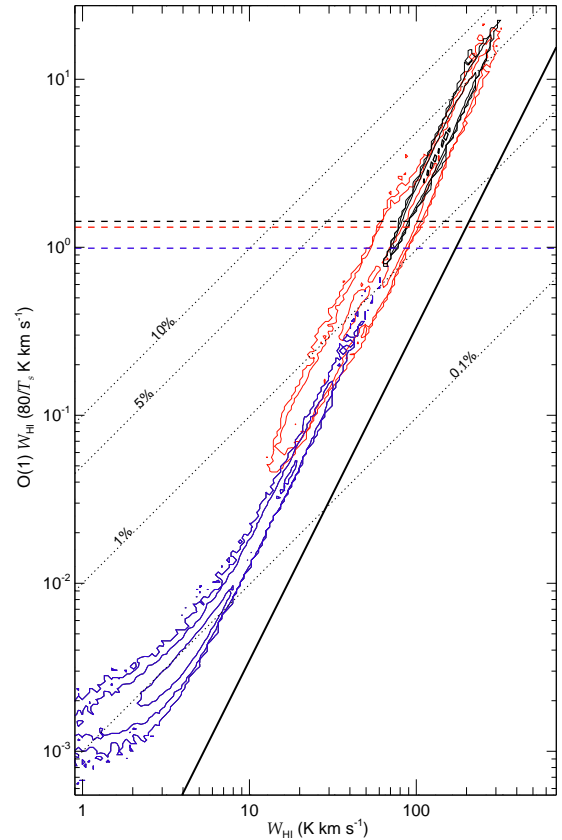}
\caption{
Contours of histogram of the first-order correction, $1/(2 \Trs) \int
\Trb^2\ \mathrm{d}v$, as a function of the optically thin solution
\wh\ for \nep, here using $\Trs = 80$~K.  Values for the three
components are overplotted in different colors: LVC(black), IVC (red),
and HVC (blue).  A line segment of slope two reveals the approximate
quadratic dependence.
The unit-slope dotted lines indicate different percentage corrections
\wh.
Horizontal colored dashed lines indicate the uncertainties
$\sigma_{\rm W_{H I}} = \sigma_{\rm N_{H I}}/C$ for the three
components in the \nep\ field, from Table~\ref{n_table}.}
\label{fig:nepW_tau}
\end{figure}

In Figure~\ref{fig:nepW_tau} we show the first-order correction from
Equation~(\ref{correct3}) as a function of the optically thin solution
\wh\ pixel by pixel in NEP, distinguishing each of the three velocity
range components.  This log$-$log plot reveals an approximate
quadratic dependence of the correction in \wh\ (such a line is
plotted), so that the fractional corrections scales as \wh,
potentially rising above the uncertainty $\sigma_{\rm N_{H I}}/C$.
The spread in this figure is consistent with expectations from simple
model spectra using Gaussian components.  The vertical height at a
given \wh\ is inversely related to the line width and the vertical
spread is also influenced by the amount of overlap of the components
in velocity.

This correction can be compared to the typical uncertainty in \nh\
discussed in Appendix~\ref{quality}.  For \nep\ a value
$\sigma_{N_{\rm H I}}$ of $0.2 \times 10^{19}$\,\cmm\ from
Table~\ref{n_table} corresponds to \wh\ $\sim~1$~K~\kms.  Any opacity
correction below this limit can be ignored.  Above this limit, LVC is
affected at the 2$-$8\,\% level; the uncertainty in the correction
will of course be less than this.  Some IVC is affected in this same
range but other IVC much less.  The HVC component is least affected by
any opacity (at the lower left the upturn relative to quadratic
relates to the positive definite nature of the correction even for low
$\Trb$).

\begin{figure}
\centering
\includegraphics[width=0.8\linewidth]{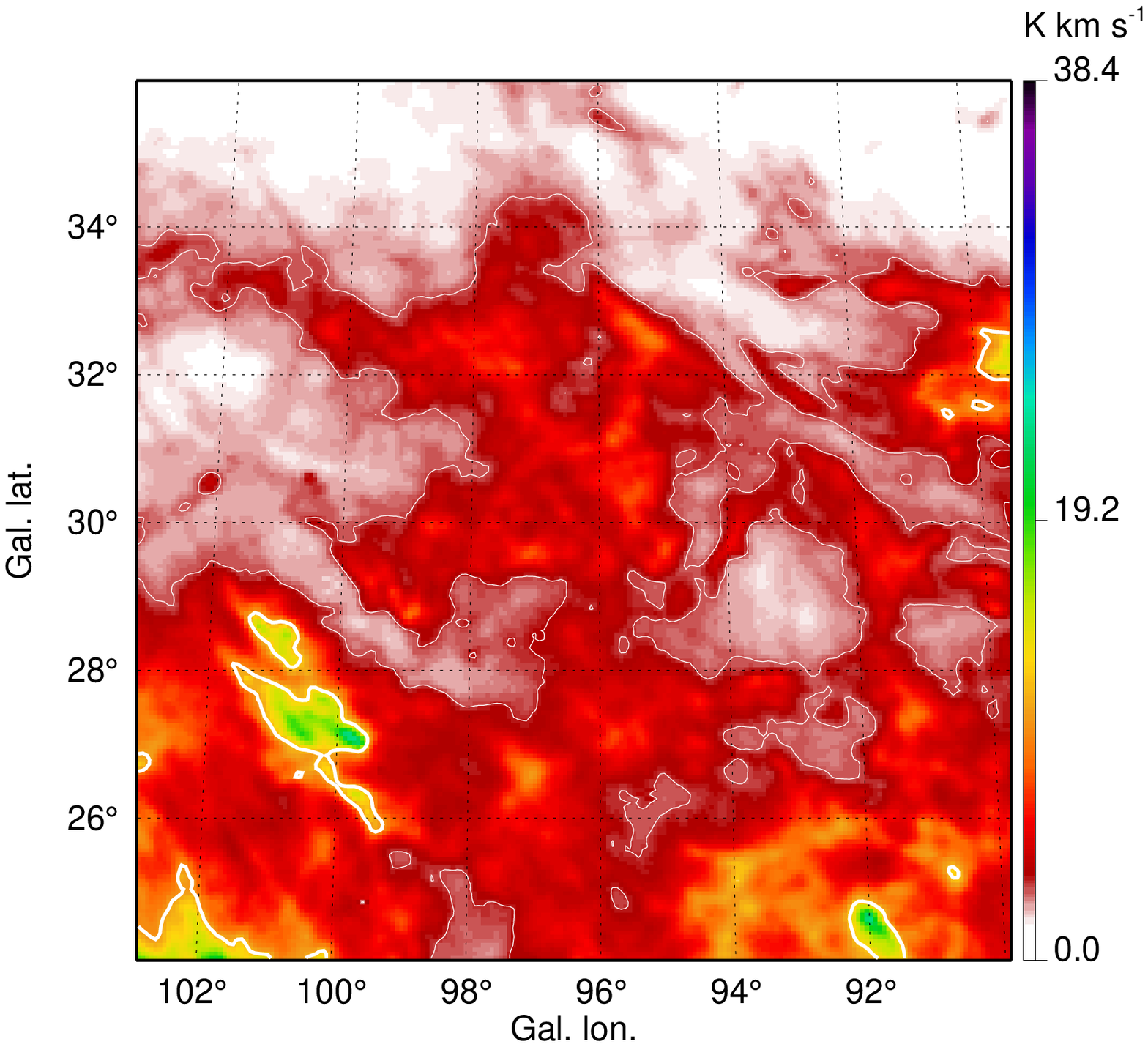}
\includegraphics[width=0.8\linewidth]{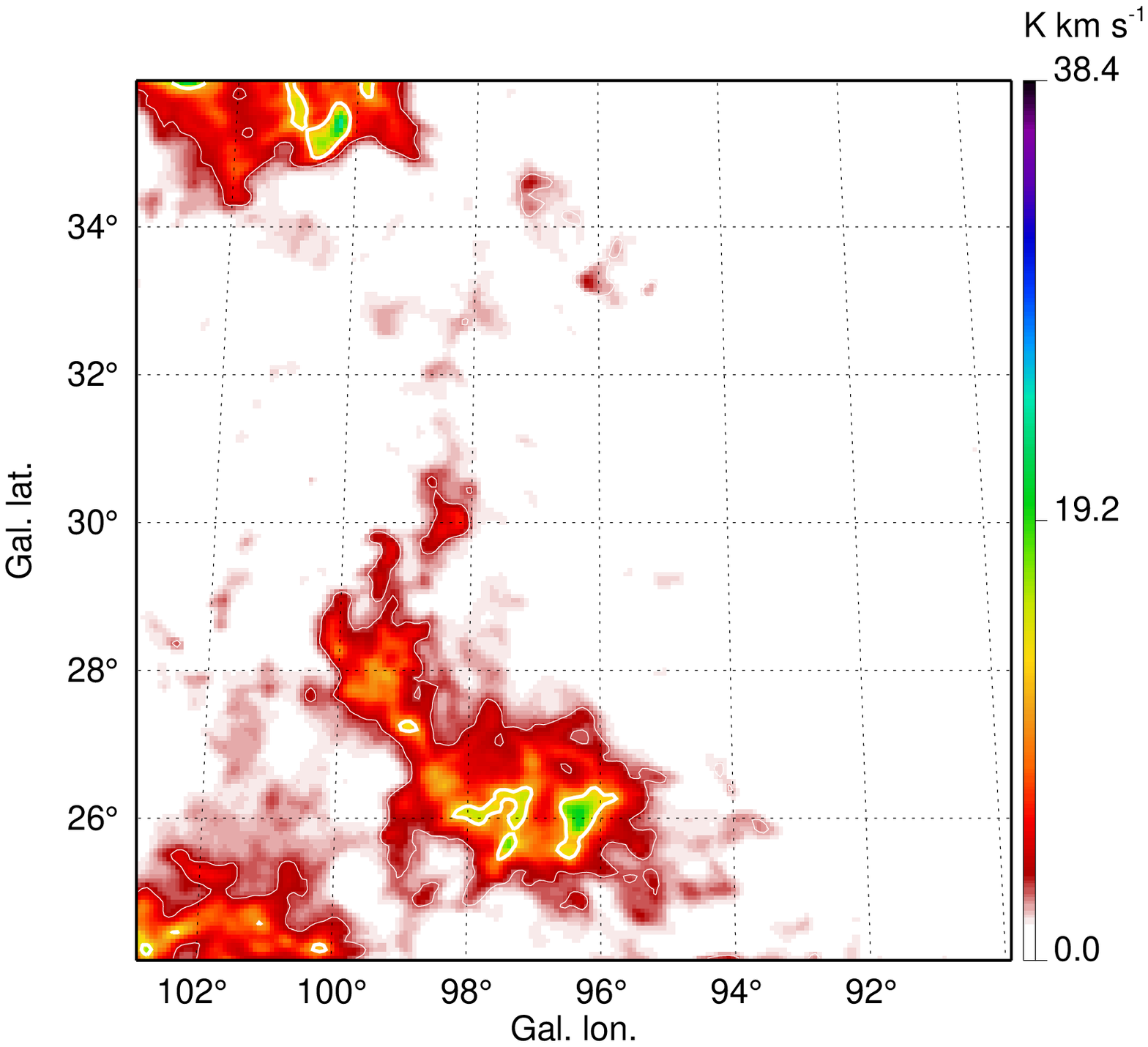}
\caption{
Map of the first order correction, $1/(2 \Trs) \int \Trb^2
\mathrm{d}v$, for \nep\ (upper: LVC, lower: IVC).  Here we set $\Trs =
80$~K.  Note that the range in the colorbar is smaller by a factor of
ten compared to that for the corresponding \wh\ maps in
Figure~\ref{fig:nepWmaps}.  The two contours show the (full)
correction as a percentage of \wh\ (2.5\,\% and 5\,\%).  The HVC
correction is minimal and so is not shown.
}
\label{fig:nepW2maps}
\end{figure}

In Figure~\ref{fig:nepW2maps} we present maps of this first-order
correction for the LVC and IVC components in NEP.  The HVC correction,
not shown, is minimal and below the noise (Figure~\ref{fig:nepW_tau}).
These appear like non-linear representations of \wh\ in
Figure~\ref{fig:nepWmaps}.  Also note the relative scales.  A few
contours show the (full) correction as a fraction of \wh.
Thus from Figures~\ref{fig:nepW_tau} and \ref{fig:nepW2maps} both LVC
and IVC components contain regions where opacity in the line is
non-negligible and thus an uncertainty arises in the derivation of
\nh, dependent on the choice of $\Trs$ and more profoundly on the
assumption of a constant $\Trs$.

\begin{figure}
\centering
\includegraphics[width=0.8\linewidth]{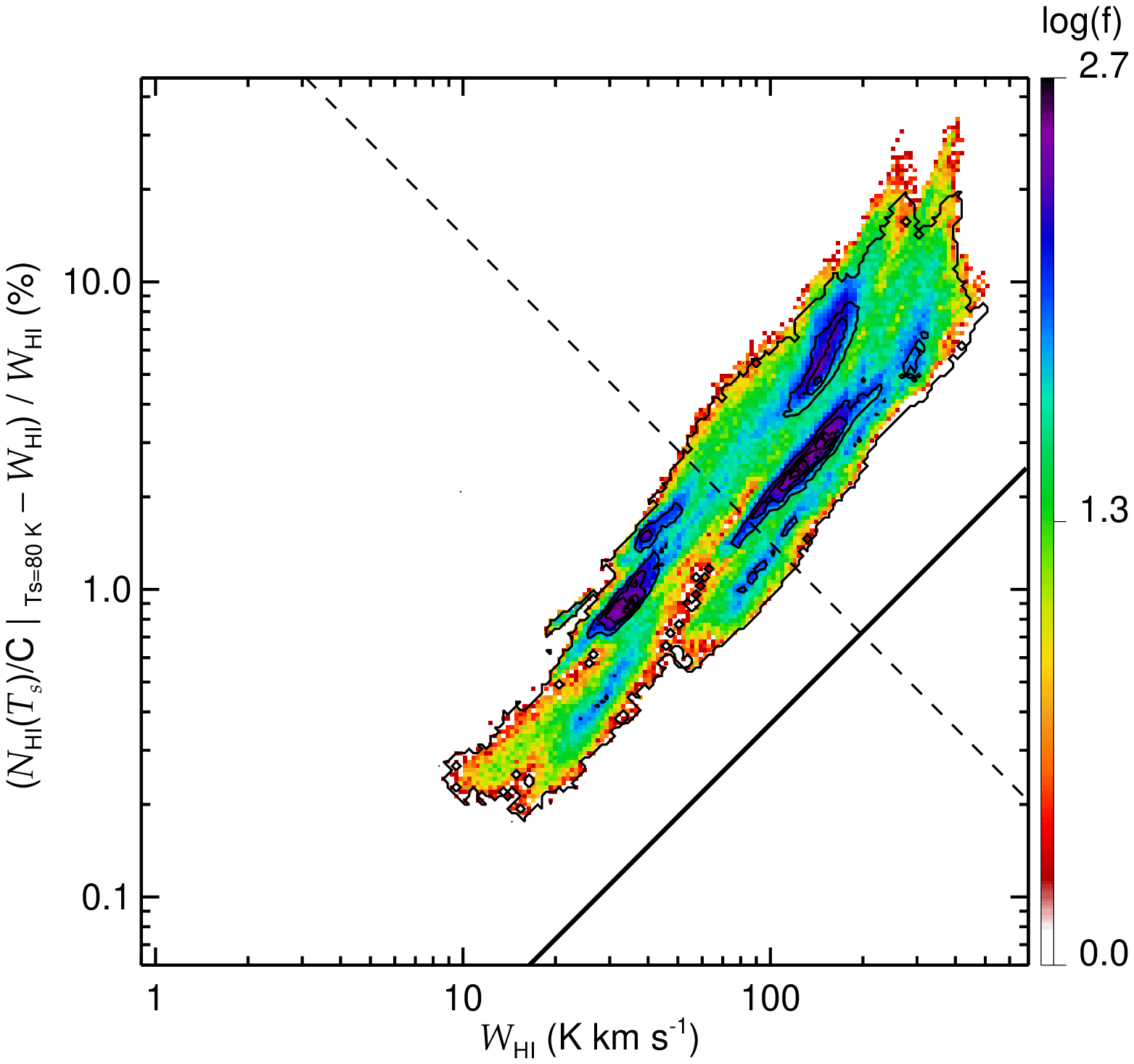}
\includegraphics[width=0.8\linewidth]{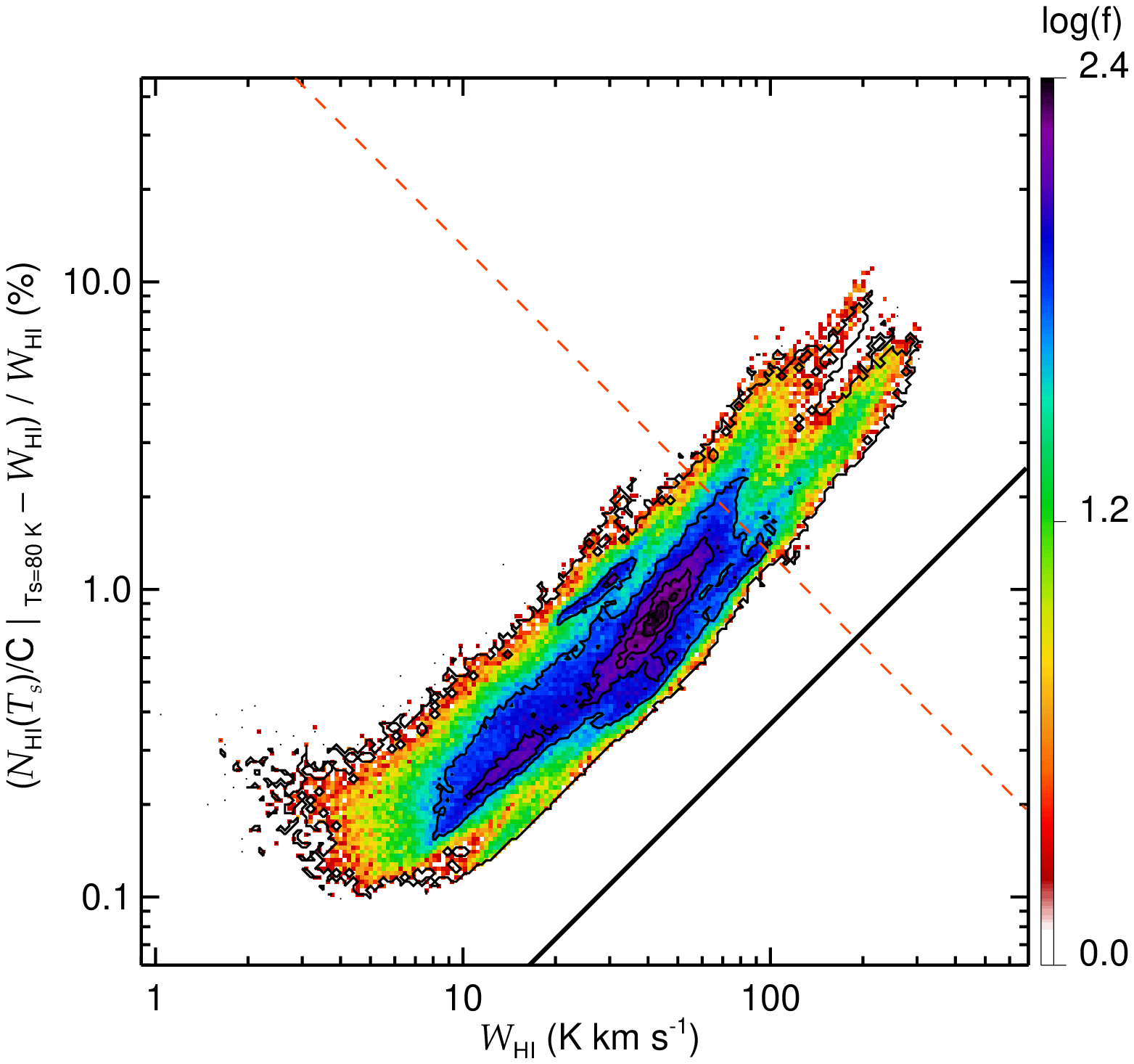}
\includegraphics[width=0.8\linewidth]{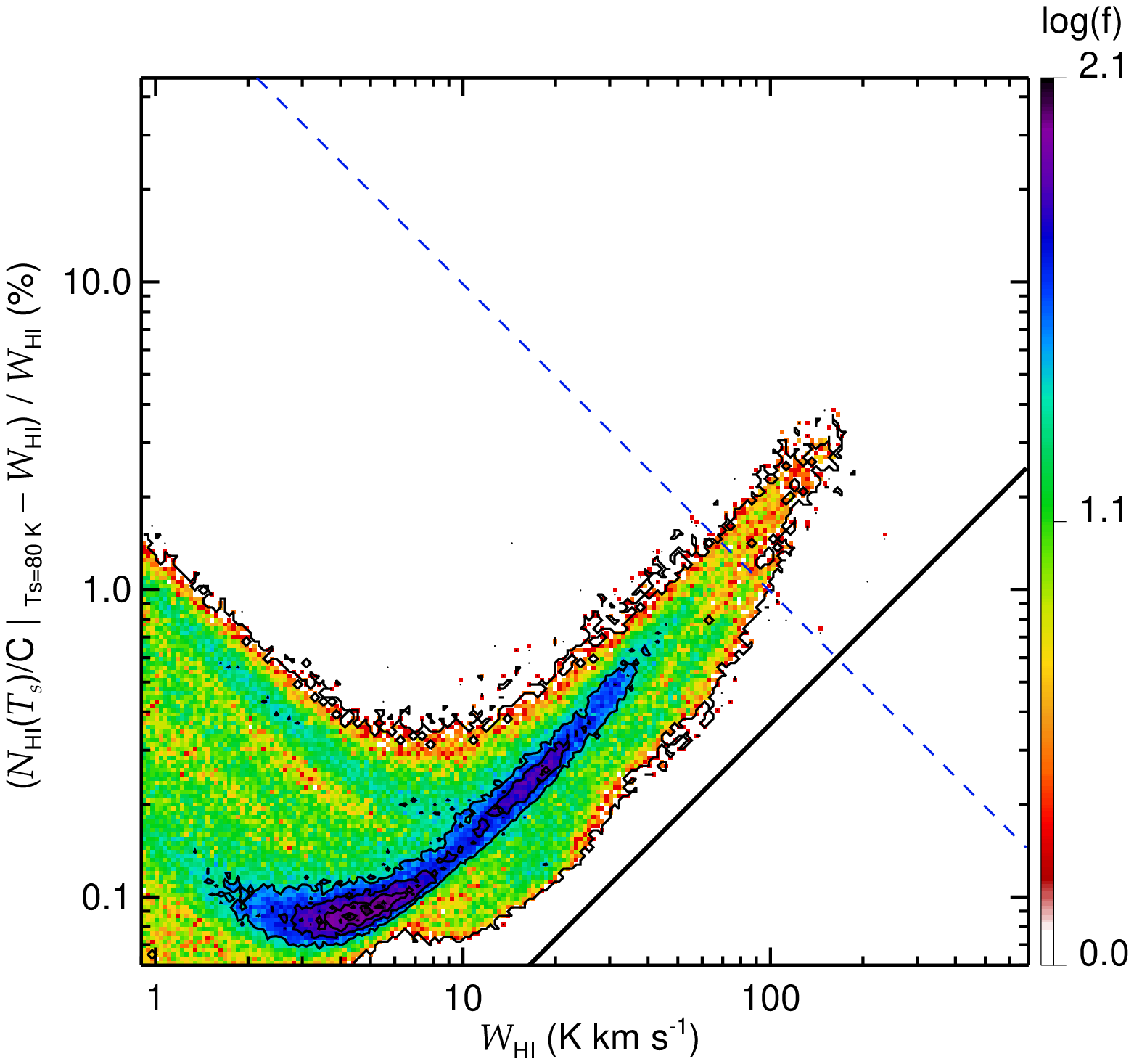}
\caption{
Histogram of full fractional correction (\nh$(\Trs)/C - $\wh)/\wh\ calculated
for $\Trs = 80$~K for all pixels in our component
\wh\ maps of all \ghigls\ fields.  Upper: LVC; middle: IVC; lower:
HVC.
A common line segment of slope one, transferred from
Figure~\ref{fig:nepW_tau}, reveals the approximate linear dependence.
The solid contours, representing the first-order correction for the
same $\Trs$, show that this approximation remains quite good even at
the highest values of \wh\ encountered.
The dashed lines, also from Figure~\ref{fig:nepW_tau}, are for typical
values of the uncertainties $\sigma_{\rm N_{H I}}/C$ in \nh$/C$ for
those components (from the NEP values in Table~\ref{n_table}).
The correction in the HVC component is rarely above this uncertainty.
IVC and LVC are progressively more affected, having typically higher
\wh, and so making an appropriate and accurate correction is more of a
concern.  
}
\label{fig:allW_tau}
\end{figure}

\subsection{Assessment using all \ghigls\ Spectra}
\label{whassess} 

Given the approximate quadratic dependence seen in
Figure~\ref{fig:nepW_tau}, it is advantageous to divide out one power
of the dependence of the opacity correction on \wh, so that
the fractional correction (\nh$(\Trs)/C - $\wh)/\wh\ is seen to be
approximately linear in \wh.  This is plotted in
Figure~\ref{fig:allW_tau} for a single $\Trs = 80$~K as a
two-dimensional histogram using the values for every pixel in our
component \wh\ maps of all fields.  The various islands, or ridges,
within the two-dimensional histogram arise because the line profiles
leading to a given \wh\ can be different: broad with low $\Trb$ and so
a smaller correction, or narrow and peaked and so more affected.

Also shown in the solid contours is the first-order correction for the
same $\Trs$.  The agreement is excellent at low \wh\ as expected and
remains quite good even at the highest values of \wh\ in these
intermediate latitude fields.  This difference is much less than the
uncertainty that arises from which value of $\Trs$ to adopt.

We have repeated this, but for $\Trs = 200$~K.  To verify that the
scaling is approximately inversely proportional to $\Trs$ for the full
correction, we have multiplied the latter correction by 200/80 and
overlaid contours on a figure like Figure~\ref{fig:allW_tau}.  Again
the agreement is excellent at low \wh\ as expected and it remains
quite good even at the highest values of \wh\ encountered.

\subsection{Impact on Applications of \nh\ Maps}
\label{applications} 

As mentioned above, maps of \nh\ have a variety of applications.  For
example, in \citet{planck2011-7.12} the \ghigls\ \nh\ maps were used
in conjunction with maps of thermal dust emission to examine regional
differences in the dust emissivity, the amount of emission per \nh.
To identify lines of sight along which there is a significant amount
of hydrogen in molecular form, for which \nh\ would no longer be
representative of the total column density, a masking procedure was
developed.  This was based empirically on the presence of emission in
excess of that expected from the correlation at lower \nh\ in a
particular field.  In retrospect the masked regions include all lines
of sight in which CO is detected using the Type-3 \Planck\ CO map
\citep{planck2013-p03a}, but the masked region is larger than this,
which is interpreted as the presence of molecular, but CO-dark, gas
for which the \hi\ diagnostic signal is therefore ``missing."
Qualitatively, there might be some ambiguity because \nh\ could be
underestimated if not corrected for opacity and indeed the
masked regions include areas with bright $\Trb$ and significant
opacity such as in Figure~\ref{fig:nepW2maps}.  However, the
\nh\ maps used were already corrected for opacity using $\Trs
= 80$~K.  In principle, one could account for the excess dust emission
by lowering $\Trs$ below 80~K on a pixel-by-pixel basis, producing
a map of $\Trs$ in the masked region.  In this solution, many lines of
sight would require $\Trs$ to conspire to be as low as the peak $\Trb$
in the spectrum.  Assuming that this is not allowed and that $\Trs =
80$~K is a reasonable value for the entire field as justified
earlier in this section, the excess dust emission cannot be accounted
for by an insufficient opacity correction.
Furthermore, the purpose of the masking was to exclude regions in
which \nh\ might not be a good tracer of total column density, and
\nh\ in the retained regions is certainly not sensitive to the choice
of $\Trs$.  Thus by excluding regions with the masking procedure, the
analysis of dust emissivity in the atomic gas is not significantly
impacted by uncertainty in the opacity correction.

However, often the full \nh\ map is desirable, as in the power
spectrum analysis.  The appropriateness and impact of using the
corrected \nh\ in such a situation can be guided by a sensitivity
analysis to gauge the robustness of the derived results when different
values of $\Trs$ are adopted.  For example, in the power spectra of
\nh\ corresponding to the analysis of \nh\ component maps in NEP in
Section~\ref{pswh} and Figure~\ref{anatomy} with $\Trs = 80$~K, if no
correction is applied then the power spectra are only minimally
steeper ($<0.02$).

\section{Assessment of Uncertainty in \nh}
\label{quality} 

As discussed by \citet{boot11}, there are a number of distinct
contributions that affect the accuracy of the spectra: noise, baseline
fitting errors, and errors in the subtraction of the stray radiation
spectrum.
Because the column density is such an important data product for
subsequent analyses, we focus here on the related question of the
uncertainty $\sigma_{\rm N_{H I}}$ in \nh\ (for this discussion, \nh\
is used interchangeably with \wh, other than the scaling parameter $C$
as in Equation~(\ref{thin})).  The results have been given in
Table~\ref{n_table}.

The following subsections summarize our understanding of the behavior
of the various contributions.
This exercise also involves various checks on the quality of the
\ghigls\ data, including a critical look for systematic errors.

\subsection{Noise in Emission-free Channels}
\label{enoise}

In emission-free channels, the noise $\sigma_{\rm ef}$ can be measured
as the rms fluctuations about zero in a spectrum or, alternatively, in
a single channel map in the cube.  The same noise can be assessed from
the width of the peak near zero in a histogram of $\Trb(x,y,v)$.  For
our basic observation and processing -- a single mapping of a field
with 4~s integration per spectrum and modified Bessel interpolation
into a cube with 0.8~\kms\ channels -- the measured noise,
$\sigma_{0}$, is typically 110~mK.  With a total of $\nmap$ repeats of
the mapping, this is found to go down as $\sigma_{0}/\sqrt{\nmap}$ as
expected.  Measured values of the typical $\sigma_{\rm ef}$ for
various \ghigls\ fields are summarized in Table~\ref{f_table}.

Because we are working with column density \nh\ here, we are
interested in how the noise increases when $\nch$ channel maps are
summed.  For $\nch$ independent emission-free channels, for a single mapping
with $\sigma_0\sim110$~mK the expected rms error in \nh\ would be
$\feffifty \times 10^{19} \sqrt{\nch/\nfifty}$~\cmm\ and for the \ghigls\
observations this can be scaled by $\sigma_{\rm ef}/\sigma_0 (< 1)$.
This error is consistent with what is found directly from the rms of
maps of the sum of emission-free channels (see also
Section~\ref{template}). 
The choice of \nfifty\ channels for the illustrative
normalization is a reasonable one, being close to the median number of
channels in the component intervals in Table~\ref{compvel_table}; it
represents a velocity interval of \vfifty\,\kms.

\subsection{Power Spectrum of the Noise}
\label{psanalysis}

As is apparent from the dots in Figure~\ref{anatomy}, the shape of the
power spectrum of the noise is not flat (i.e., white) at $k>
0.1$\,\iarcm, but instead has a characteristic decrease at the largest $k$
that arises from correlations induced by the modified Bessel function
interpolation of each observed \hi\ spectrum onto the grid
(Section~\ref{mapping}).  We have verified the precise form of this
decrease through simulations, starting with a white noise channel map
that is subsequently spatially filtered (convolved) to correspond to
the modified Bessel function interpolation used in the gridding of our
data.

When we sum a number of emission-free channels in our observed cube
and compute its power spectrum,
at high $k$ the shape of this power spectrum is the same regardless of
the number of channels.  Furthermore, this same shape is seen in the
\nh\ data itself (Figure~\ref{anatomy}).  Therefore the noise can be
quantified simply by a ``noise template," $N(k)$, and a scale factor,
$\eta$.  This is reflected in the full parameterized model of the
one-dimensional power spectrum in Equation~(\ref{plaw}).  We use the
fitted level $\eta$ as a quantitative diagnostic of the noise (from
all sources) in the map.  We denote as $\sigma_{\rm ps}$ the estimate
of the rms noise in a map that is derived using this power-spectrum
analysis. From this complementary approach estimates of
$\sigma_{N_{\rm H I}}$ for the component \nh\ maps of the \ghigls\
fields have been entered in Table~\ref{n_table}.

\begin{figure}
\centering
\includegraphics[width=1.0\linewidth]{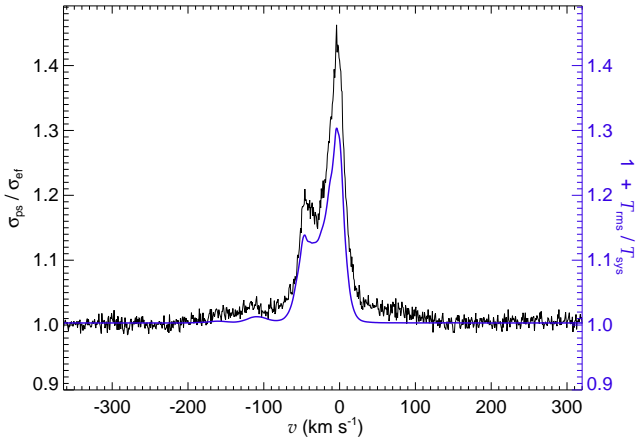}
\caption{
Estimate of the noise in each channel of the \nep\ cube from the
power-spectrum analysis, $\sqrt{\eta} = \sigma_{\rm ps}/\sigma_{\rm
ef}$ (Appendix~\ref{template}).  At the ends of the spectrum
$\sqrt{\eta}\sim1$, as expected for primarily emission-free channels.
However, where there is signal the estimated noise is somewhat larger.
To model the contribution of noise arising from the 21-cm emission
itself, we use the approximation $1 + {\rm rms}\, \Trb(v)/20~{\rm K}$,
based on the ${\rm rms}\, \Trb$ measured in each channel map
(Appendix~\ref{tnoise}).  The resulting smooth curve (right axis)
captures some basic features of the increase in noise.  }
\label{fig:nep_noisepower}
\end{figure}

\subsubsection{Noise Template}
\label{template}

In this paper the noise template used is that derived from
emission-free end channels.  This can be quantified by $\sigma_{\rm
ef}$, the rms noise of the inverse Fourier transform of this noise
template.  Because the same template, simply scaled by $\eta$, is used
in the power-spectrum analysis we have $\sigma_{\rm ps} = \sqrt{\eta}
\sigma_{\rm ef}$.

As an example, power-spectra of the NEP cube can be modelled on a
channel by channel basis using Equation~(\ref{plaw}) to find $\eta$.
The resulting spectrum $\sqrt{\eta} = \sigma_{\rm ps}/\sigma_{\rm ef}$
is shown in Figure~\ref{fig:nep_noisepower}.  This $\sqrt{\eta}$
spectrum approaches unity in the velocity ranges $-350 < v <
-250$~\kms\ and $175 < v < 275$~\kms\ because these are the ranges of
the typical end channels used to remove baselines.  The end channels
used vary from spectrum to spectrum across the field.  In these
velocity ranges over 99\,\% of the pixels in each channel have been
used to determine a baseline.  This also demonstrates that the noise
template is defined consistently.

As another application, we made \nh\ maps from an increasing number of
channels of a simulated noise cube and found the expected behavior
$\eta\propto\nch$, or $\sigma_{\rm ps} \propto \sqrt{\nch}$.  Fitting
baselines has a notable effect on the accumulation of noise.  To
demonstrate this we fit baselines in this noise cube, using data where
the emission-free end channels would normally be, and subtracted them.
 
For \nh\ maps made from channels where baselines were fit the noise
$\eta$ first increases with $\nch$ but eventually reaches a maximum
and turns over when $\nch$ is comparable to the range used to fit the
baselines, the exact point depending on the degree of the polynomial
baseline being fit.  This effect is a result of the correlations
introduced by the baseline fitting.  We examined the \nep\ cube in the
same way and found a similar behavior with increasing $\nch$.

For the channels where there is signal, away from where the baselines
are defined, the noise is somewhat larger than simply $\sigma_{\rm
ef}$ (Figure~\ref{fig:nep_noisepower}).  This will affect the noise in
\nh\ maps too; the noise estimated from the sum of emission-free
channels will always be an underestimate.

\subsection{Noise in the Line Emission}
\label{tnoise}

The 21-cm line emission itself increases the noise by a factor
approximately $1 + \Trb(v)/\Trsys$ \citep{boot11} and because the GBT
21-cm receiver noise is so low ($\Trsys \sim 20$~K) the increase can
be significant even in the \ghigls\ fields.

This additional noise will not be uniform across the channel map, but
the size of the effect can be demonstrated using a representative
$\Trb(v)$.  For each channel map we adopted the rms value, ${\rm
rms}\, \Trb(v)$, noting that this spectrum is the sum in quadrature of
the mean and standard deviation spectra seen in
Figure~\ref{fig:avg_std}.  The approximate increase in the noise
spectrum predicted on this basis, $1 + {\rm rms}\, \Trb(v)/20~{\rm
K}$, is shown for \nep\ in Figure~\ref{fig:nep_noisepower}.  It
captures basic features of the increase seen in the actual noise
spectrum.

\subsection{Baseline Errors}
\label{bnoise}

Fitting and subtracting a baseline introduces errors in \nh\ through
uncertainties in the fitted coefficients.  Baseline errors add to the
increased noise apparent in the shoulders adjacent to the
emission-free channels in Figure~\ref{fig:nep_noisepower} and to the
channels with more significant emission as well.

This will propagate to uncertainty in \nh\ maps too.  Unlike the noise
in the line emission, baseline errors are similar for adjacent
channels in a spectrum and so accumulate as $\nch$.  \citet{boot11}
showed that these amount to an uncertainty in \nh\ typically $\fbfifty
\times 10^{19} (\nch/\nfifty)$~\cmm, comparable to the noise estimates
above for $\nch \sim \nfifty$ but with a different $\nch$ dependence.
This was found by comparing independent XX and YY polarization cubes.
These cubes have common stray radiation and so their difference
(XX$-$YY) reflects the quality of baselines plus thermal noise (which
can be seen increasing where there is a signal).

The assumption made in both \citet{planck2011-7.12} and \citet{boot11}
is that the noise in the difference and the noise in the sum are
identical.  This is not the entire story, as we demonstrated by
examination of the power spectra for a single observation of a
subfield of NEP.  We produced maps of \wh\ from both XX+YY and XX$-$YY
cubes.  The power-spectrum analysis yielded slightly larger values of
$\sigma_{\rm ps}$ in the XX$+$YY case, by up to 25\,\%.

\subsection{Errors in the Stray Radiation Spectrum}
\label{snoise}

\citet{boot11} showed that there is a significant error contribution
to \nh\ from the uncertainty in the predicted stray radiation spectrum
that is subtracted.  This is an error that can vary slowly over a
field in the course of mapping and like the baseline removal could
have a non-linear effect on the scaling parameter of the noise power
spectrum.

An empirical approach to determining this uncertainty is to compare
the reproducibility of \nh\ derived from cubes made from different
independent mappings of a field (up to three in our survey).  This
includes all sources of error but is dominated by the accumulated
baseline errors and the imperfect stray radiation subtraction.  In
principle it would be possible to schedule observations to make
repeated maps with virtually the same stray radiation. However, in
practice because of our exploitation of flexible scheduling the
mapping of (parts of) any field took place at different times of day,
on different days, and at different times of year, and so generally
the stray radiation and the estimated correction are different in
different maps of the same field.  Indeed, this was used by
\citet{boot11} to study the properties of the stray radiation and to
calibrate the amplitude of the sidelobes in the all-sky response of
the GBT.
A corollary is that because the stray radiation can now be predicted
in advance, future observations could be scheduled for times at which
the stray radiation is minimized.  The data reported in this paper for
the faint N1 field approach this ideal.

For every \ghigls\ field in addition to the observed emission-line
cube we have a predicted stray radiation cube from which we can
compute \nh$_{\rm stray}$.  Using the above difference approach,
\citet{boot11} showed that the error contribution to \nh\ is of order
0.07~\nh$_{\rm stray}$.

Because the errors in two repeated observations $i$ and $j$ are
independent, and ultimately the two are averaged together to form
$\left<N_{\rm H}\right> = (N_{{\rm H},i} + N_{{\rm H},j})/2$, the
estimator of interest in assessing errors in \nh\ is the dispersion
(the standard deviation about the mean) of the map of $\Delta N_{\rm
H} = (N_{{\rm H},i} - N_{{\rm H},j})/2$, or for the complementary
approach being used here $\sigma_{\rm ps}$ from the power-spectrum
analysis of that difference map.

Again using the power-spectrum analysis to find $\sigma_{\rm ps}$ we
are able to look at not only the noise in the repeat observation
difference maps, but also the noise in their sum.  The disagreement
between the results is somewhat greater than in the analysis of XX and
YY used to study baseline errors.  The power-spectrum analysis yielded
$\sigma_{\rm ps}$ values for the sum roughly 50\,\% larger than for
the difference.

\section{Comparison with Data from EBHIS}
\label{ebhis}

A thorough comparison of our \ghigls\ data with that from the LAB
survey was performed in \citet{boot11}.  Here we compare our data with
that from the Effelsberg-Bonn \hi\ Survey (EBHIS)
\citep{floe10,wink10,kerp11,wink12} with the goal of testing the
quality of the \ghigls\ data and the data reduction procedures,
especially the correction for stray radiation.
The EBHIS collaboration has provided unpublished data cubes and stray
radiation cubes for the \nep\ field that has been used for many
illustrations in this paper, another large field centred on SPIDER but
including portions of POL, SP, SPC, and UMA (these all relate to the
NCP loop and so we use the designation \ncpeb\ here), and DRACO.

\subsection{EBHIS Data and \ghigls\ Modifications}
\label{ebhisdata}

\begin{figure*}
\centering
\begin{tabular}{cc}
%
\includegraphics[width=6.5cm, keepaspectratio]{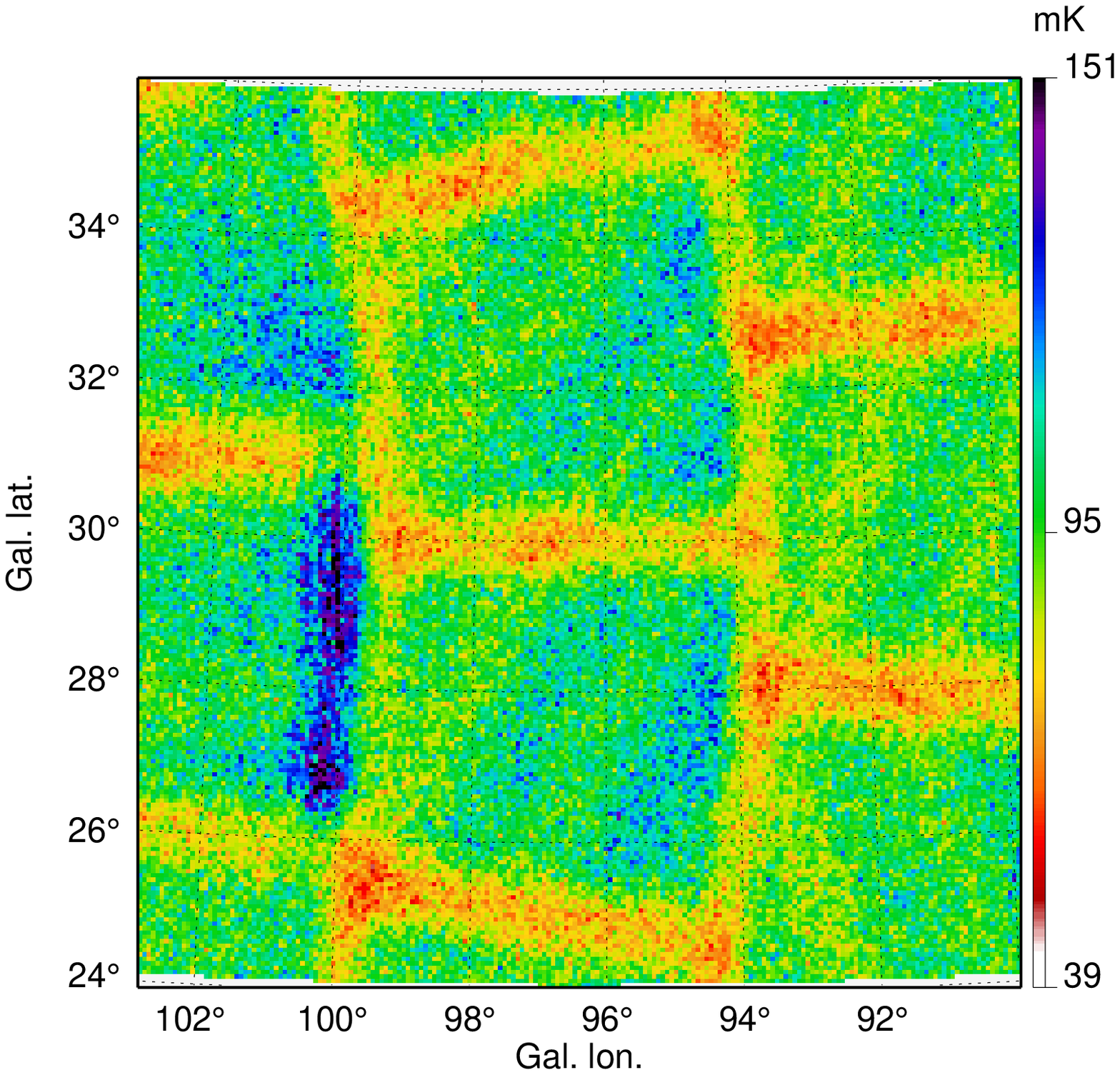} &
\includegraphics[width=6.5cm, keepaspectratio]{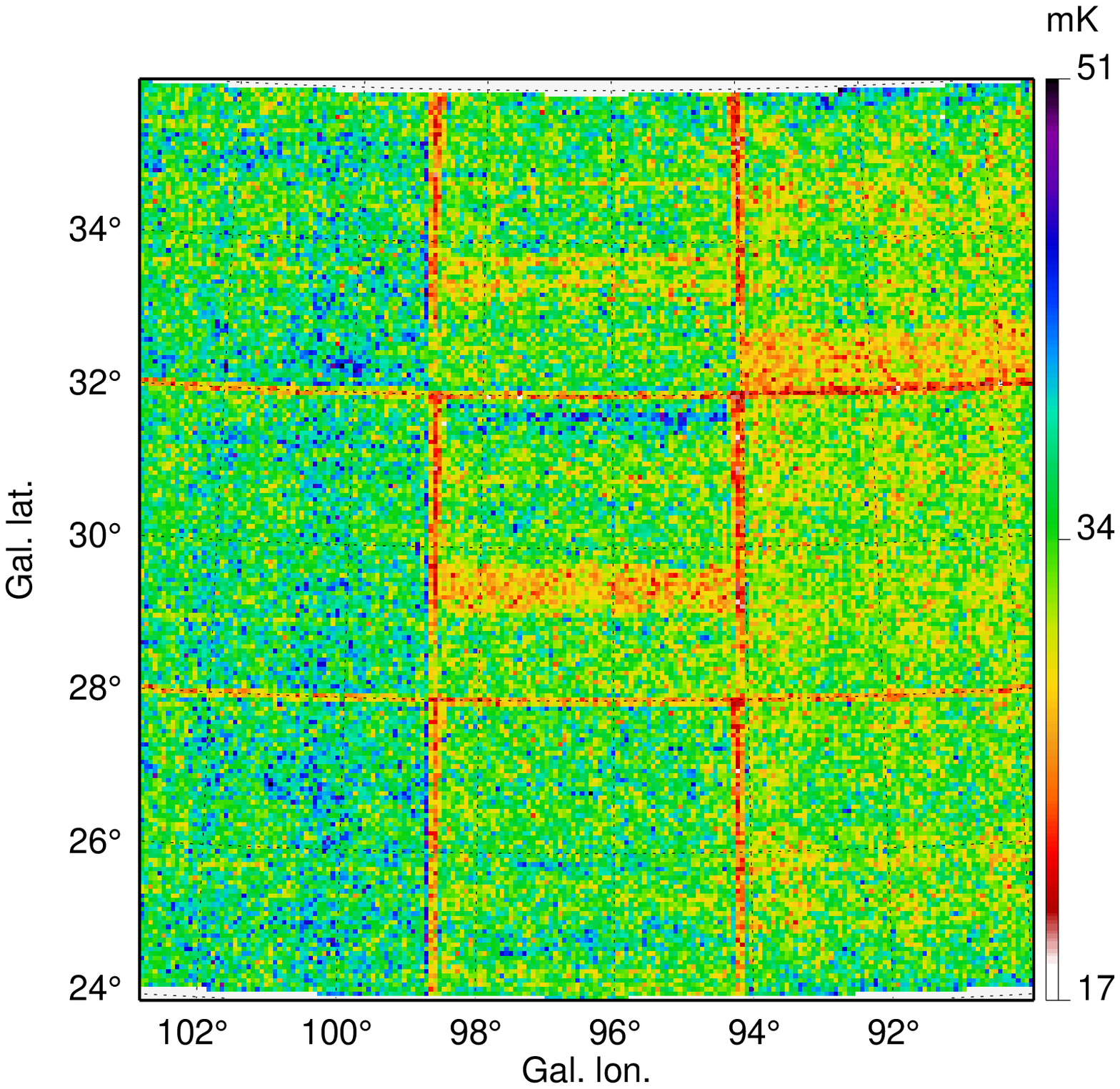} \\
%
\includegraphics[width=6.5cm, keepaspectratio]{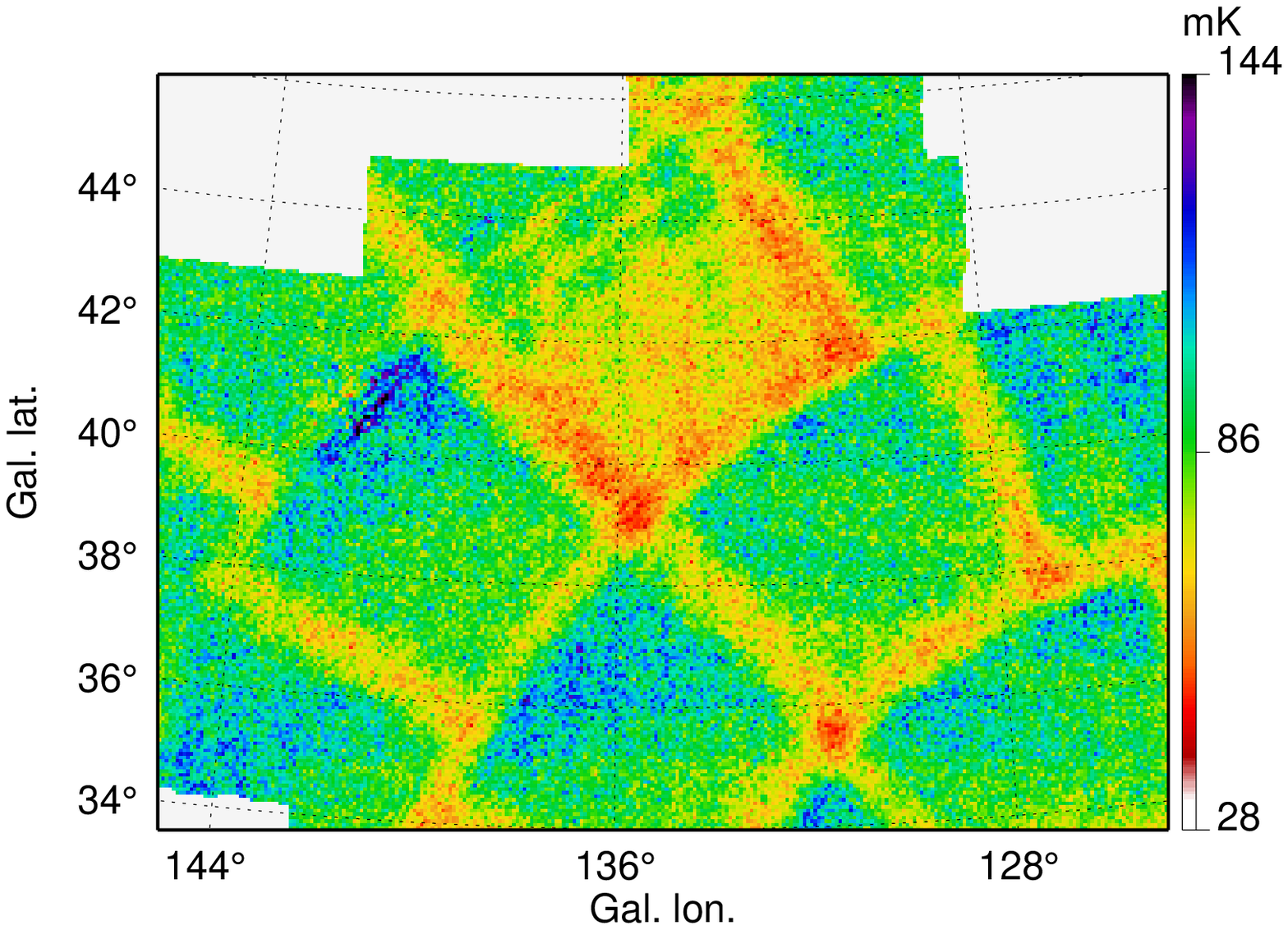} &
\includegraphics[width=6.5cm, keepaspectratio]{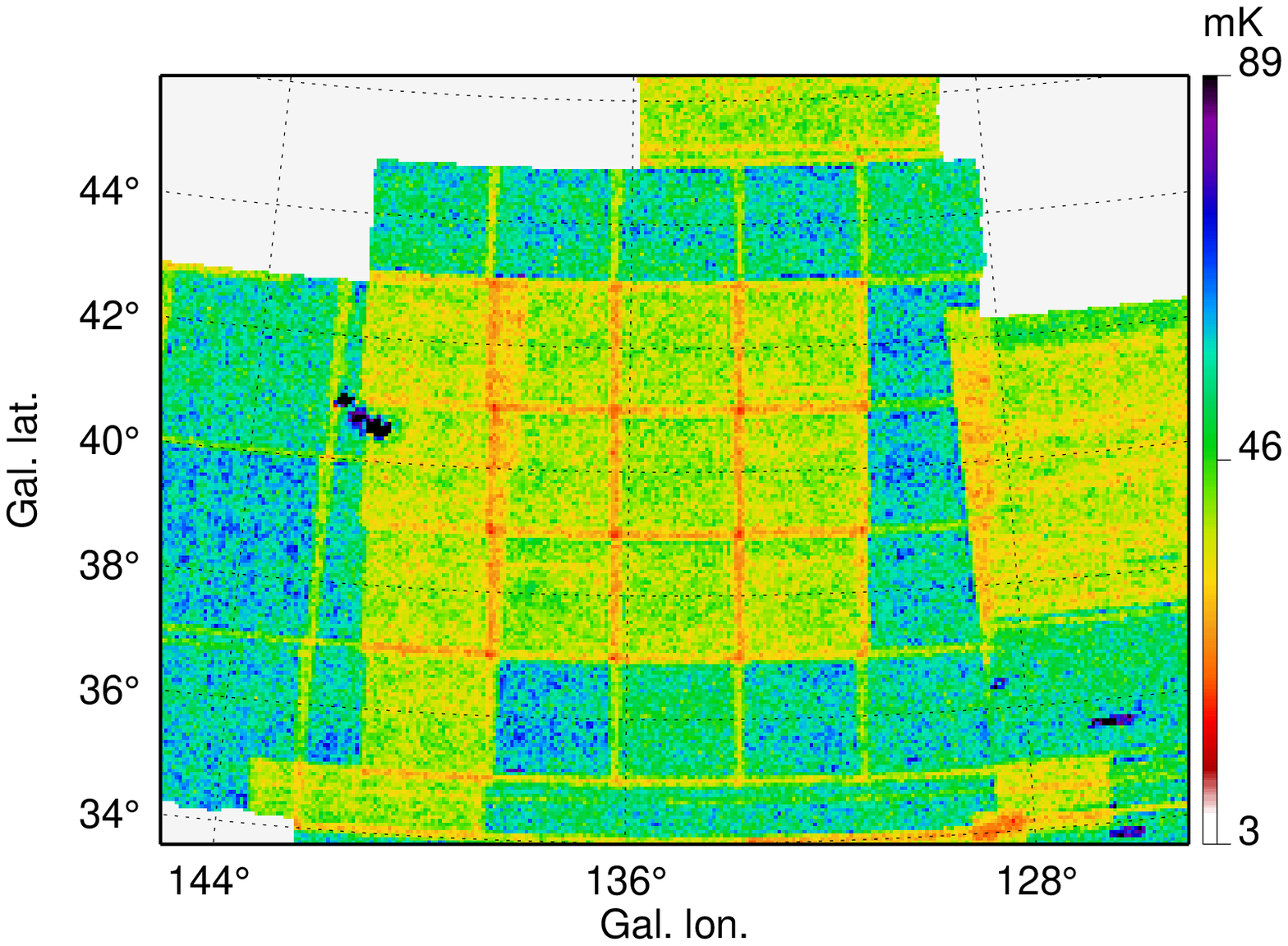} \\ 
%
\includegraphics[width=6.5cm, keepaspectratio]{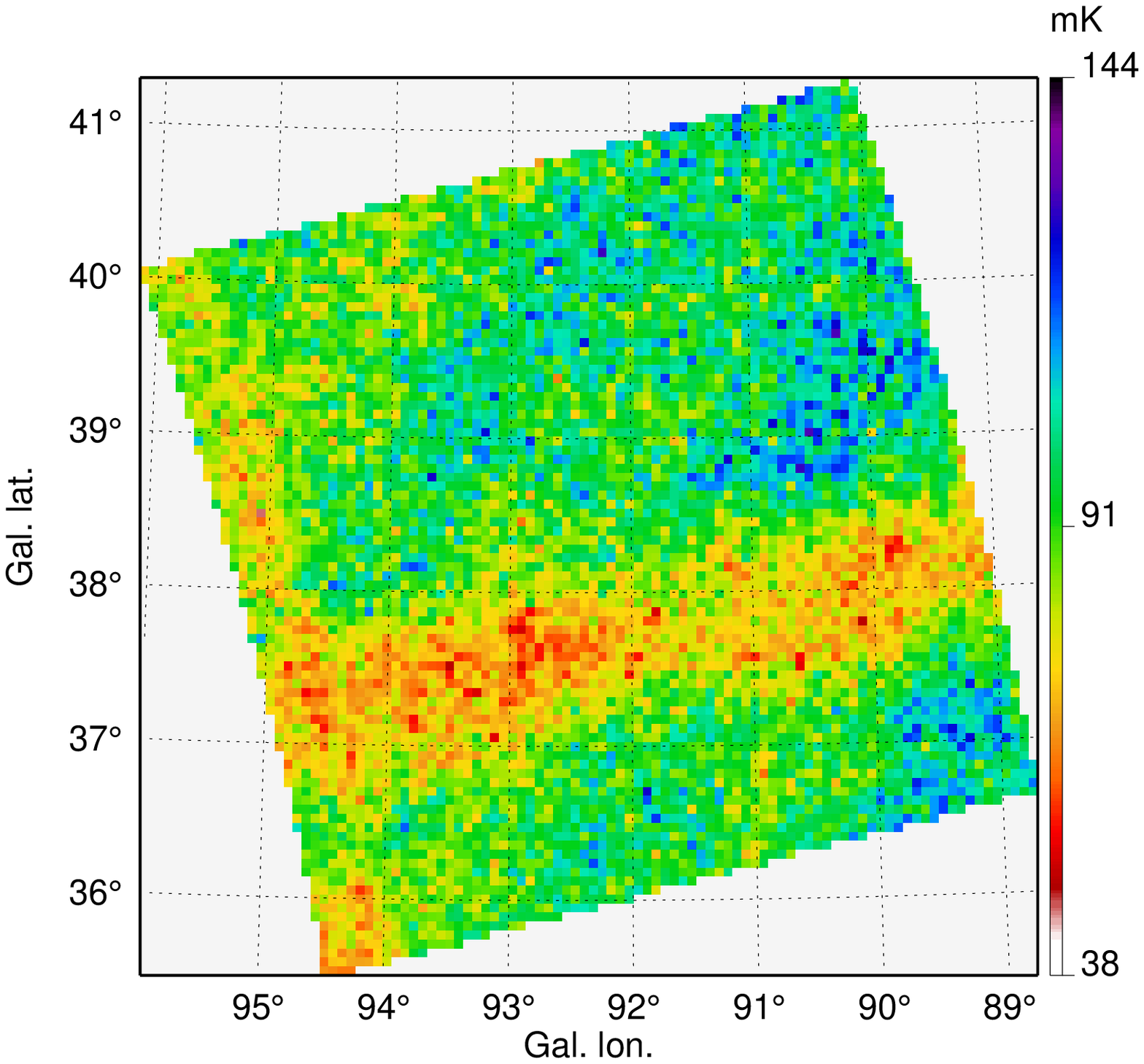} &
\includegraphics[width=6.5cm, keepaspectratio]{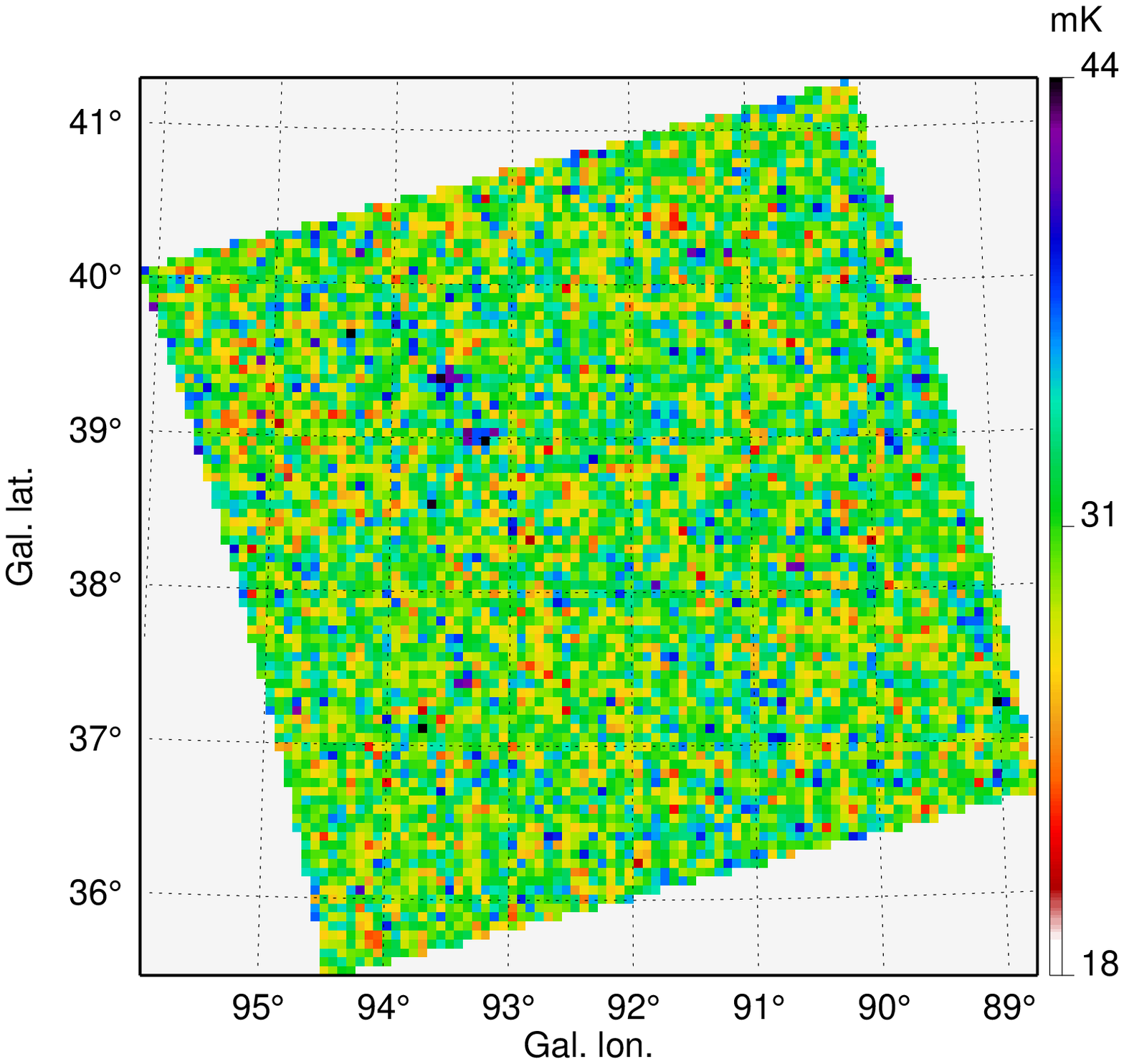} \\
\end{tabular}
\caption{
Left: Noise map for EBHIS produced from the rms in emission-free
channels in each spectrum.  Right: Noise map for the \ghigls\ data
adjusted to the same resolution and same gridding.
Top to bottom respectively: NEP, \ncpeb, and DRACO.
EBHIS maps have been cropped to match the available \ghigls\ coverage.
Large scale patterns in the noise reflect the different observing
strategies.  Note that the angular scale and the central value and
range of the colorbar are different for each map.  In \ncpeb\
emission in the M81 -- M82 system is sometimes challenging for both
\ghigls\ and EBHIS, causing some artifacts within about a degree of
$(l, b) = (142\deg, 41\deg)$.
}
\label{fig:efchannelebhis}
\end{figure*}

The EBHIS data have an angular resolution (FWHM) $\sim10\farcm5$
\citep{wink10} on a Galactic SFL grid with $3\farcm2564$ pixels (we
regridded to the \ghigls\ 3\farcm5) and a channel spacing of $\Delta v
= 1.288$~\kms.
The noise maps for all three cubes are presented in the first column
of Figure~\ref{fig:efchannelebhis} where the EBHIS maps have been
cropped to match the coverage of the equivalent \ghigls\ field.
The noise in each map pixel is set by the rms of emission-free
channels in the corresponding spectrum.  The mean noise is
$\sigma_{\rm ef} \sim95$~mK (NEP), $\sim86$~mK (\ncpeb), and
$\sim90$~mK (DRACO).

For a direct comparison with EBHIS, the \ghigls\ data cubes first need
to be convolved to the slightly lower resolution of the EBHIS survey.
Following the convolution of each channel in the cubes, the \ghigls\
data were regridded both spatially (cubic convolution interpolation
using the IDL routine $interpolate$) and along the velocity axis
(linear interpolation).  These steps of convolution and regridding
reduce the original noise of the \ghigls\ data for NEP, \ncpeb, and
DRACO from 68~mK, 75/105~mK,\footnote{
Values for SPIDER but the flanking fields are similar.
} 
and 61~mK (Table~\ref{f_table}) by about a factor 2 to an average rms
noise 34~mK, 40/55~mK, and 31~mK, respectively.
The rms noise maps for the modified \ghigls\ data are shown in the
right column of Figure~\ref{fig:efchannelebhis}.

The scanning and map-making strategies are reflected in the various
geometric (grid) patterns (called \GPs\ below) that appear in these
noise maps.  In addition, spectra with very few emission-free end
channels due to the presence of galaxies result in noise peaks in these
maps.

\subsection{Relative Calibration}
\label{Wratiomapsebhis}

Prior to detailed comparisons between EBHIS and \ghigls, we first
determine if correction by a scale factor is necessary.  This is done
by comparing $\Trb$ along every common line of sight, fitting the
scatter plots using the anticipated linear model:
\begin{equation}
\mathrm{EBHIS} = a  \times \mathrm{GHIGLS} + b  \,.
\end{equation}
This is repeated for scatterplots of \wh\ as well.  Both correlations
are very good.

The scale factor (slope) $a$ should be close to unity, but reflects the
different methods by which the spectra have been calibrated.  In the
ideal case, there would be no offset $b$.  However, inconsistencies
remaining after the stray radiation and/or baseline corrections can
introduce an offset and this can in turn influence the slope.  These
inconsistencies and noise will have the largest relative effect where
$\Trb$ is small.  To ensure that the derived scale factor is not
influenced unduly we limit the data to $\Trb > T_{\mathrm{limit}}$.
The regression is repeated on each field for many values of
$T_{\mathrm{limit}}$ while also varying the velocity range over which
the data are fit.  Not surprisingly the velocity components in which
the spectra have the largest range in $\Trb$ yield the most robust
fits.

We find that $a$ and $b$ vary slightly as a function of
$T_{\mathrm{limit}}$ and velocity component and that in general the
derived slopes and offsets are anti-correlated.  With the assumption
that errors from removal of stray radiation and baseline will result
in both over- and under-corrections in both datasets, the most
robust slopes would be those with the lowest absolute offsets.  For
each velocity component we look for the $T_{\mathrm{limit}}$ that
satisfies this criterion and also check for the relative constancy of
the slope over a neighboring range in $T_{\mathrm{limit}}$.
The scale factors derived from the $\Trb$ and \wh\ analyses are
consistent with one another when there is a suitably large range in
$\Trb$ (or \wh; e.g., Table~\ref{n_table}) and likewise the results
from the three velocity components in each of the three fields
examined are also consistent.  The dispersion of the different
estimates is 0.030.  We adopt a scale factor EBHIS/\ghigls $ = 1.009
\pm 0.011.$
Before comparing the spectral data in the analyses below, we made the
minor adjustment of all EBHIS survey data to the \ghigls\ scale using
this scale factor.

The EBHIS and \ghigls\ calibrations are consistent to within 1\,\%.
From \citet{boot11} the \ghigls\ calibration has a formal uncertainty
of 1\,\% and there we concluded conservatively that ``our calibration
does not have systematic errors that exceed a few percent."
We recall that our previous comparison with the LAB survey over all
\ghigls\ fields resulted in a scale factor LAB/\ghigls\ =
$1.0288\pm0.0012$ \citep{boot11}.

\subsection{Mean and Individual Spectra}
\label{averagespectrumebhis}

\begin{figure}
\centering
\includegraphics[width=1.0\linewidth]{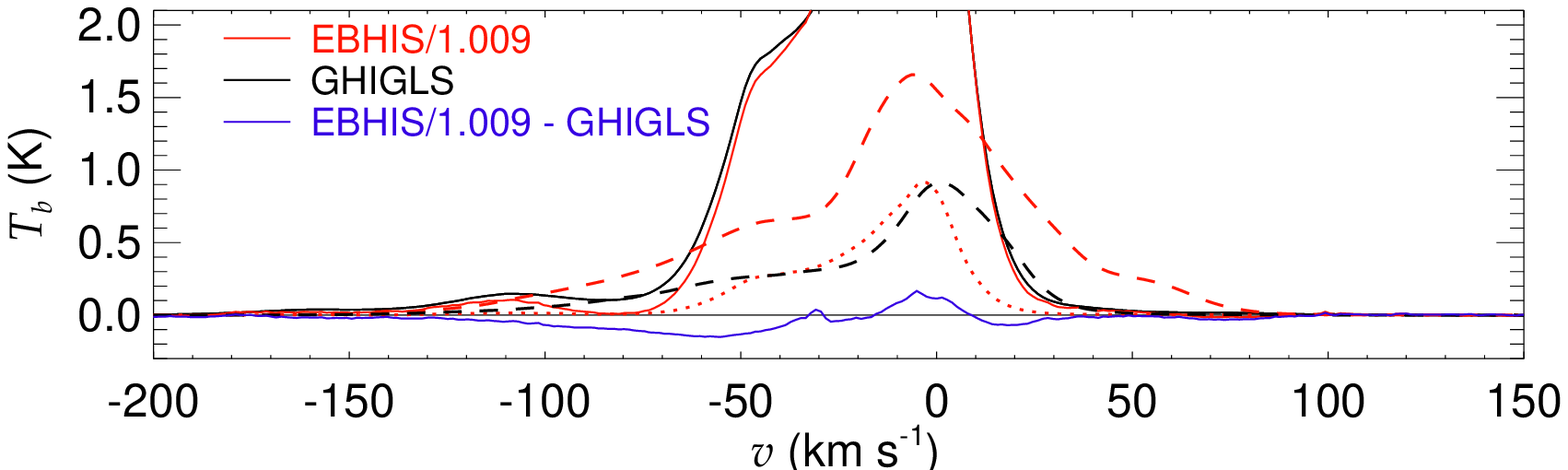}
\includegraphics[width=1.0\linewidth]{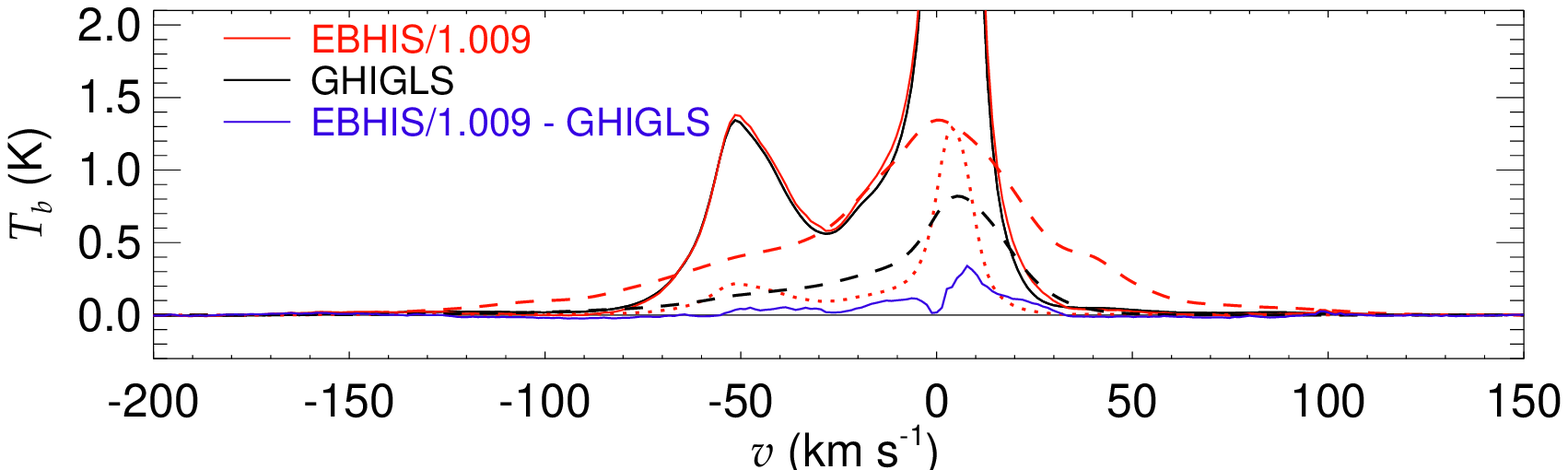}
\includegraphics[width=1.0\linewidth]{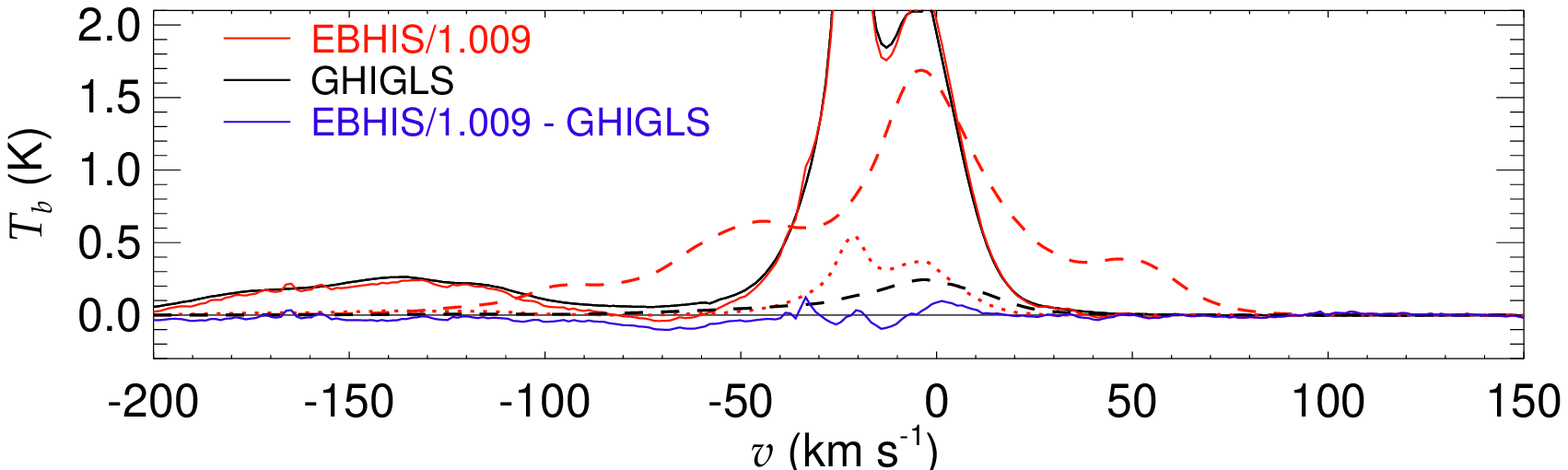}
\caption{
Top panel: Average \ghigls\ spectrum for NEP (black), average scaled
EBHIS spectrum (red), residual scaled \EBHISmGBT\ (blue).  Average \ghigls\
spectrum peaks at 5.7~K.  
Broken lines show the average spectra of the
predicted stray radiation that has been removed from the data; dashed
for \ghigls\ (black) and dashed and dotted (red) for EBHIS far and near side lobes,
respectively.
Following panels: Average spectrum for \ncpeb\ peaking at 8.1~K and for
DRACO peaking (in the IVC) at 3.8~K.
}
\label{fig:avgspectrumebhis}
\end{figure}

As shown by the average spectra for EBHIS and (convolved and
regridded) \ghigls\ in each field in
Figure~\ref{fig:avgspectrumebhis}, and the difference spectrum, there
is remarkable agreement between the two surveys.  Also shown are the
average spectra of the predicted stray radiation that have been
subtracted in producing these spectra.  For EBHIS, the stray radiation
spectrum is calculated in two parts, for the near and far side lobes
(\citealp{kalb10})).  The GBT has an unblocked aperture, but there is
a significant spillover sidelobe from the secondary reflector
\citep{boot11}.  Obviously, the stray radiation is substantial in
fields such as surveyed by \ghigls\ so that without its removal the
spectra would be quite different and there would be no agreement
between the surveys.  The spectral extent of the stray radiation is
larger for the far sidelobes where Doppler effects are more
significant, well into HVC velocities in the case of EBHIS.

\begin{figure}
\centering
\includegraphics[width=1.0\linewidth]{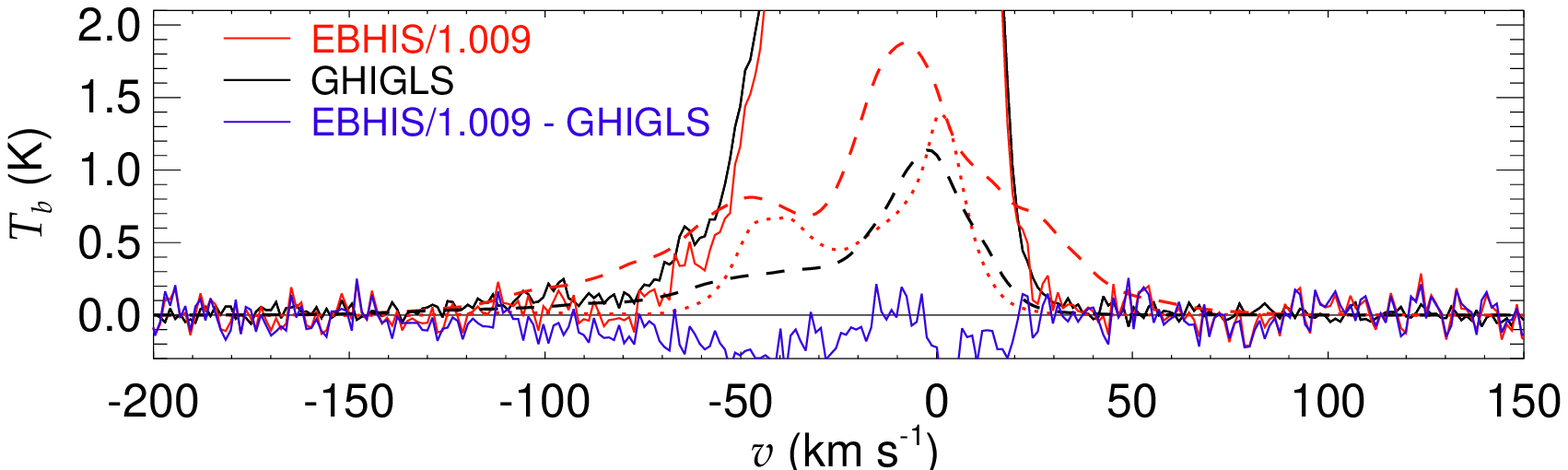}
\includegraphics[width=1.0\linewidth]{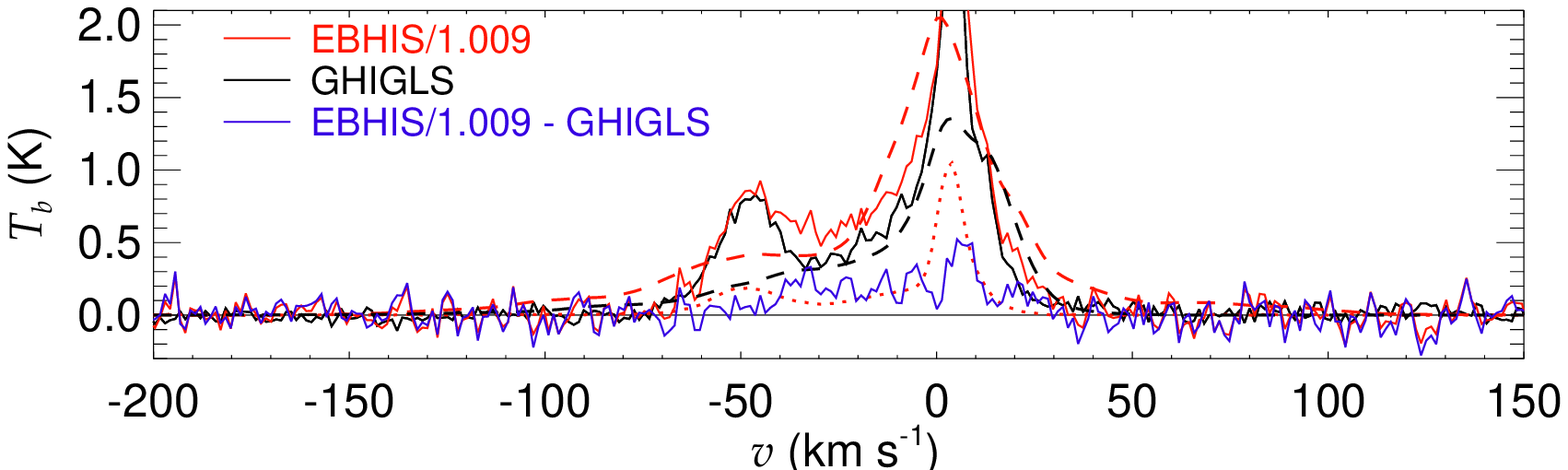}
\includegraphics[width=1.0\linewidth]{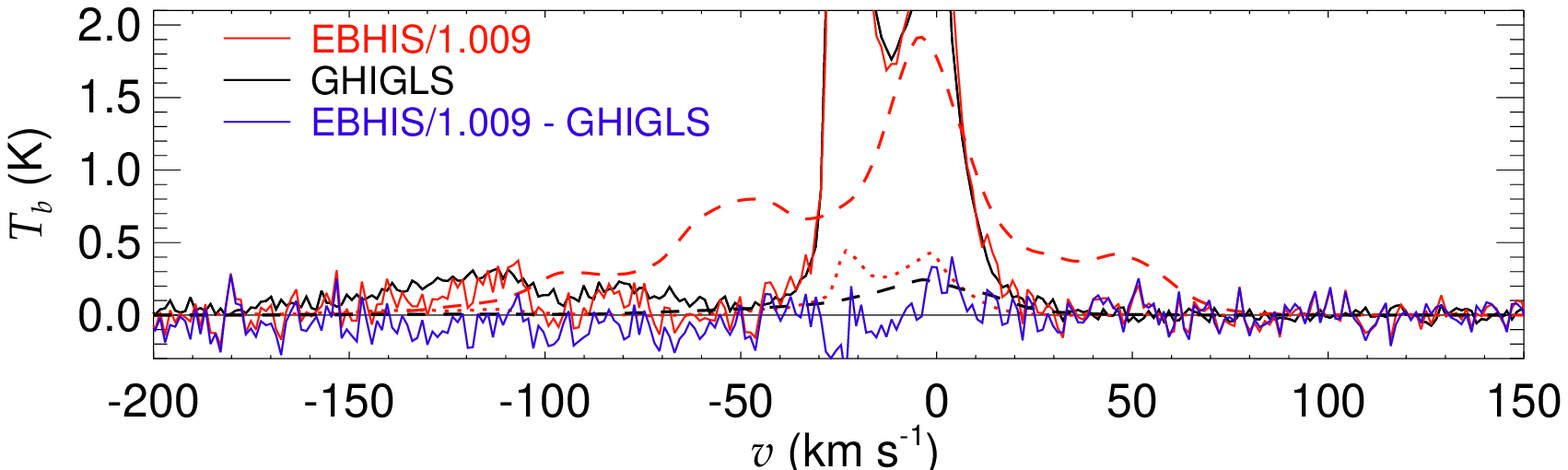}
\caption{
Same as Figure~\ref{fig:avgspectrumebhis} but for individual spectra.
Top panel: NEP spectrum at $(99\pdeg7, 27\pdeg0)$ peaking at 18.7~K.
Following panels: \ncpeb\ spectrum at $(137\pdeg6, 38\pdeg1)$ peaking
at 3.2~K and DRACO spectrum at $(90\pdeg0, 38\pdeg7)$ peaking (in the
IVC) at 17.7~K.
}
\label{fig:singlespectrumebhis}
\end{figure}

Individual spectra, though noisier, show similar features.
Figure~\ref{fig:singlespectrumebhis} compares an EBHIS spectrum with
the corresponding (convolved and regridded) \ghigls\ spectrum for
three distinct lines of sight.  The \nep\ spectrum has strong relative
emission in the LVC range.  The \ncpeb\ spectrum was selected for its
relatively large \ghigls\ predicted stray radiation spectrum.  The
DRACO spectrum has strong relative emission in the IVC range.  The
difference spectra confirm that the overall agreement is very good.

\subsection{Comparisons of Maps of \wh}
\label{Wcontributionsebhis}

In this section we look for evidence in the \ghigls\ spectra for
errors that might be attributable to uncertainties in the baselines
that were fit and subtracted or to imperfect predictions of the stray
radiation spectrum that has been subtracted.  The EBHIS survey
provides an independent basis for this assessment.

Each of these two contributions is somewhat correlated from one
channel to the next and thus any errors are more easily seen in the
line integral \wh.  Additionally, the velocity dependence of these
components suggests that any \wh\ comparison should be made over
restricted ranges of velocity (e.g., using LVC, IVC, and HVC
components as defined in Table~\ref{compvel_table}).  Consequently we
produce a series of \wh\ maps corresponding to these velocity
components.  

In the analysis below of these maps we find some evidence
for errors relating to the subtraction of baselines and stray
radiation. These errors are at a low level, consistent with the slight increase
of $\sigma_{N_{\rm H I}}$ over the value expected from noise in the line
emission alone (Appendix~\ref{quality}).  Overall the \ghigls\ and EBHIS
maps are in good agreement.

\begin{figure*}
\centering
\includegraphics[width=14.0cm]{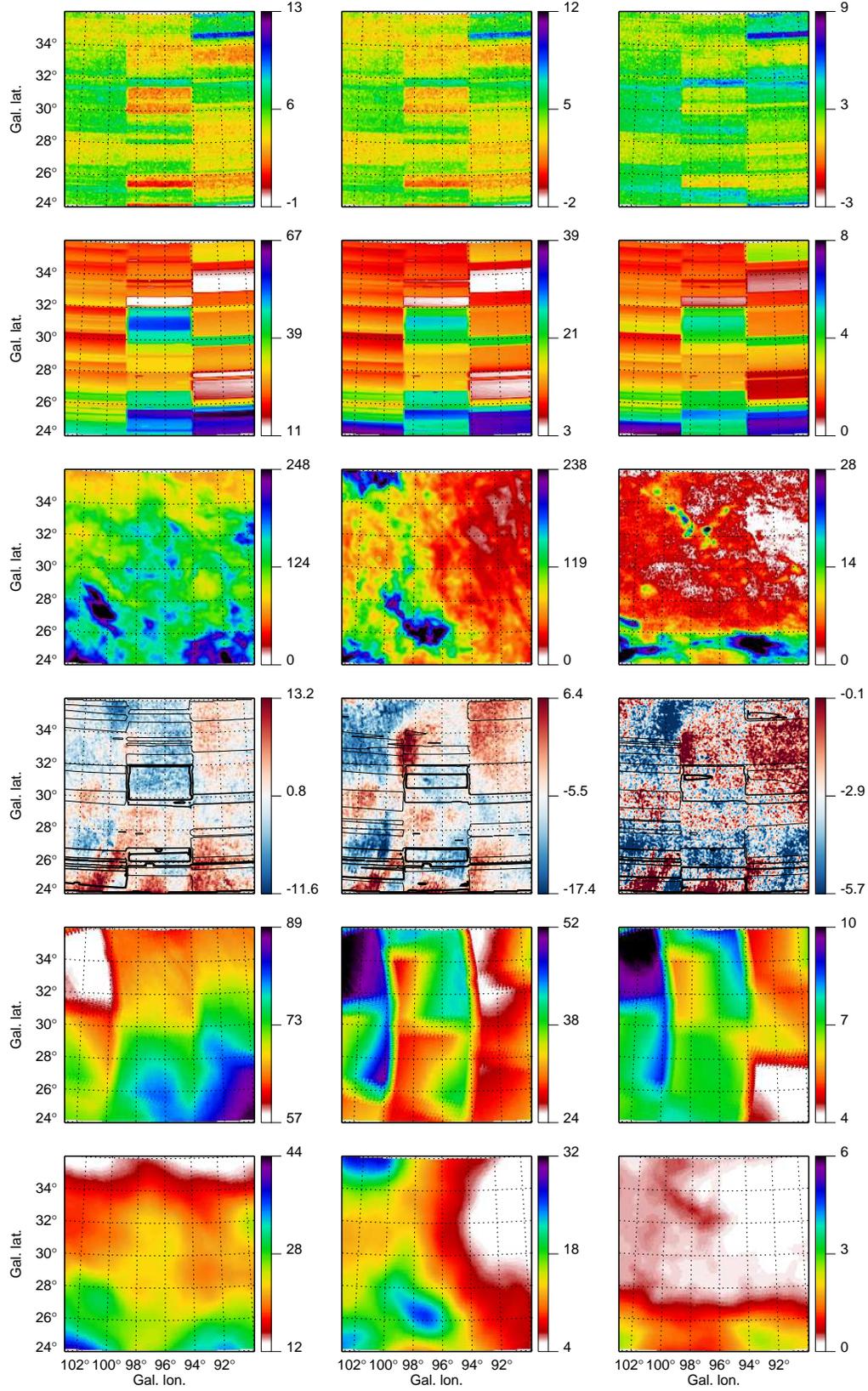}
\caption{
\wh-related maps of the NEP field (see text in Section~\ref{Deb} for
details).  Columns (left to right) are for integrals over the LVC,
IVC, and HVC velocity intervals.  Rows (top to bottom) are from the
following data: \ghigls\ baselines (subtracted in producing \ghigls\
\wh), \ghigls\ stray radiation (subtracted), \ghigls\ \wh\ (as in
Figure~\ref{fig:nepWmaps}), \wh\ residual from scaled \EBHISmGBT,
EBHIS far sidelobe (subtracted in producing EBHIS \wh), and EBHIS near
sidelobe stray radiation (subtracted).
On the residual map are overlaid contours from the \ghigls\ stray
radiation map (from thinnest to thickest at 25\,\%, 50\,\%, and 75\,\%
of the maximum value) for reference and orientation.
}
\label{fig:WcomponentsNEP}
\end{figure*}

\begin{figure*}
\centering
\includegraphics[width=18.5cm]{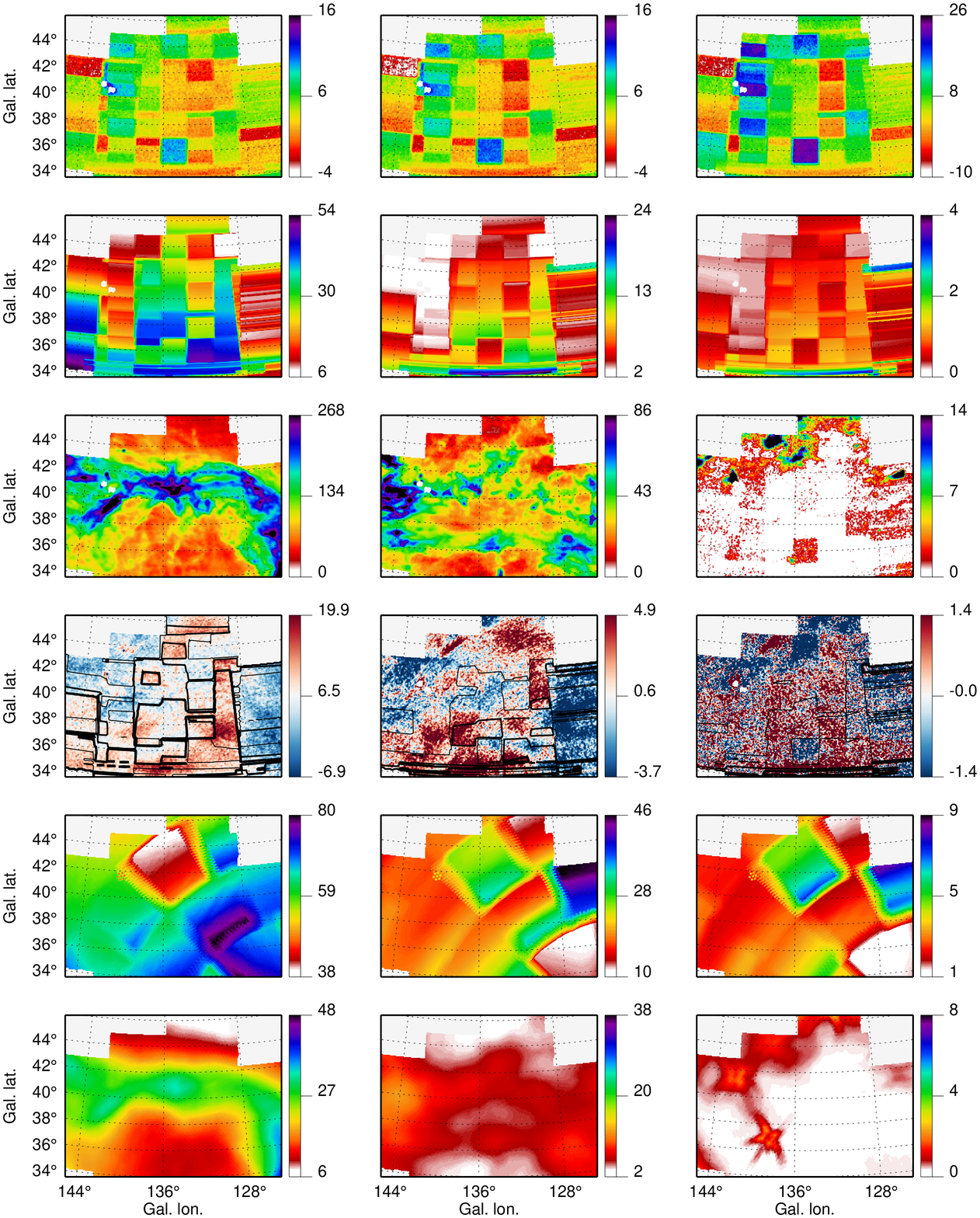}
\caption{
Same as Figure~\ref{fig:WcomponentsNEP} for \ncpeb.  In the
\ghigls-related maps, pixels are excluded where it was not possible to
fit a simple baseline because of emission from other galaxies present
in the (frequency-switched) spectrum.
Pixels near M81 and M82 where it is not possible to fit a baseline are
masked and not used in subsequent analysis.  A useful indicator of
scale is the pattern of the EBHIS multibeam system that appears
imprinted at one position in the maps of the far sidelobe emission in
row five.
}
\label{fig:WcomponentsSPIDER}
\end{figure*}

\subsubsection{Data}
\label{Deb}

Figure~\ref{fig:WcomponentsNEP} shows the data available for this
investigation. The field is \nep.
The columns (left to right) are for integrals over the LVC, IVC, and
HVC velocity intervals.
Unless otherwise indicated, in all of the panels the range for the
colorbar extends from the minimum to maximum of the data values and is
usually quite different from one panel to another.

The first row contains the \wh\ contributions from the baselines that
have been fit and subtracted in producing the \ghigls\ cubes.
The second row shows the contributions from the predicted stray
radiation for \ghigls\ that has also been subtracted.
In the third row are the \wh\ maps from \ghigls\ already seen in
Figure~\ref{fig:nepWmaps}. The range of the colorbar starts at zero
and extends to the 99\,\% percentile.
For the residual maps in the fourth row we have first computed the
\wh\ maps from the EBHIS spectra, adjusted as above to the \ghigls\
scale, and then subtracted the \ghigls\ \wh\ maps.  Here the colorbar
is centered on the mean and has a total range that is 10\,\%\ (20\,\%
for HVC) of that in the \wh\ map above it.  A different color table
has been used to emphasize the positive and negative excursions.
The fifth and sixth rows show the contributions from the predicted
stray radiation for the far and near sidelobes that have been
subtracted in producing the EBHIS cubes.  The near sidelobe stray
radiation map is like a blurred version of the original \wh\ map in
row three and is therefore roughly correlated with the signal.  Note
that the dynamic range spanned by the colorbar is set to be the same
as for the far sidelobe map, revealing that the corrections for the
near sidelobe emission are generally much smaller.

Figure~\ref{fig:WcomponentsSPIDER} contains the corresponding
\wh-related maps for \ncpeb.

Even more so than in Figure~\ref{fig:efchannelebhis} there are different
\GPs\ that are obvious in rows one and two and in rows five and six.
The \GPs\ are related to the scanning strategies of the respective
surveys and how the observation blocks were organized and scheduled.
Given their origins the \GPs\ are not correlated with the astronomical
\wh\ signal in row three, except as already noted for the EBHIS near
sidelobe stray radiation in row six.

A strategy for identifying sources of systematic effects/errors is to
search for corresponding telltale patterns like the \GPs\ in the
residual maps, i.e., is there a ``smoking gun" revealing anything
amiss with any of these corrections?

\subsubsection{Residual Map}
\label{Reb}

As can be seen from the relative dynamic ranges of the maps, the
baseline and stray radiation corrections that have been subtracted
from the \ghigls\ and EBHIS data are significant.  To the extent that
they have been accurately calculated -- and because the data from the
two telescopes have been set to a common calibration scale -- the
residual maps should appear structureless and centred on zero,
reflecting solely the combined \wh\ noise.  This is not quite the
case.

The residual map is centered on zero for the LVC component in NEP and
the IVC and HVC components in \ncpeb, and there are broad areas in
other maps where this is the case too.  But when the average residual
spectrum in Figure~\ref{fig:avgspectrumebhis} is systematically
non-zero over a considerable velocity range, for example the negative
residual in the IVC and HVC components in NEP, then this will also be
the case for the mean of the \wh\ residual map.

On the residual map we show contours from the \ghigls\ stray radiation
maps to highlight the patchwork of largely rectangular patterns
relating to the observing blocks (the \GPs) for reference and orientation.  In
size and orientation the \GPs\ for EBHIS are quite different from
those for \ghigls\ (see also Figure~\ref{fig:efchannelebhis}).

The residual map is defined as scaled \EBHISmGBT.  \ghigls\ has been
produced from the original data minus the corrections in the top two
rows and so if a correction there were too large it would produce an
excess in the residual map.  Similarly EBHIS has been produced from
the original data minus the corrections in the bottom two rows and so
if a correction there were too large it would produce a deficit in the
residual map.

\subsubsection{Noise in the Line Emission}
\label{Neb}

The noise in the line emission of the two surveys, of order
$\sigma_{\rm ef} \times [1.0 + \Trb(v)$/(20\,K)] (see
Appendix~\ref{tnoise}), accumulates as $\sqrt{\nch} \Delta v$ to
contribute fluctuations $\delta W \sim 0.9$\,\Kkms\ in the residual
maps for NEP (0.6 to 1.3\,\Kkms\ depending on the field and velocity
component, the larger values in the more extensive HVC
intervals).\footnote{
In the \ghigls\ \wh\ maps themselves the fluctuations would be $\sim
0.3$\,\Kkms\ for NEP (0.2 to 0.7\,\Kkms\ overall).
}
These fluctuations are small compared to the range shown in the
residual maps, but do contribute to the graininess everywhere.  Note
that the noise in the line emission would not produce any systematic
offset in the maps.

\subsubsection{Errors Related to Fitted Baselines}
\label{Beb}

The \ghigls\ baselines tend to be fairly stable within a given
scheduled observing block, and this is reflected in the rectangular
\GPs\ in the first row of maps. The measured standard deviation within
a rectangle, typically 0.5\,\Kkms\ in NEP (0.3 to 1.1\,\Kkms\
overall), contributes to the graininess in the residual maps.  The
measured standard deviation in large smooth areas of the residual maps
is typically 1.5\,\Kkms\ in NEP (1.5 to 1.9\,\Kkms\ overall).  Our
interpretation is that this can be accounted for by the noise in the
line emission and the baseline fluctuations, including those of EBHIS
which we have not attempted to quantify.  However, we note that there are
low-amplitude striped patterns in the EBHIS and residual maps that can
be discerned along the distinctively oriented scan lines of EBHIS,
with a spacing related to the multibeam system, for example in the IVC
component in NEP and the LVC and IVC components in \ncpeb.  The
measured peak-to-peak amplitude of the pattern is small, of order
2\,\Kkms.

The \ghigls\ baselines do vary from observing block to
observing block.  Just as the \ghigls\ baselines are fairly stable
within the rectangular \GPs, any errors in the mean offsets within
these rectangles are likely to be correlated, which could produce a
corresponding pattern in the residual map.  As discussed below, the
magnitude of such rectangle to rectangle systematic errors is of the
same order as the above-mentioned fluctuations and therefore not
normally readily discerned in the residual maps given the range of
values therein.
However, the stray radiation correction and the range in the residual
maps generally decrease steadily from LVC to HVC, making HVC \wh\ maps
the most favourable for investigating the baseline errors.

A special case to examine is the large 60 deg$^2$ region in
\ncpeb\ ranging in $l$ from 125\deg\ and 145\deg\ and in $b$ from
34\deg\ to 40\deg\ in which there is virtually no HVC emission
detectable (see the column of HVC maps in
Figure~\ref{fig:WcomponentsSPIDER}).  In the \ghigls\ \wh\ map there
are clear \GPs\ relating to the observation blocks used, whereas
the EBHIS \wh\ map (not shown) is featureless, and so this rectangular
pattern appears (in reverse) in the residual map.  Within rectangles in
the \ghigls\ \wh\ map the standard deviation is 0.5\,\Kkms\ whereas
the standard deviation for the region as a whole is 1.0\,\Kkms, from
which we deduce that the typical dispersion in the mean amplitudes of
rectangles is 0.9\,\Kkms.  We also note that even with these clearly
visible \GPs\ in the \ghigls\ \wh\ map, the standard
deviation is still slightly smaller that it is in the smooth EBHIS
\wh\ map (1.1\,\Kkms).
The \ghigls\ stray radiation correction map has the same \GPs, but its
overall low level ($0.7 \pm 0.3$\,\Kkms) and the fact that it is
uncorrelated with the observed pattern in \wh\ make it an unlikely
source.  On the other hand there is a weak positive correlation of the
pattern in \wh\ with that in the baseline correction map.  Thus we
conclude that small observing block to observing block errors in the
baseline correction with dispersion of about 1\,\Kkms\ are the likely
source of the pattern; this would correspond to a rectangle to
rectangle dispersion in mean offsets in $\Trb$ of 8\,mK over this
entire HVC velocity interval.  The origin of this is unknown but could
be related to the general problem of interpolating a low-order
function over a large range.  Even when the function is
well-constrained by the data in many emission-free end channels, the
large range over which the function is interpolated introduces an
error.
For completeness, we note that the average \wh\ in this HVC region is
0.04\,\Kkms\ for \ghigls\ and 0.34\,\Kkms\ for EBHIS.  This small
difference of 0.3\,\Kkms, the mean of the residual map in this region,
is in accord with the difference in the average spectrum (like
Figure~\ref{fig:avgspectrumebhis}); this difference could be due to
small errors in baselines and/or in the stray radiation corrections.

Where there is faint but significant HVC emission detected by \ghigls\
in large regions of NEP above 27\deg, and also in DRACO, we find a
negative residual, i.e., EBHIS underestimates this emission by a few
\Kkms.  We interpret this as a result of the different sensitivities
of the two surveys combined with modelling of the baselines.  The
\ghigls\ residual baseline is fit iteratively with a third-degree
polynomial \citep{boot11}, whereas the EBHIS baseline is determined
using an iterative Gaussian smoothing technique \citep{wink10} applied
over their entire bandwidth of 100~MHz.  Note that the EBHIS data were
observed with a frequency-switching of 3~MHz, but due to the gain
varying as a function of frequency, the data were not reduced using
the frequency-switching technique \citep{wink10}; thus the need for a
more sophisticated algorithm to remove the baseline.
Empirically, the sign of the offset indicates that the lower
sensitivity (higher noise) of the EBHIS spectra may result in a fitted
baseline that eliminates part of what in \ghigls\ is detected as HVC
emission rather than elevated baseline.  Likewise, the \ghigls\ data
baseline fitting could be removing signal that with more sensitive
observations would lead to the detection of resolved structure.

As mentioned, baseline fitting procedures are not as successful at
determining and removing a baseline when the emission from galaxies
dramatically reduces the number of emission-free channels and/or
significantly dominates the shaping of the baseline.

\subsubsection{Stray Radiation}
\label{SReb}

In Figure~\ref{fig:avgspectrumebhis} at HVC velocities the average
residual spectrum contains only low frequency oscillations, consistent
with a low-order polynomial used to model baselines.  This can be
contrasted with the sharper oscillations in the LVC and IVC ranges,
which must derive from errors other than the baseline.  This sharper
structure, also discernible in the residuals of individual spectra in
Figure~\ref{fig:singlespectrumebhis}, will affect the LVC and IVC \wh\
residual maps.  The residuals for these components are larger than for
the HVC.  Furthermore, because of the larger dynamic range of the
stray radiation corrections in the LVC and IVC velocity intervals,
these intervals are the most favorable for looking for any errors
arising because of uncertainties in the stray radiation corrections.

For \ghigls\ there is no direct spatial correlation between the stray
radiation and \wh\ because of the unblocked geometry of the GBT
design.  Instead, for intermediate latitude fields such as these the
amount of stray radiation is strongly influenced by the time at which
any given observation is made because of how the offset spillover
sidelobe beam pattern \citep{boot11} overlaps (or not) with the
stronger emission near the Galactic plane.  This results in a
rectangular pattern in the stray radiation map relating to the
observational blocks used.  The stray radiation corrections for EBHIS
are also significant and the maps have quite different morphological
structure.

There are some smoking guns.  In the NEP residual map for IVC in
Figure~\ref{fig:WcomponentsNEP} there are triangular shapes that are
clearly anti-correlated in amplitude with the EBHIS far sidelobe stray
radiation correction.  To a lesser extent there is a positive
correlation discernible for LVC.  In the \ncpeb\ residual map for LVC
in Figure~\ref{fig:WcomponentsSPIDER} there are rectangular shapes
with a clear positive correlation in amplitude with the \ghigls\ stray
radiation correction.  This persists somewhat in the IVC.

Attempting some quantification and motivated by this suggestion of
multiplicative factors, we sought to reduce the standard deviation of
the residual maps by removing correlations with various combinations
of the stray radiation maps.
For example, in the case of IVC in NEP, a factor $-0.4$ of the EBHIS
far sidelobe correction is indicated (in the case of LVC the factor is
$+0.2$).  However, while this lowers the standard deviation of the
residual map from 4.6 to 3.7\,\Kkms\ it is clearly not the whole
story.  There is still a rectangular pattern in the revised residual map that
reveals the \ghigls\ observational scheduling blocks, but it is not
multiplicatively related at a significant level ($> -0.1$).  Telltale
signs of the \GPs\ of both \ghigls\ and EBHIS stray
radiation corrections in the revised residual maps indicate
clearly that there are substantial additive and/or subtractive errors
too, which make any multiplicative factor uncertain.  Accounting for
these errors is even more important than the multiplicative changes to
reduce the standard deviation in the residual to a level about
1\,\Kkms\ that could be expected from baseline fitting errors and line
noise alone.
If we suppose that the entire 4.6\,\Kkms\ dispersion is roughly
equally apportioned to \ghigls\ and EBHIS (thus 3.2\,\Kkms\ each),
then this amounts to about a 10\,\% fraction of the mean stray
radiation corrections (an even small fraction in the case of LVC).
However, we feel that it is probably better to think of the errors in
absolute terms rather than as a fractional error.

For our \ncpeb\ LVC example, a factor $+0.2$ of the \ghigls\ stray
radiation correction is indicated (the stray radiation correction has
been overestimated).  In this case this appears to be much larger than
the estimated \wh\ stray radiation uncertainty of a factor 0.07 found
by \citet{boot11}.  But again the standard deviation of the residual
map is only lowered from 4.5 to 3.6\,\Kkms\ and there are still
telltale signs of the schedule-related \GPs\ of both \ghigls\ and
EBHIS stray radiation corrections in the original and revised residual
maps.

We have carried out a similar comparative study in DRACO.  In the
\ghigls\ observations of this field, which is a relatively faint
extension off the upper right of the much larger NEP field, the stray
radiation corrections are fortuitously at the low end of the values
encountered in the NEP observations.  However, this makes it more
difficult to spot any errors in these corrections.  On the other hand
the EBHIS far sidelobe stray radiation correction is quite large and
its pattern is discernible particularly on the IVC residual map.  But
because the standard deviation of the residual map is already so low
(2.6\,\Kkms) it is difficult to draw any definitive conclusions,
except of course that the corrections applied in both surveys are very
good.  This can be appreciated as well in the agreement of the mean
spectra in Figure~\ref{fig:avgspectrumebhis}.

\section{Comparison with Data from GASS}
\label{gass}

Here we compare our data with that from the southern hemisphere-based
\hi\ Galactic All-Sky Survey (GASS) \citep{mccl09,kalb10,kalb15}
available in the GASS~III data release repository.\footnote{
\url{www.astro.uni-bonn.de/hisurvey/}
}
Because of the latitude coverage of the GASS data and our choice of
primarily northern circumpolar fields for the surveys with the GBT
ACS, we have only one field available for this comparison, MC.  In
extracting the GASS cube we adopted an optimal modified Bessel
function mapping of the data onto a $4\farcm8$ grid using the gridding
parameters recommended by \citet{mang07}.  This is similar but not
identical to our modified Bessel function mapping of the \ghigls\
spectra onto a GLS grid (see Section~\ref{mapping}).  The downloaded
GASS cube has angular resolution $\sim16$\arcmin\ (modified from the
original average $14\farcm4$ due to our gridding choice) on a
Cartesian Equatorial grid, 0.82~\kms\ channels, and an
empirically-determined noise of $\sigma_{\rm ef} \sim47$~mK.

The \ghigls\ data cube was convolved to this lower resolution and
regridded in all dimensions.  This reduces the noise from the original
83~mK (Table~\ref{f_table}) to an effective noise 17~mK.

Following the same analysis as in Appendix~\ref{Wratiomapsebhis},
these GASS data appear to be brighter than \ghigls\ by a factor
GASS/\ghigls\ $= 1.062 \pm 0.019.$ This is in agreement with the
conclusion arrived at independently by \citet{kalb15} who compared
GASS to EBHIS and LAB (LDS).

\begin{figure}
\centering
\includegraphics[width=1.0\linewidth]{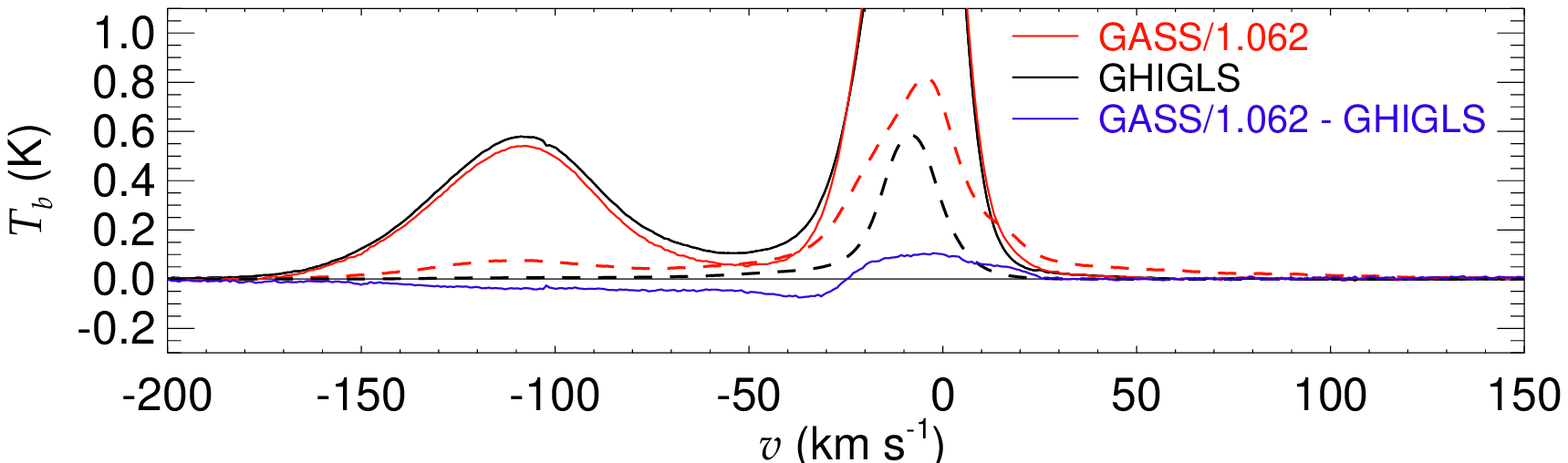}
\includegraphics[width=1.0\linewidth]{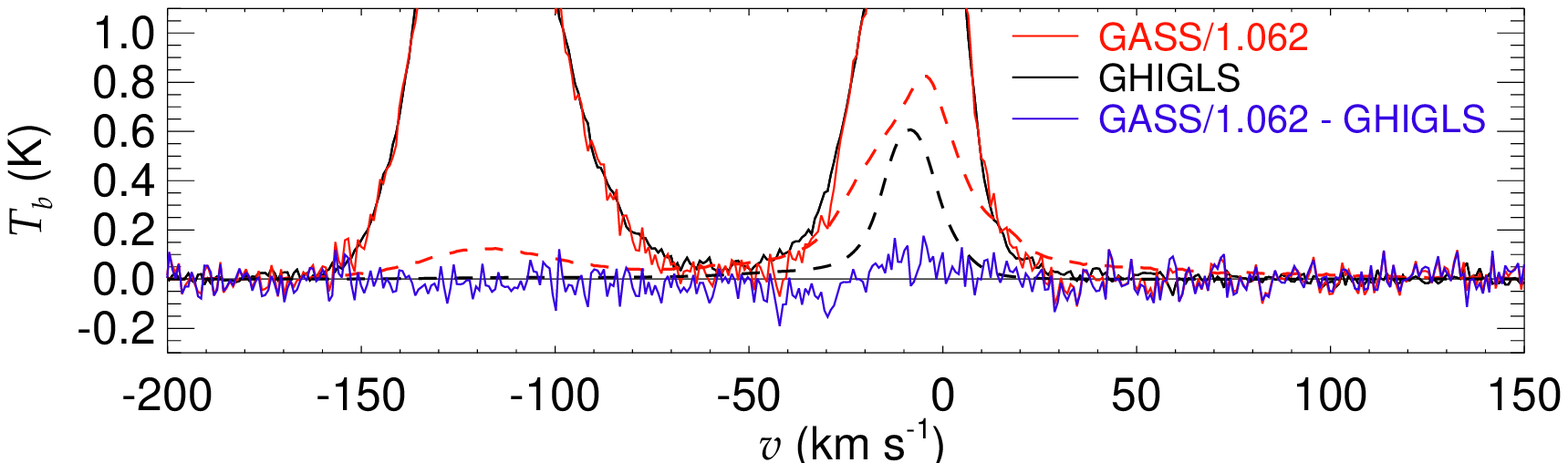}
\caption{
Top panel: Average \ghigls\ spectrum for MC (black), average scaled
GASS spectrum (red), residual \GASSmGBT\ (blue).  Dashed line spectra
are the average stray radiation predictions that have been removed
from the data.  The MC average spectrum peaks at 3.2~K.
Bottom panel: Individual MC spectrum at $(59\pdeg4, -81\pdeg5)$ with
relatively strong HVC emission peaking at 2.5~K (LVC peaks at 3.0~K).
Note the different $\Trb$ range as compared to
Figures~\ref{fig:avgspectrumebhis} and \ref{fig:singlespectrumebhis}.
}
\label{fig:spectrumgass}
\end{figure}

On adjusting the GASS data to the \ghigls\ scale we again find
remarkable agreement as demonstrated in the spectra in
Figure~\ref{fig:spectrumgass}.  Also shown there are the stray
radiation corrections that have been applied to the \ghigls\ and the
GASS data.

\begin{figure*}
\centering
\includegraphics[width=18.5cm]{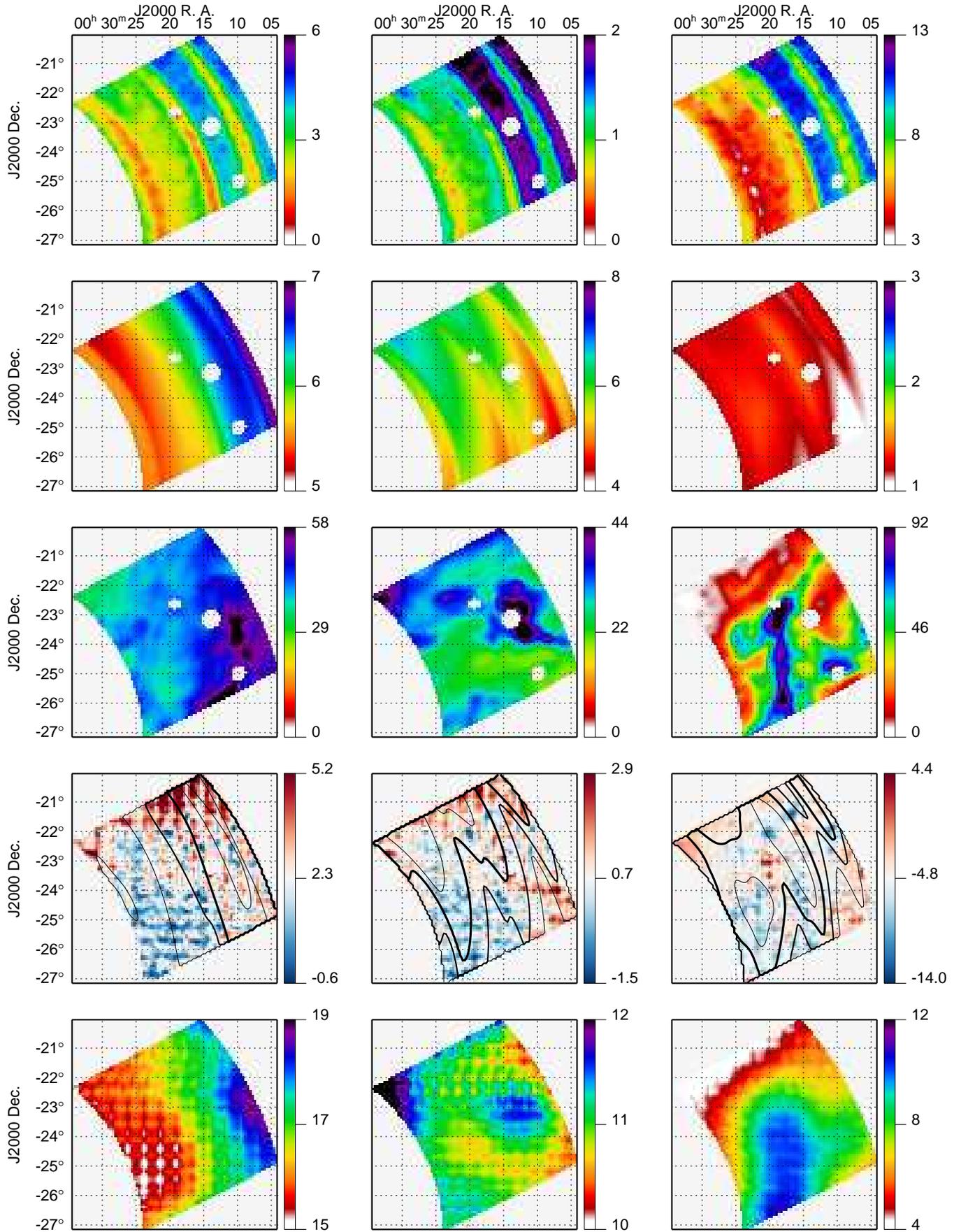}
\caption{
Same as Figure~\ref{fig:WcomponentsNEP} for MC and comparison with
GASS rather than EBHIS.  There is only one stray radiation correction
cube for GASS.  Note the areas masked because of the effects of the
galaxies MCG-04-02-003, NGC~0045, and NGC~0024 (from left to right).
}
\label{fig:WcomponentsMC}
\end{figure*}

As in Appendix~\ref{Wcontributionsebhis} we produced a series of \wh\
maps for each velocity component (Table~\ref{compvel_table}) and
present them in Figure~\ref{fig:WcomponentsMC}.  The residual maps
confirm the overall good agreement of the two data sets.

In the HVC part of the spectrum (the largest range, adopting
$-153.4$\,\kms\ to $-32.9$\,\kms\ from Table~\ref{compvel_table}) the
residual in the average spectrum (Figure~\ref{fig:spectrumgass}) is
negative and therefore so is the residual map for that component,
which has a mean $-4.8$\,\Kkms.  The \ghigls\ stray radiation
correction is small by comparison and there is no clear imprint of the
morphology of the GASS stray radiation correction in the \wh\ residual
map.  This suggests that the origin of the offset might be in the
modelling of the baselines.  The GASS instrumental baseline was
removed by an iterative procedure \citep{kalb10,kalb15} using either a
9th- or 11th-order polynomial over a much larger bandwidth than used
for \ghigls\ for which a third-degree polynomial could be used
\citep{boot11}.  With the lower sensitivity of the GASS measurements
the fitting of the baseline between the main emission peaks in the
profile might include what in \ghigls\ is detected as emission, thus
raising the baseline and lowering the apparent signal, consistent with
the sign of the offset.

Over the IVC range (the smallest, $-32.9$\,\kms\ to $-8.8$\,\kms) the
residual in the average spectrum changes sign producing a net effect
consistent with the near-zero mean ($+0.7$\,\Kkms) in the \wh\
residual map.  The dynamic range in both stray radiation corrections
is small and in the residual map there is no telltale sign of errors
in these corrections.  However, the location and abruptness of the
sign change in the residual spectra is suggestive of small errors in
the stray radiation corrections, rather than in the baselines.

Over the LVC range ($-8.8$\,\kms\ to $+48.3$\,\kms) the positive
residual in the average spectrum results in a slightly positive mean
($+2.3$\,\Kkms) in the \wh\ residual map.  There are again no
unequivocal signs of errors in the stray radiation corrections that
have been applied.

In the GASS data, including the noise map and the stray radiation
corrections and propagating to the \wh\ residual maps, there is a slight
hatching pattern related to the orientation of the GASS scans and the
multibeam system, but this has no bearing on the other considerations
here.








\end{document}